\documentstyle[aps,amsfonts,amssymb,amsmath]{revtex}

\newcounter{bean}
\newcommand{\blist}{\begin{list}{(\roman{bean})}{\usecounter{bean}}}
\newcommand{\elist}{\end{list}}

\newcommand{\mysection}[1]{\section{#1}
                           \setcounter{equation}{0}\setcounter{pro}{0}}
\newtheorem{pro}{Lemma}

\newcommand{\bea}{\begin{eqnarray}} 
\newcommand{\eea}{\end{eqnarray}}
\newcommand{\beann}{\begin{eqnarray*}} 
\newcommand{\eeann}{\end{eqnarray*}}
\newcommand{\ba}{\begin{array}}
\newcommand{\ea}{\end{array}}
\newcommand{\ben}{\begin{enumerate}} 
\newcommand{\een}{\end{enumerate}}
\newcommand{\bit}{\begin{itemize}} 
\newcommand{\eit}{\end{itemize}}

\newcommand{\cA}{{\cal A}}

\newcommand{\cC}{{\cal C}}
\newcommand{\cD}{{\cal D}}

\newcommand{\cF}{{\cal F}}

\newcommand{\cH}{{\cal H}}

\newcommand{\cO}{{\cal O}}
\newcommand{\cP}{{\cal P}}

\newcommand{\cT}{{\cal T}}

\newcommand{\5}{\bar }  
\newcommand{\6}{\partial } 
\newcommand{\7}{\hat } 
\newcommand{\4}{\tilde }

\newcommand{\sfrac}[2]{\mbox{$\frac{{#1}}{{#2}}$}\,}
\newcommand{\f}[3]{{f_{#1#2}}^{#3}}

\newcommand{\Ii}{{{\mathrm i}}}

\newcommand{\vep}{\varepsilon}

\newcommand{\da}{{\dot\alpha}} 
\newcommand{\dbe}{{\dot\beta}}
\newcommand{\dg}{{\dot\gamma}}

\newcommand{\then}{\Rightarrow} 
\newcommand{\LRA}{\Leftrightarrow} 

\newcommand{\dr}{\raise.3ex\hbox{$\stackrel{\leftarrow}{\delta}$}{}}
\newcommand{\dl}{\raise.3ex\hbox{$\stackrel{\rightarrow}{\delta}$}{}}

\newcommand{\af}{{\mathrm af}}

\newcommand{\gh}{{\mathrm gh}}

\newcommand{\lieg}{{\mathfrak g}_{{\mathrm YM}}}
\newcommand{\liefull}{{\mathfrak g}}
\newcommand{\TR}{{\mathrm tr}_\Lindex}
\newcommand{\TRprime}{{\mathrm tr}_{\Lindex'}}
\newcommand{\Space}{{\mathfrak F}}
\newcommand{\CONST}{{\mathfrak E}}
\newcommand{\Wspace}{{\mathfrak W}}
\newcommand{\inv}{{\mathfrak T}_{{\mathrm inv}}}

\newcommand{\om}{{\overline{m}}}
\newcommand{\um}{{\underline{m}}}

\newcommand{\unk}{{\underline{k}}}

\newcommand{\hatlambda}{\5{\7\lambda}}
\newcommand{\hatpsi}{\,\5{\!\7\psi}}
\newcommand{\hatchi}{\,\5{\!\7\chi}}

\newcommand{\proof}[1]{{\bf Proof of lemma \ref{#1}: }}
\newcommand{\comment}{{\bf Comment: }}
\newcommand{\comments}{{\bf Comments: }}
\newcommand{\QED}{\bigskip\\}
\newcommand{\lra}{\leftrightarrow}
\newcommand{\rankg}{{\mathrm rank}(\lieg)}
\newcommand{\free}{{{\mathrm free}}}
\newcommand{\Abel}{{{\mathrm Abel}}}
\newcommand{\Lth}{\theta_{{\mathrm L}}}
\newcommand{\Kill}{\delta}
\newcommand{\CC}{{\mathrm c.c.}}

\newcommand{\ext}{s_{{\mathrm ext}}}
\newcommand{\gammaext}{\gamma_{{\mathrm ext}}}
\newcommand{\soneext}{s_{{\mathrm ext},1}}
\newcommand{\gauge}{s}
\newcommand{\lie}{s_{{\mathrm lie}}}
\newcommand{\susy}{s_{{\mathrm susy}}}
\newcommand{\curv}{s_{{\mathrm curv}}}

\newcommand{\FD}{A}          
\newcommand{\Index}{a}       
\newcommand{\Jndex}{b}       
\newcommand{\Kdex}{c}        
\newcommand{\Ldex}{d}        
\newcommand{\Tindex}{\tau} 
\newcommand{\Windex}{I}      
\newcommand{\indec}{s}       
\newcommand{\ifree}{{i_{\mathrm f}}}
\newcommand{\jfree}{{j_{\mathrm f}}}
\newcommand{\kfree}{{k_{\mathrm f}}}
\newcommand{\lfree}{{l_{\mathrm f}}}
\newcommand{\gindex}{M}      
\newcommand{\hindex}{N}      
\newcommand{\iindex}{P}      
\newcommand{\Lindex}{r}      
\newcommand{\Pindex}{\Gamma} 

\begin{document}

\title{
\begin{flushright}
\normalsize{
MPI-MIS-105/2002\\
hep-th/0212070\vspace{4ex}
}
\end{flushright}
Extended BRST cohomology, consistent
deformations and anomalies of four-dimensional
supersymmetric gauge theories
}

\author{
Friedemann Brandt
}

\address{ 
Max-Planck-Institute for Mathematics in the Sciences,
Inselstra\ss e 22-26, D-04103 Leipzig, Germany}

\maketitle

\begin{abstract}
The local cohomology of an extended BRST differential which
includes global N=1 supersymmetry and Poincar\'e transformations is 
completely and explicitly computed
in four-dimensional supersymmetric gauge theories
with super-Yang-Mills multiplets, 
chiral matter multiplets and linear multiplets containing
2-form gauge potentials.
In particular we determine to first order all 
N=1 supersymmetric and Poincar\'e invariant
consistent deformations
of these theories 
that preserve the N=1 supersymmetry algebra
on-shell modulo gauge transformations, and all Poincar\'e invariant
candidate gauge and supersymmetry anomalies.
When the Yang-Mills gauge group is semisimple and no linear
multiplets are present, we find that all such deformations 
can be constructed from standard superspace integrals and
preserve the supersymmetry transformations in a formulation
with auxiliary fields,
and the candidate anomalies are exhausted by supersymmetric 
generalizations of the well-known chiral anomalies.
In the general case there are 
additional deformations and candidate anomalies 
which are
relevant especially to the deformation of free
theories and the general classification of
interaction terms in supersymmetric
field theories.
\end{abstract}

\vspace{10ex}

\begin{center}
\begin{tabular}{cl}
&Contents\\[4pt]
\ref{intro}& Introduction \dotfill \pageref{`i'}\\
\ref{BRST}& Field content, Lagrangian, 
            extended BRST transformations\ldots\pageref{`ii'}\\
\ref{DESCENT} & Cohomological problem and descent equations 
                \dotfill \pageref{`iii'}\\
\ref{CALC} & Change of variables \dotfill \pageref{`iv'}\\
\ref{algebra}& Gauge covariant algebra \dotfill  \pageref{`v'}\\
\ref{ALL}   & Computation of the cohomology \dotfill \pageref{`vii'}\\
            & 
              Cohomology for small ghost numbers \dotfill \pageref{`vi'}\\
            & 
              Elimination of trivial pairs \dotfill \pageref{`vii.1'}\\
            & 
              Decomposition of the cohomological problem
                  \dotfill \pageref{`vii.2'}\\
            & 
              Lie algebra cohomology \dotfill \pageref{`vii.3'}\\
            & 
              Supersymmetry algebra cohomology \dotfill \pageref{`vii.4'}\\
            & 
              Completion and result of the computation
               \dotfill \pageref{`vii.5'}\\
\ref{DEFOS} & Consistent deformations, counterterms and anomalies
              \dotfill \pageref{`viii'}\\
\ref{negative} & Remarks on the cohomology in negative ghost numbers
              \dotfill \pageref{`ix0'}\\
\ref{otherform} & Other formulations of supersymmetry \dotfill \pageref{`ix'}\\
            & 
              Formulation with auxiliary fields
              \dotfill \pageref{`ix.1'} \\
            & 
              Formulation with linearly realized supersymmetry
              \dotfill \pageref{`ix.2'} \\
\ref{otheraction} & Other Lagrangians \dotfill \pageref{`x'}\\
\ref{conclusion} & Conclusion \dotfill \pageref{`xi'}\\
\ref{conv} & Conventions and useful formulae \dotfill \pageref{`A'}\\
\ref{proofs} & Proofs of the lemmas \dotfill \pageref{`D'}
\end{tabular}
\end{center}

\newpage


\mysection{Introduction}\label{`i'}\label{intro}

This work closes a gap in the analysis of
four-dimensional globally supersymmetric gauge theories
by deriving completely the local cohomology 
of an extended BRST differential which includes
N=1 supersymmetry and the Poincar\'e symmetries
in addition to the standard ingredients related to
the gauge symmetries and the field equations.
We analyse theories with super-Yang-Mills multiplets
for all compact gauge groups,
chiral matter
multiplets in linear representation of these groups,
and linear multiplets containing 2-form gauge potentials.
In particular this includes 
super-Yang-Mills theories with 
arbitrarily many Abelian gauge fields, and free supersymmetric theories with
any number of vector gauge fields and 2-form gauge potentials.
The extended BRST differential involves thus
constant ghosts for
N=1 supersymmetry and the Poincar\'e symmetries,
ghost fields for the gauge symmetries,
ghost-for-ghost fields for the reducible gauge symmetries
of the 2-form gauge potentials,
and antifields for the field equations, Noether identities
and reducibility relations between the Noether operators.

Until now the cohomology under study was only
examined for super-Yang-Mills theories
(with chiral matter multiplets but without linear multiplets)
in the restricted space of functionals with integrands of mass
dimension four or smaller than four\footnote{We refer here to
the dimension assignments given in Eq. (\ref{dimensions}).} 
and ghost numbers zero or one
\cite{White:1992ai,Maggiore:1996gr}. The
restrictions on the space of functionals were
motivated by applications in the context of renormalization
of power counting renormalizable theories.
These applications are no longer the only arena
of interest for BRST cohomological investigations:
local BRST cohomology is now applied
also in nonrenormalizable and effective field theories 
\cite{Gomis:1996jp,Weinberg:1996kr}, in the analysis of
local conservation laws (characteristic cohomology) \cite{BBH1},
and in particular in the study of consistent deformations of 
classical field theories \cite{BH}%
\footnote{The classification of consistent deformations 
is particularly interesting for free theories because it yields the
possible interaction terms that can be added to these theories
in a manner consistent with
the gauge symmetries, N=1 supersymmetry and Poincar\'e invariance.}.
Our analysis covers these applications because
we shall compute the cohomology in the
space of all local functionals, without restrictions on the
dimension or ghost number.

The paper has been organized as follows.
In section \ref{BRST} we specify the theories under study
and the extended BRST differential. 
In section \ref{DESCENT} we define the cohomological
problem and relate it to other useful
cohomological groups.
In section \ref{CALC} we introduce suitable variables 
to compute the cohomology efficiently.
In section \ref{algebra} we discuss
the extended BRST transformations of these variables and derive
a related graded commutator algebra which is of crucial importance
for the cohomology. The cohomology and the main steps
of the computation are presented in section \ref{ALL}, 
the  results most important for algebraic renormalization, anomalies
and consistent deformations
in section \ref{DEFOS}.
In section \ref{negative} we comment on the  
cohomology in negative ghost numbers.
In section \ref{otherform} we show that our results
do not depend on the formulation of supersymmetry used
here, and in section \ref{otheraction} we discuss to which extend
they depend on the Lagrangian.
The main text ends with a brief conclusion in section \ref{conclusion}
and is supplemented by
two appendices, the first of which contains conventions and notation
(in particular a list of frequently used functions and operators
can be found here), while the second outlines 
proofs of lemmas given in the main text.

\mysection{Field content, Lagrangian, extended BRST transformations}
\label{`ii'}\label{BRST}

We denote the super Yang-Mills multiplets by
$A_\mu^i, \lambda_\alpha^i$, the
chiral multiplets by $\varphi^\indec, \chi_\alpha^\indec$,
and the linear multiplets by
$\phi^\Index, B_{\mu\nu}^\Index, \psi_\alpha^\Index$.
The index $i$ of the super Yang-Mills multiplets
refers to a basis of a reductive (= semisimple plus Abelian)
Lie algebra $\lieg$ whose semisimple part (if any) is compact
($\lieg$ is otherwise arbitrary; its semisimple or
Abelian part may vanish).
We shall assume that this basis of $\lieg$ has been
chosen such that the Cartan-Killing metric
on the semisimple part of $\lieg$ is proportional to the unit matrix.
The index $\indec$ of the chiral multiplets refers 
to a representation of $\lieg$ (which may be trivial, see below). The index
$\Index$ of the
linear multiplets is not related to $\lieg$.
The Yang-Mills ghost fields are denoted by $C^i$, the
ghost fields of the $B^\Index_{\mu\nu}$ by $Q_\mu^\Index$
and the corresponding ghost-for-ghost fields by $R^\Index$.
Furthermore we introduce antifields for all these fields and
constant ghosts $c^\mu$, $c^{\mu\nu}$
and $\xi^\alpha$ for global spacetime translations, Lorentz transformations
and supersymmetry transformations, respectively:
\begin{align}
\mbox{fields:}\quad \{\Phi^\FD\}=\{&
A_\mu^i,\lambda_\alpha^i,\5\lambda_\da^i,C^i,
\phi^\Index,B_{\mu\nu}^\Index,\psi_\alpha^\Index,\5\psi_\da^\Index,
Q_\mu^\Index,R^\Index,
\varphi^\indec,\5\varphi_\indec,\chi^\indec_\alpha,
\5\chi_{\indec\da}
\}
\nonumber\\
\mbox{antifields:}\quad \{\Phi^*_\FD\}=\{&
A^{*\mu}_i,\lambda^{*\alpha}_i,\5\lambda^{*\da}_i,C^*_i,
\phi^*_\Index,B^{*\mu\nu}_\Index,\psi^{*\alpha}_\Index,
\5\psi^{*\da}_\Index,
Q^{*\mu}_\Index,R^*_\Index,
\varphi^*_\indec,\5\varphi^{*\indec},\chi^{*\alpha}_\indec,
\5\chi^{*\indec\da}\}
\nonumber\\
\mbox{constant ghosts:}\quad \{&c^\mu,\xi^\alpha,\5\xi^\da,c^{\mu\nu}\}.
\nonumber\end{align}
$\lambda^i_\alpha$, $\psi^\Index_\alpha$ and $\chi^\indec_\alpha$ are
the components of complex Weyl spinor fields, 
$\5\lambda^i_\da$, $\5\psi^\Index_\da$ and $\5\chi_{\indec\da}$
denote their complex conjugates.
$\varphi^s$ are complex scalar fields,
$\5\varphi_s$ their complex conjugates.%
\footnote{For a complex field, the field and its
complex conjugate are treated as independent variables (instead of
the real and imaginary part).
}
$A_\mu^i$, $\phi^\Index$,
$B_{\mu\nu}^\Index$, $C^i$, $Q_\mu^\Index$ and $R^\Index$
are real fields. The Poincar\'e ghosts $c^\mu$ and $c_\mu{}^\nu$
are real, the supersymmetry ghosts $\xi^\alpha$ are constant
complex Weyl spinors, $\5\xi^\da$ is the complex conjugate of
$\xi^\alpha$. According to our conventions, the antifield $\5\Phi^*$
of the complex conjugate of a field $\Phi$
is related to the complex conjugate $\overline{\Phi^*}$
of the antifield of $\Phi$
according to:
\bea
\bar\Phi^*=-\overline{\Phi^*}.
\eea
In particular
the antifields of real fields are thus purely
imaginary.
$B_{\mu\nu}^\Index$, $B^{*\mu\nu}_\Index$ and 
$c^{\mu\nu}$
are antisymmetric in their spacetime indices:
\[
B_{\mu\nu}^\Index=-B_{\nu\mu}^\Index,\quad
B^{*\mu\nu}_\Index=-B^{*\nu\mu}_\Index,\quad
c^{\mu\nu}=-c^{\nu\mu}.
\]
The Grassmann parities $|\Phi^\FD|$ of the fields and
constant ghosts are
(one has $|\Phi^\FD|=|\5\Phi^\FD|$):
\beann
&&|A_\mu^i|=|\phi^\Index|=|B_{\mu\nu}^\Index|=|\varphi^\indec|
=|R^\Index|=|\xi^\alpha|=0,
\\
&&|\lambda^i_\alpha|=|\psi_\alpha^\Index|=|\chi_\alpha^\indec|=
|C^i|=|Q^\Index_\mu|=|c^\mu|=|c_\mu{}^\nu|=1.
\eeann
The ghost numbers of the fields and constant ghosts are
\beann
&&\gh(A_\mu^i)=\gh(\lambda^i_\alpha)=\gh(B_{\mu\nu}^\Index)=
\gh(\phi^\Index)=\gh(\psi_\alpha^\Index)=\gh(\varphi^\indec)
=\gh(\chi_\alpha^\indec)=0
\\
&&\gh(C^i)=\gh(Q^\Index_\mu)=\gh(c^\mu)=\gh(c_\mu{}^\nu)=1
\\
&&\gh(R^\Index)=2.
\eeann
The Grassmann parity of an antifield is opposite to the
Grassmann parity of the corresponding field, and the
ghost numbers of a field and its antifield add up to $-1$,
\[
|\Phi^*_\FD|=|\Phi^\FD|+1\quad\mbox{(mod 2)},\quad
\gh(\Phi^*_\FD)=-1-\gh(\Phi^\FD).
\]
To avoid the writing of indices we shall
occasionally use the notation $\varphi$, $\chi$,
$\phi$, $B_{\mu\nu}$, $\psi$, $\5\psi$, 
$Q_\mu$, $R$, $\5\varphi^*$, $\5\chi^*$ for
``column vectors'' with
entries $\varphi^\indec$, \dots ,
$\5\chi^{*\indec}$. 
Analgously $\varphi^*$, $\chi^*$,
$\phi^*$, $B^{*\mu\nu}$, $\psi^*$, $\5\psi^*$, 
$Q^{*\mu}$, $R^*$, $\5\varphi$, $\5\chi$
denote ``row vectors''
with entries $\varphi^*_\indec$, \dots ,
$\5\chi_\indec$. 
Transposition of such vectors
is denoted by $(\ )^t$.
The Lie algebra $\lieg$ is represented on
$\varphi$ and $\chi$ by antihermitian matrices\footnote{Our
analysis covers arbitrary
antihermitian representations $\{T_i\}$, including
trivial ones. Actually the
use of antihermitian representations is not essential and only made
to simplify the notation. All results hold analogously also
for other representations.} $T_i$ with real
structure constants $\f ijk$,
\beann
T^\dagger_i=-T_i,\quad [T_i,T_j]=\f ijk T_k\ .
\eeann
We shall compute the cohomology 
explicitly for the following simple Lagrangians
(the results for more general Lagrangians are discussed
in section \ref{otheraction}):
\begin{align}
L=&-\sfrac 14 \Kill_{ij}F^i_{\mu\nu} F^{j\mu\nu}
    +\sfrac{\Ii}2 \Kill_{ij}(\nabla_\mu\5\lambda^i\5\sigma^\mu\lambda^j
                         -\5\lambda^i\5\sigma^\mu \nabla_\mu\lambda^j)
\nonumber\\
   &+\sfrac 12\6_\mu\phi^t\6^\mu\phi
    -\sfrac 12H_\mu^t H^\mu
     +\sfrac{\Ii}{2}(\6_\mu\5\psi^t\5\sigma^\mu\psi
                    -\5\psi^t\5\sigma^\mu\6_\mu\psi)
\nonumber\\
   &+\nabla_\mu\5\varphi \nabla^\mu\varphi
     +\sfrac{\Ii}{4}(\nabla_\mu\5\chi\5\sigma^\mu\chi
                     -\5\chi\5\sigma^\mu \nabla_\mu\chi)
\nonumber\\
   &+\sfrac 12\Kill^{ij} 
    (\5\varphi T_i \varphi)(\5\varphi T_j \varphi)
    +\5\varphi T_i\chi\lambda^i
    -\5\lambda^i\5\chi T_i\varphi
\label{L}\end{align}
where the field strengths $F^i_{\mu\nu}$ and $H^\mu$ and the
covariant derivatives $\nabla_\mu$ are:
\begin{align}
&F^i_{\mu\nu}=\6_\mu A^i_\nu-\6_\nu A^i_\mu+\f jki A^j_\mu A^k_\nu
\label{F}\\
&H^{\mu}=\sfrac 12\vep^{\mu\nu\rho\sigma}\6_\nu B_{\rho\sigma}
\label{H}\\
&\nabla_\mu \lambda^i=\6_\mu \lambda^i+\f jki A^j_\mu \lambda^k,\quad
\nabla_\mu \5\lambda^i=\6_\mu \5\lambda^i+\f jki A^j_\mu \5\lambda^k
\label{nablalambda}\\
&\nabla_\mu\varphi=\6_\mu \varphi+A^i_\mu T_i\varphi,\quad
\nabla_\mu\5\varphi=\6_\mu \5\varphi-A^i_\mu\5\varphi T_i
\label{nablavarphi}\\
&\nabla_\mu\chi=\6_\mu \chi+A^i_\mu T_i\chi,\quad
\nabla_\mu \5\chi=
\6_\mu \5\chi-A^i_\mu\5\chi T_i.
\label{nablachi}\end{align}

The extended BRST transformations of the fields and constant
ghosts are:
\begin{align}
\ext A^i_\mu&=\6_\mu C^i+\f jki A_\mu^j C^k
  +\7c^\nu\6_\nu A^i_\mu
  -c_\mu{}^\nu A^i_\nu
  -\Ii\xi\sigma_\mu\5\lambda^i+\Ii\lambda^i\sigma_\mu\5\xi
\label{sA}\\
\ext \lambda^i_\alpha&=-\f jki C^j\lambda^k_\alpha
  +\7c^\mu\6_\mu\lambda^i_\alpha
  -\sfrac 12 c^{\mu\nu}(\sigma_{\mu\nu}\lambda^i)_\alpha
  -\Kill^{ij}\xi_\alpha (\5\varphi T_j\varphi
                    +\lambda^*_j\xi+\5\xi\5\lambda^*_j)
  +(\sigma^{\mu\nu}\xi)_\alpha F^i_{\mu\nu}
\label{slambda}\\
\ext \phi&=\7c^\mu\6_\mu \phi+\xi\psi+\5\psi\5\xi
\label{sphi}\\
\ext B_{\mu\nu}&=\6_\nu Q_\mu-\6_\mu Q_\nu
  +\7c^\rho\6_\rho B_{\mu\nu}
  -c_\mu{}^\rho B_{\rho\nu}-c_\nu{}^\rho B_{\mu\rho}
  +2\xi\sigma_{\mu\nu}\psi-2\5\psi\5\sigma_{\mu\nu}\5\xi
\label{sB}\\
\ext \psi_\alpha&=\7c^\mu\6_\mu \psi_\alpha
  -\sfrac 12 c^{\mu\nu}(\sigma_{\mu\nu}\psi)_\alpha
  +(\sigma^\mu\5\xi)_\alpha(H_\mu-\Ii\6_\mu\phi)
\label{spsi}\\
\ext \varphi&=-C^i T_i\varphi+\7c^\mu\6_\mu\varphi+\xi\chi
\label{svarphi}\\
\ext \chi_\alpha&=-C^i T_i\chi_\alpha
  +\7c^\mu\6_\mu \chi_\alpha
  -\sfrac 12 c^{\mu\nu}(\sigma_{\mu\nu}\chi)_\alpha
  -4\xi_\alpha \5\xi\5\chi^*
  -2\Ii(\sigma^\mu\5\xi)_\alpha \nabla_\mu\varphi
\label{schi}\\
\ext C^i&=\sfrac 12\f kji C^jC^k+\7c^\mu\6_\mu C^i-2\Ii\xi\sigma^\mu\5\xi A^i_\mu
\label{sC}\\
\ext Q_\mu&=\Ii\6_\mu R
  +\7c^\nu\6_\nu Q_\mu -c_\mu{}^\nu Q_\nu
  -2\Ii\xi\sigma^\nu\5\xi B_{\mu\nu}
  +2\Ii\xi\sigma_\mu\5\xi\phi
\label{sQ}\\
\ext R&=\7c^\mu\6_\mu R-2\xi\sigma^\mu\5\xi Q_\mu
\label{sR}\\
\ext c^\mu&=
  c_\nu{}^\mu c^\nu+2\Ii\xi\sigma^\mu\5\xi
\label{sc}\\
\ext c_\nu{}^\mu&=
  -c_\nu{}^\rho c_\rho{}^\mu
\label{scmunu}\\
\ext \xi^\alpha&=
  \sfrac 12c^{\mu\nu}(\xi\sigma_{\mu\nu})^\alpha
\label{sxi}\\
\ext \5\xi^\da&=
  -\sfrac 12c^{\mu\nu}(\5\sigma_{\mu\nu}\5\xi)^\da
\label{s5xi}
\end{align}
where
\begin{equation}
\7c^\mu=c^\mu-x^\nu {c_\nu}^\mu.
\label{hatxi}
\end{equation}
The transformations of the complex conjugate fields are obtained
from those given above according to
\[
\ext \5\Phi=(-)^{|\Phi|}\ \overline{\ext \Phi}.
\]
The extended BRST transformations of the antifields
are obtained according to:
\begin{align}
\ext \Phi^*_\FD&=\frac{\7\6^R L_{{\mathrm ext}}}{\7\6\Phi^\FD}
\label{sAF}\\
L_{{\mathrm ext}}&=L-(\ext \Phi^N)|_{\Phi^*=0}\Phi^*_N
+\sfrac 12 \Kill^{ij}(\lambda^*_i\xi+\5\xi\5\lambda^*_i)
                 (\lambda^*_j\xi+\5\xi\5\lambda^*_j)
+4\chi^*\xi\,\5\xi\5\chi^*
\label{Lext}
\end{align}
where $\7\6^R/\7\6\Phi^\FD$ is the Euler-Lagrange right derivative
with respect to $\Phi^\FD$. The extended BRST transformations
of derivatives of the fields and antifields are obtained
from those of the fields and antifields simply by
prolongation, i.e., by using
\[
{}[\ext ,\6_\mu]=0.
\]
By construction $\ext$ squares to zero
on all fields, antifields and constant ghosts,
\[
\ext^2=0.
\]
\comment
The terms in (\ref{sC}), (\ref{sQ}) and (\ref{sc}) 
which are bilinear in the supersymmetry ghosts
reflect that the commutators of supersymmetry
transformations contain gauge transformations and spacetime translations.
In addition these commutators contain
terms which vanish only on-shell. This is reflected
by the antifield dependent
terms in the extended BRST transformations 
(\ref{slambda}) and (\ref{schi}).
Schematically one has
\[
{}[\mathrm{susy\ transformation},\mathrm{susy\ transformation}]
\approx
\mathrm{translation}+\mathrm{gauge\ transformation}
\]
where $\approx$ is equality on-shell, defined according to
\[
X\approx Y\quad:\LRA\quad
X-Y=\sum_k Z^{\mu_1\dots\mu_k \FD}\6_{\mu_1}\dots\6_{\mu_k}\,
\frac{\7\6^R L}{\7\6\Phi^\FD}\ .
\]
[Here $X$, $Y$ and $Z^{\mu_1\dots\mu_k \FD}$ may depend on the
fields and their derivatives;
$Z^{\mu_1\dots\mu_k \FD}\6_{\mu_1}\dots\6_{\mu_k}\,
{\7\6^R L}/{\7\6\Phi^\FD}$ vanishes on-shell, i.e., for
all solutions of the fields equations, because
$L$ does not depend on ghosts].
The Poincar\'e, supersymmetry and gauge transformations
form thus an ``open algebra'' according to standard terminology.
By introducing additional fields
one may ``close'' and simplify the algebra, but this is
irrelevant to the cohomology,
see section \ref{otherform}.

\mysection{Cohomological problem and descent equations}
\label{`iii'}\label{DESCENT}

The primary goal is the determination
of the local cohomological groups $H^{g,4}(\ext|d)$, i.e.,
the cohomology of $\ext$ modulo the exterior derivative $d=dx^\mu\6_\mu$ 
in the space of
local 4-forms with ghost numbers $g$.
Unless differently specified, the term ``local $p$-forms''
is in this paper reserved for exterior $p$-forms
$dx^{\mu_1}\dots dx^{\mu_p}\omega_{\mu_1\dots\mu_p}$
(the differentials are treated as Grassmann odd quantities) where
$\omega_{\mu_1\dots\mu_p}$ can depend 
on the fields, antifields, their derivatives, constant ghosts and
explicitly on the spacetime coordinates such that the overall
number of derivatives of fields and antifields is
finite, without further restriction on this number or on
the order of derivatives that may occur%
\footnote{When dealing with a more complicated Lagrangian than
(\ref{L}) (especially with an effective Lagrangian),
one may have to adapt the
definition of local forms to the Lagrangian, see
section \ref{otheraction}.}.
The precise
mathematical setting of the cohomological problem
is made in the jet spaces associated
to the fields and antifields, see \cite{report} (the constant
ghosts are just added as local coordinates of these jet spaces).
The cocycles of
$H^{g,4}(\ext|d)$ are local 4-forms with ghost number $g$, denoted
by  $\omega^{g,4}$, which are
$\ext$-closed up to $d$-exact forms $d\omega^{g+1,3}$
where $\omega^{g+1,3}$ is a local 3-form with ghost number $g+1$,
\bea
\ext\omega^{g,4}+d\omega^{g+1,3}=0.
\label{p1}
\eea
Coboundaries of $H^{g,4}(\ext|d)$ are local 4-forms
$\omega^{g,4}=\ext\omega^{g-1,4}+d\omega^{g,3}$,
where $\omega^{g-1,4}$ and $\omega^{g,3}$ are
local forms with form-degree and ghost number
indicated by their superscripts.

As usual, (\ref{p1}) implies descent equations for $\ext$ and $d$
which relate $H^{*,4}(\ext|d)$ to $H(\ext)$
and $H(\ext+d)$, the cohomologies of $\ext$ and $(\ext+d)$ in the
space of local forms
(see section
9 of \cite{report} for a general discussion).
In the present case these relations are very direct.
To describe them precisely, we define the space $\Space$
of local functions (0-forms) that depend on the $c^\mu$ and $x^\mu$
only via the combinations $\7c^\mu=c^\mu-x^\nu c_\nu{}^\mu$, 
and the space $\CONST$ of polynomials in the constant ghosts,
\bea
\Space&:=&\{f(\7c^\mu,\xi^\alpha,\5\xi^\da,c^{\mu\nu},
\Phi^\FD,\Phi^*_\FD,
\6_\mu\Phi^\FD,\6_\mu\Phi^*_\FD,
\6_\mu\6_\nu\Phi^\FD,\6_\mu\6_\nu\Phi^*_\FD,\dots)\}
\\
\CONST&:=&\{f(c^\mu,\xi^\alpha,\5\xi^\da,c^{\mu\nu})\}.
\eea
$\Space$ is mapped by $\ext$ to itself ($\ext\Space\subseteq\Space$), 
because the $\ext$-transformations of all fields,
antifields, their derivatives and of the variables
$\7c^\mu,\xi^\alpha,\5\xi^\da,c^{\mu\nu}$ are
contained in $\Space$: $\ext$ 
acts on all these variables
according to $\ext=\7c^\mu\6_\mu+\dots$ where the nonwritten terms
do not involve $c^\mu$ or $x^\mu$. In particular, 
this holds for
$\ext\7c^\mu$:
\bea
\ext\7c^\mu=\7c^\nu\6_\nu\7c^\mu+2\Ii\xi\sigma^\mu\5\xi
=-\7c^\nu{c_\nu}^\mu+2\Ii\xi\sigma^\mu\5\xi.
\label{s7c}
\eea
Since $\CONST$ is also mapped by $\ext$ to itself 
($\ext\CONST\subseteq\CONST$), both the cohomology
$H(\ext,\Space)$ of $\ext$ in $\Space$ and 
the cohomology $H(\ext,\CONST)$ of $\ext$ in $\CONST$
are well-defined.
The relation between $H^{*,4}(\ext|d)$, $H(\ext+d)$ and $H(\ext,\Space)$ 
can now be described as follows:
\begin{pro}\label{proDescent}
$H(\ext,\Space)$, $H(\ext+d)$ and
$H^{*,4}(\ext|d)\oplus H(\ext,\CONST)$ are isomorphic:
\bea
H^g(\ext,\Space)\simeq
H^g(\ext+d)\simeq
H^{g-4,4}(\ext|d)\oplus H^g(\ext,\CONST),
\label{iso}
\eea
where the degree $g$ in $H^g(\ext,\Space)$, $H^g(\ext,\CONST)$ and
$H^{g,4}(\ext|d)$
is the ghost number, while in
$H^g(\ext+d)$ it is the sum of the ghost number and the
form-degree.
The representatives of $H^g(\ext+d)$
can be obtained from
those of $H^g(\ext,\Space)$
by substituting $c^\mu+dx^\mu$ for $c^\mu$,
the
representatives of $H^{g,4}(\ext|d)$ are the
4-forms contained in the 
representatives of $H^{g+4}(\ext+d)/H^{g+4}(\ext,\CONST)$.

\end{pro}

\comments
a)
We shall compute $H(\ext,\Space)$ and
derive $H^{*,4}(\ext|d)$ from it
according to the lemma. The representatives of $H^{g+4}(\ext+d)$
are the solutions of the descent equations
(see proof of the lemma).

b)
The relations between $H^{*,4}(\ext|d)$, $H(\ext+d)$ and 
$H(\ext,\Space)$ are much simpler than
the relations between $H^{*,4}(\gauge|d)$, $H(\gauge+d)$ and 
$H(\gauge)$ where $\gauge$ is the standard (non-extended)
BRST differential for the theories under study. 
The reason is that $\ext$ and $\ext+d$ are
directly related because $\ext$ contains the spacetime translations
($\ext+d$ arises on all fields and antifields
from $\ext$ by substituting $c^\mu+dx^\mu$ for $c^\mu$).
As a consequence $H(\ext,\Space)$ contains already
the complete structure of the descent equations for
$\ext$ and $d$.
In contrast,
$\gauge$ and $\gauge+d$ are truly different
and the descent equations for $\gauge$ and $d$ are
only contained in $H(\gauge+d)$ but not in 
$H(\gauge)$. Hence, $H(\ext,\Space)$ is
more similar to $H(\gauge+d)$ than to $H(\gauge)$.
In particular, $\7c^\mu$ plays in $H(\ext,\Space)$ a role
similar to $dx^\mu$ in $H(\gauge+d)$ (apart from the fact
that $\ext\7c^\mu$ does not vanish, in contrast
to $(\gauge+d) dx^\mu$). This role of $\7c^\mu$ is also similar to
the role of the diffeomorphism ghosts in gravitational theories,
see \cite{Brandt:1990et,BBHgrav}.

c) As a side remark, which is related to the
previous comment, I note that 
$H(\ext)$ (the cohomology of $\ext$
in the space of all local forms rather than only in $\Space$) and
$H^{*,p}(\ext|d)$ for all $p>0$ (rather than only for $p=4$) 
can also be directly derived from
$H(\ext,\Space)$.
The nontrivial representatives of $H(\ext)$ are linear combinations
of the nontrivial representatives of $H(\ext,\Space)$ with
coefficients that are ordinary differential forms
$\omega(dx,x)$ on ${\mathbb{R}}^4$
(independent of fields, antifields or constant
ghosts) which can be assumed not to be $d$-exact.
The nontrivial representatives of
$H^{*,p}(\ext|d)$ for $p>0$ are linear combinations of
nontrivial representatives of $H(\ext)$ with
form-degree $p$ (``solutions with
a trivial descent'') and of the $p$-form part contained
in nontrivial representatives of $H(\ext+d)/H(\ext,\CONST)$ (``solutions
with a non-trivial descent'').
This can be proved as analogous results in
Einstein-Yang-Mills theory, see section 6 of \cite{BBHgrav}
(the role of the space $\cA$ in \cite{BBHgrav} is now
taken by $\Space$, the role of the diffeomorphism ghosts
by the $\7c^\mu$). Note that for $p=4$ the statement on 
$H^{*,p}(\ext|d)$ is in agreement with lemma \ref{proDescent}
because 4-forms $\omega(dx,x)$ are $d$-exact in ${\mathbb{R}}^4$.

d) 
A gauge fixing need not be specified because it
does not affect the cohomology (see, e.g., sections
2.6 and 2.7 of \cite{report}).

\mysection{Change of variables}\label{`iv'}\label{CALC}

To compute $H(\ext,\Space)$ we shall follow a strategy \cite{ten,jet}
based on new jet coordinates $u^\ell$, $v^\ell$ and $w^\Windex$
which satisfy
\bea
\ext u^\ell=v^\ell,\quad \ext w^\Windex=r^\Windex(w).
\label{susw}
\eea
The important requirement here is that
$r^\Windex(w)$ is a function of the $w$'s only.
To construct such jet coordinates we use operations
$\6^+_+$, $\6^+_-$, $\6^-_+$, $\6_-^-$ which are
defined as follows \cite{sugra}:
let $Z^m_n$ be a Lorentz-irreducible multiplet of
fields or antifields with $m$ undotted and $n$ dotted spinor indices,
\[
Z_n^m\equiv
\{Z_{\alpha_1\cdots\alpha_n}^{\da_1\cdots\da_m}\}\ ,\quad
Z_{\alpha_1\cdots\alpha_n}^{\da_1\cdots\da_m}=
Z_{(\alpha_1\cdots\alpha_n)}^{(\da_1\cdots\da_m)}\ ;
\]
we define $\6^+_+$, $\6^+_-$, $\6^-_+$, $\6_-^-$ and $\Box$ according to
\bea \6_+^+Z_n^m&\equiv&
\{\6_{(\alpha_0}^{(\da_0}
Z_{\alpha_1\cdots\alpha_n)}^{\da_1\cdots\da_m)}\}
\nonumber\\
\6_+^-Z_n^m&\equiv&
\{m\, \6^{}_{\da_m(\alpha_0}
Z_{\alpha_1\cdots\alpha_n)}^{\da_1\cdots\da_m}\}
\nonumber\\
\6_-^+Z_n^m&\equiv&
\{n\, \6^{\alpha_n(\da_0}_{}
Z_{\alpha_1\cdots\alpha_n}^{\da_1\cdots\da_m)}\}
\nonumber\\
\6_-^-Z_n^m&\equiv&
\{mn\, \6^{\alpha_n}_{\da_m}
Z_{\alpha_1\cdots\alpha_n}^{\da_1\cdots\da_m}\}
\nonumber\\
\Box Z_n^m&\equiv&
\{\6_\mu\6^\mu Z_{\alpha_1\cdots\alpha_n}^{\da_1\cdots\da_m}\}
=
\{\sfrac 12 \6_{\beta\dbe}\6^{\dbe\beta} 
Z_{\alpha_1\cdots\alpha_n}^{\da_1\cdots\da_m}\}.
\label{bas4}\eea
Using these operations and the notation
\[
A\equiv\{A^i{}^{\da}_\alpha\},\ 
B^{(+)}\equiv\{\sigma^{\mu\nu}_{\alpha\beta}B^\Index_{\mu\nu}\},\ 
B^{(-)}\equiv\{\5\sigma^{\mu\nu\da\dbe}B^\Index_{\mu\nu}\},\ 
H\equiv\{H^\Index{}^{\da}_\alpha\},\ 
Q\equiv\{Q^\Index{}^{\da}_\alpha\}\quad etc,
\]
we define the following jet variables $u^\ell,v^\ell,w_{(0)}^\Windex$:
\begin{align}
\{u^\ell\}=&\bigcup_{p=0}^\infty\bigcup_{q=0}^\infty
\left(
\Box^p(\6_+^+)^q A\cup
\Box^p\6_-^-A
\right.
\nonumber\\[-2.2ex]
&\hspace{3.4em}
\cup\Box^p(\6_+^+)^qC^{*}
\nonumber\\
&\hspace{3.4em}
\cup\Box^p(\6_+^+)^qA^{*}\cup
\Box^p(\6_+^+)^q\6^+_-A^{*}\cup
\Box^p(\6_+^+)^q\6^-_+A^{*}
\nonumber\\
&\hspace{3.4em}
\cup\Box^p(\6_+^+)^q\lambda^{*}\cup
\Box^p(\6_+^+)^q\6^+_-\lambda^{*}\cup
\Box^p(\6_+^+)^q\5\lambda^{*}\cup
\Box^p(\6_+^+)^q\6^-_+\5\lambda^{*}
\nonumber\\
&\hspace{3.4em}
\cup\Box^p(\6_+^+)^qB^{(+)}\cup
\Box^p(\6_+^+)^qB^{(-)}\cup
\Box^p(\6_+^+)^q\6^-_+B^{(-)}
\nonumber\\
&\hspace{3.4em}
\cup\Box^p(\6_+^+)^qQ\cup
\Box^p\6_-^-Q
\nonumber\\
&\hspace{3.4em}
\cup\Box^p(\6_+^+)^qR^*
\nonumber\\
&\hspace{3.4em}
\cup\Box^p(\6_+^+)^qQ^*\cup
\Box^p(\6_+^+)^q\6^+_-Q^*\cup
\Box^p(\6_+^+)^q\6^-_+Q^*
\nonumber\\
&\hspace{3.4em}
\cup\Box^p(\6_+^+)^q\6^-_+B^{*(-)}
\nonumber\\
&\hspace{3.4em}
\cup\Box^p(\6_+^+)^q\phi^*
\nonumber\\
&\hspace{3.4em}
\cup\Box^p(\6_+^+)^q\psi^{*}\cup
\Box^p(\6_+^+)^q\6^+_-\psi^{*}\cup
\Box^p(\6_+^+)^q\5\psi^{*}\cup
\Box^p(\6_+^+)^q\6^-_+\5\psi^{*}
\nonumber\\
&\hspace{3.4em}
\cup\Box^p(\6_+^+)^q\varphi^*\cup
\Box^p(\6_+^+)^q\5\varphi^*
\nonumber\\
&\hspace{3.4em}
\cup\left.
\Box^p(\6_+^+)^q\chi^{*}\cup
\Box^p(\6_+^+)^q\6^+_-\chi^{*}\cup
\Box^p(\6_+^+)^q\5\chi^{*}\cup
\Box^p(\6_+^+)^q\6^-_+\5\chi^{*}
\right)
\label{u}\\
\{v^\ell\}=&\{\ext u^\ell\}
\label{v}\\
\{w_{(0)}^\Windex\}=&\{C,R,\7c^\mu,c^{\mu\nu}\ (\mu<\nu),
\xi^\alpha,\5\xi^\da\}
\nonumber\\
&\cup\bigcup_{q=0}^\infty
\left((\6_+^+)^q\6^-_+A\cup
(\6_+^+)^q\6^+_-A\cup
(\6_+^+)^q\lambda\cup
(\6_+^+)^q\5\lambda
\right.
\nonumber\\[-2.2ex]
&\hspace{3em}
\cup(\6_+^+)^qH\cup
(\6_+^+)^q\phi\cup
(\6_+^+)^q\psi\cup
(\6_+^+)^q\5\psi
\nonumber\\
&\hspace{3em}
\cup\left.
(\6_+^+)^q\varphi\cup(\6_+^+)^q\5\varphi\cup
(\6_+^+)^q\chi\cup
(\6_+^+)^q\5\chi
\right).
\label{w0}\end{align}

\begin{pro}\label{proBasis}
The $u$'s, $w_{(0)}$'s and the linearized $v$'s
form a basis of the vector space (over ${\mathbb C}$) spanned 
by the fields, antifields, all their independent derivatives and the
$\7c^\mu,c^{\mu\nu},\xi^\alpha,\5\xi^\da$. 
\end{pro}

This implies that
the $u$'s, $v$'s and $w_{(0)}$'s can be used as new
jet coordinates substituting for the fields, antifields,
their derivatives and
constant ghosts. They do not have the desired
quality (\ref{susw}) but can be extended to
variables with this quality by means of an
algorithm given in \cite{jet}:

\begin{pro}\label{proW}
The algorithm described in section 2 of \cite{jet} completes
the $w_{(0)}^\Windex$ to local functions $w^\Windex$ 
such that (\ref{susw}) holds.
\end{pro}

(\ref{susw}) will allow us to compute the cohomology solely
in terms of the $w$'s (see section \ref{pairs}). 
We introduce the following notation
for them:
\begin{align}
\{w^\Windex\}=&\{\7C,\7R,\7c^\mu,c^{\mu\nu}\ (\mu<\nu),
\xi^\alpha,\5\xi^\da\}\cup\{\7T^\Tindex\}
\nonumber\\
\{\7T^\Tindex\}=&
\bigcup_{q=0}^\infty
\left((\7\nabla_+^+)^q\7F^{(+)}\cup
(\7\nabla_+^+)^q\7F^{(-)}\cup
(\7\nabla_+^+)^q\7\lambda\cup
(\7\nabla_+^+)^q\hatlambda
\right.
\nonumber\\[-2.2ex]
&\hspace{2.5em}
\cup(\7\nabla_+^+)^q\7H\cup
(\7\nabla_+^+)^q\phi\cup
(\7\nabla_+^+)^q\7\psi\cup
(\7\nabla_+^+)^q\hatpsi
\nonumber\\
&\hspace{2.5em}
\cup\left.
(\7\nabla_+^+)^q\varphi\cup
(\7\nabla_+^+)^q\5\varphi\cup
(\7\nabla_+^+)^q\7\chi\cup
(\7\nabla_+^+)^q\hatchi
\right).
\label{w}\end{align}
with an obvious correspondence to the 
variables in (\ref{w0}) [$\7F^{(\pm)}$ corresponds to 
$\6^\mp_\pm A$]. 
The $\7T$'s may be called generalized tensor fields
because their antifield independent parts are
ordinary gauge covariant tensor fields. In addition
they contain terms which depend on antifields such
that $\ext w^\Windex$ contains no
terms that vanish on-shell.
The $\7\nabla_\mu$ can be viewed as generalizations of the ordinary
covariant derivatives $\nabla_\mu$ and are related to the
latter as follows: the antifield independent part of
$\7\nabla_\mu f(\7T)$ coincides on-shell
with $\nabla_\mu$ acting on the antifield independent part of
$f(\7T)$:
\bea
\Big[\7\nabla_\mu f(\7T)\Big]_{\Phi^*=0}\approx 
\nabla_\mu \Big[f(\7T)\Big]_{\Phi^*=0}.
\label{nablaformula}
\eea
For instance,
the antifield independent part of 
$\7\nabla_\mu\7\nabla_\nu\phi$ is
$(\6_\mu\6_\nu-\sfrac 14\eta_{\mu\nu}\Box)\phi$.
The explicit form of the $\7T$'s in terms of the
original variables (fields, antifields, their derivatives
and the constant ghosts)
is somewhat involved. Fortunately we need not compute
them explicitly to perform the cohomological analysis because their
existence is guaranteed by lemma \ref{proW}, and their
extended BRST transformations can be directly obtained from
the transformations of the $w_{(0)}$'s 
as we shall see in the following section.
Nevertheless, for later purpose
and to illustrate the structure of the new variables,
we list a few $w$'s explicitly: 
\begin{align}
\7C^i&=C^i+A^i_\mu \7c^\mu
\label{7C}\\
\7R&=R+\Ii Q_\mu \7c^\mu+\sfrac{\Ii}2 B_{\mu\nu} \7c^\nu\7c^\mu
\label{7R}\\
\7\lambda^i_\alpha&=\lambda^i_\alpha
-\sfrac{\Ii}4 \Kill^{ij}\7c^\mu(\sigma_\mu\5\lambda^*_j)_\alpha
\label{7lambda}\\
\7F^{i(+)}_{\alpha\beta}&=\sigma^{\mu\nu}_{\alpha\beta}F^i_{\mu\nu}
-2\Kill^{ij}\xi^{}_{(\alpha}\lambda^*_{\beta)j}
+\sfrac 23\Kill^{ij}\7c^\mu\sigma_{\mu\nu\alpha\beta}A^{*\nu}_j
-\sfrac 16 \Kill^{ij}\7c^\mu\7c^\nu \sigma_{\mu\nu\alpha\beta}C^*_j
\label{7F+}\\
\7F^{i(-)}_{\da\dbe}&=\5\sigma^{\mu\nu}_{\da\dbe}F^i_{\mu\nu}
-2\Kill^{ij}{\5\xi}^{}_{(\da}\5\lambda^*_{\dbe)j}
+\sfrac 23\Kill^{ij}\7c^\mu\5\sigma_{\mu\nu\da\dbe}A^{*\nu}_j
-\sfrac 16 \Kill^{ij}\7c^\mu\7c^\nu \5\sigma_{\mu\nu\da\dbe}C^*_j
\label{7F-}\\
\7\psi_\alpha^\Index&=\psi_\alpha^\Index-\sfrac{\Ii}4
\delta^{\Index\Jndex}\7c^\mu(\sigma_\mu\5\psi^{*}_\Jndex)_\alpha
\label{7psi}\\
\7H_\mu^\Index&=H_\mu^\Index
+\sfrac 34\delta^{\Index\Jndex}
(\psi^{*}_\Jndex\sigma_\mu\5\xi-\xi\sigma_\mu\5\psi^{*}_\Jndex)
- \vep_{\mu\nu\rho\sigma}\delta^{\Index\Jndex}(
\sfrac 12\7c^\nu B^{*\rho\sigma}_\Jndex
+\sfrac 16 \7c^\nu\7c^\rho Q^{*\sigma}_\Jndex
-\sfrac {\Ii}{24}\7c^\nu\7c^\rho\7c^\sigma R^{*}_\Jndex)
\label{7H}\\
\7\nabla_\mu\phi^\Index&=\6_\mu\phi^\Index
+\sfrac 14\delta^{\Index\Jndex}
(\Ii\xi\sigma_\mu\5\psi^{*}_\Jndex+\Ii\psi^{*}_\Jndex\sigma_\mu\5\xi
-\7c_\mu\phi^{*}_\Jndex)
\label{76phi}\\
\7\chi_\alpha&=\chi_\alpha-\sfrac{\Ii}2\7c^\mu(\sigma_\mu\5\chi^*)_\alpha
\label{7chi}\\
\7\nabla_\mu\varphi&=\nabla_\mu\varphi
+\sfrac{\Ii}2 \xi\sigma_\mu\5\chi^*
-\sfrac 14\7c_\mu\5\varphi^*.
\label{76varphi}\end{align}
$\7F^{i(+)}$ and $\7F^{i(-)}$ are the Lorentz irreducible
parts of generalized Yang-Mills field strengths given by
\begin{align}
\7F^i_{\mu\nu}&=
\sfrac 12 (\7F^{i(+)}_{\alpha\beta}\sigma_{\mu\nu}^{\alpha\beta}
+\7F^{i(-)}_{\da\dbe}\5\sigma_{\mu\nu}^{\da\dbe})
\nonumber\\
&=F^i_{\mu\nu}+ \Kill^{ij}(\xi\sigma_{\mu\nu}\lambda^*_j
  +\5\lambda^*_j\5\sigma_{\mu\nu}\5\xi
  +\sfrac 13 \7c^{}_\mu A^*_{j\nu}-\sfrac 13 \7c^{}_\nu A^*_{j\mu}
  -\sfrac 16 \7c_\mu\7c_\nu C^*_j).
\label{7F}\end{align}

\mysection{Gauge covariant algebra}
\label{`v'}\label{algebra}

By construction the extended BRST transformations of the $w$'s 
can be expressed solely in terms of the $w$'s again, see eq. (\ref{susw}).
As explained in \cite{ten}, this is related to
a graded commutator algebra
which is realized on the $\7T$'s. The cohomology of $\ext$
can be interpreted as the cohomology associated 
with this algebra
(similiar to Lie algebra cohomology -- in fact one may
view it as a
generalization of Lie algebra cohomology, see remark at the end of this
section). We shall now discuss the extended BRST transformations
of the $w$'s and the corresponding algebra because
these will be of crucial
importance for the solution of the cohomological problem under study.

The extended BRST transformations of the $w$'s 
can be directly obtained 
from the extended BRST transformations 
of the $w_{(0)}$'s:

\begin{pro}[\cite{jet}]\label{prosw}
The extended BRST transformations of the $w$'s are given by
$\ext w^\Windex=r^\Windex(w)$ with $r^\Windex$
the same function as in
$\ext w_{(0)}^\Windex=r^\Windex(w_{(0)})+O(1)$ where
$O(1)$ collects all terms that are at least linear in the $u$'s and $v$'s.
\end{pro}

This lemma is very useful because it allows one to derive the
extended BRST transformations of the $w$'s without having to
compute these variables and their extended BRST transformations explicitly
(as remarked and demonstrated in section \ref{CALC},
the explicit structure of the $w$'s is quite involved).
The proof of the lemma was given in \cite{jet} and will not be
repeated here. It is a consequence of
the algorithm 
used to construct the $w$'s (see lemma \ref{proW}).
For later purpose and to illustrate
the lemma let us apply it to derive
$\ext\7C^i$. We start from $\ext C^i$ given by (\ref{sC}) and use
that (\ref{sA}) gives $\6_\mu C^i=\Ii\xi\sigma_\mu\5\lambda^i
-\Ii\lambda^i\sigma_\mu\5\xi-\7c^\nu\6^{}_{[\nu} A^i_{\mu]}+O(1)$. This yields
\[
\ext C^i
  =\sfrac 12\f kjiC^jC^k
  +\Ii\7c^\mu(\xi\sigma_\mu\5\lambda^i-\lambda^i\sigma_\mu\5\xi)
  +\7c^\mu\7c^\nu\6_\mu A^i_\nu+O(1).
\]
Applying now lemma \ref{prosw} we conclude
\bea
\ext \7C^i
  =\sfrac 12\f kji\7C^j\7C^k
  +\Ii\7c^\mu(\xi\sigma_\mu\hatlambda{}^i-\7\lambda^i\sigma_\mu\5\xi)
  +\sfrac 12\7c^\mu\7c^\nu\7F^i_{\mu\nu}.
\label{s7C}
\eea
When one verifies this result directly
using (\ref{7C}) and the extended BRST transformations
given in section \ref{BRST}, one finds
that (\ref{s7C}) 
actually contains no antifield dependent terms, i.e.,
all antifield dependent terms (coming from 
$\hatlambda{}^i$, $\7\lambda^i$ and
$\7F^i_{\mu\nu}$) 
cancel out exactly on the right hand side of (\ref{s7C}).
 
Analogously one can derive $\ext\7R$ starting
from $\ext R$ given by eq. (\ref{sR}) and
using that (\ref{sQ}) gives
$\6_\mu R=\Ii\7c^\nu\6_{[\nu} Q_{\mu]}-2\xi\sigma_\mu\5\xi\phi+O(1)$
and that 
(\ref{sB}) gives
$\6_{[\nu} Q_{\mu]}=
-\sfrac 12\7c^\rho\6_{[\rho} B_{\mu\nu]}
-\xi\sigma_{\mu\nu}\psi+\5\psi\5\sigma_{\mu\nu}\5\xi+O(1)$.
One obtains
\[
\ext R
  =-2\7c^\mu\xi\sigma_\mu\5\xi\phi
  -\Ii\7c^\mu\7c^\nu(\xi\sigma_{\mu\nu}\psi-\5\psi\5\sigma_{\mu\nu}\5\xi)
  -\sfrac {\Ii}2\7c^\mu\7c^\nu\7c^\rho\6_\mu B_{\nu\rho}
  +O(1).
\]
Using also (\ref{H}) in the form
$\6_{[\mu}B_{\nu\rho]}=(1/3)\vep_{\mu\nu\rho\sigma}H^\sigma$,
one concludes
\bea
\ext\7R
  =-2\7c^\mu\xi\sigma_\mu\5\xi\phi
  -\Ii\7c^\mu\7c^\nu(\xi\sigma_{\mu\nu}\7\psi
  -\hatpsi\5\sigma_{\mu\nu}\5\xi)
  -\sfrac{\Ii}6\7c^\mu\7c^\nu\7c^\rho\vep_{\mu\nu\rho\sigma}\7H^\sigma.
\label{s7R}
\eea
Again, the
antifield dependent terms on the right hand side cancel out exactly.

To derive the graded commutator algebra
we proceed as in \cite{ten} and use that $\{w^\Windex\}$
decomposes into subsets
of variables with ghost numbers 2, 1 and 0, respectively.
Those with ghost number 2 are the $\7R^\Index$, those with
ghost number 0 are the $\7T^\Tindex$ and those with ghost number 1
are the $\7C^i$, $\7c^\mu$, $c^{\mu\nu}$, $\xi^\alpha$,
$\5\xi^\da$ which we denote collectively by $\cC^\gindex$ now,
\bea
\{\cC^\gindex\}
=\{\7C^i,\7c^\mu,c^{\mu\nu}\ (\mu<\nu),\xi^\alpha,\5\xi^\da\}.
\label{CCs}
\eea
Since $\ext\7T^\Tindex$ has ghost number 1 and since there are no $w$'s with
negative ghost numbers, we conclude from eq. (\ref{susw}) that
$\ext\7T^\Tindex$ is a linear combination
of the $\cC$'s with 
coefficients that are functions
of the $\7T$'s:
\bea
\ext\7T^\Tindex=\cC^\gindex R^\Tindex_\gindex(\7T).
\eea
Moreover, since the $\7T^\Tindex$ are independent variables,
we can define operators $\Delta_\gindex$ on the
space of functions of the $\7T$'s through
\bea
\Delta_\gindex :=R^\Tindex_\gindex(\7T)\,\frac{\6}{\6\7T^\Tindex}\ .
\eea
Using these operators, we can express the extended BRST transformation
of any function of the $\7T$'s according to
\bea
\ext f(\7T)=\cC^\gindex\Delta_\gindex f(\7T)\ .
\label{sf(T)}
\eea
By construction the $\Delta_\gindex$ are graded derivations
acting on the space of functions of the $\7T$'s. For
these graded derivations we introduce the following notation:
\bea
\{\Delta_\gindex\}=\{\delta_i,\7\nabla_\mu,l_{\mu\nu},\cD_\alpha,\5\cD_\da\},
\label{Deltas}\eea
so that
\bea
\ext f(\7T)=(\7C^i\delta_i+\7c^\mu\7\nabla_\mu
  +\sfrac 12c^{\mu\nu}l_{\mu\nu}+\xi^\alpha \cD_\alpha
  +\5\xi^\da\5\cD_\da)f(\7T).
\label{s7T}\eea
The $\delta_i$ and $l_{\mu\nu}$ are
linearly realized 
on the generalized tensor fields and represent the
Lie algebras of the Yang-Mills gauge group and
the Lorentz group as indicated by the 
indices of the $\7T$'s, for instance:
\beann
&&\delta_i\7F^{j(+)}=-\f ikj\7F^{k(+)},\quad 
\delta_i\phi=0,\quad  \delta_i\varphi=-T_i\varphi,
\\
&&l_{\mu\nu}\7\lambda^i_\alpha
=-(\sigma_{\mu\nu}\7\lambda^i)_\alpha,\quad 
l_{\mu\nu}\phi=0,\quad 
l_{\mu\nu}\7\nabla_\rho\phi=\eta_{\rho\nu}\7\nabla_\mu\phi
-\eta_{\rho\mu}\7\nabla_\nu\phi.
\eeann
In contrast, $\7\nabla_\mu$, $\cD_\alpha$ and $\5\cD_\da$
are nonlinearly realized, see comment b) below.

\begin{pro}\label{prosW}
The graded commutator algebra of the graded derivations (\ref{Deltas})
reads
\bea
&&[\7\nabla_\mu,\7\nabla_\nu]=-\7F_{\mu\nu}^i\delta_i
\nonumber\\
&&[\cD_\alpha,\7\nabla_\mu]=\Ii(\sigma_\mu\hatlambda{}^i)_\alpha\delta_i,\quad
[\5\cD_\da,\7\nabla_\mu]=-\Ii(\7\lambda^i\sigma_\mu)_\alpha\delta_i
\nonumber\\
&&\{\cD_\alpha,\5\cD_\da\}=-2\Ii\sigma^\mu_{\alpha\da}\7\nabla_\mu,\quad
\{\cD_\alpha,\cD_\beta\}=\{\5\cD_\da,\5\cD_\dbe\}=0
\nonumber\\
&&[\delta_i,\7\nabla_\mu]=[\delta_i,\cD_\alpha]=[\delta_i,\5\cD_\da]=0
\nonumber\\
&&[l_{\mu\nu},\7\nabla_\rho]=
\eta_{\rho\nu}\7\nabla_\mu-\eta_{\rho\mu}\7\nabla_\nu,\quad
[l_{\mu\nu},\cD_\alpha]=-\sigma_{\mu\nu\alpha}{}^\beta \cD_\beta,\quad
[l_{\mu\nu},\5\cD_\da]=\5\sigma_{\mu\nu}{}^\dbe{}_\da\5\cD_\dbe
\nonumber\\
&&[\delta_i,\delta_j]=\f ijk\delta_k,\quad
[l_{\mu\nu},l_{\rho\sigma}]=\eta_{\rho\nu}l_{\mu\sigma}
+\eta_{\sigma\nu}l_{\rho\mu}-(\mu\lra\nu),\quad
[\delta_i,l_{\mu\nu}]=0.
\label{alg}
\eea
\end{pro}

\comments a) Notice that (\ref{alg}) is not a graded Lie 
algebra because $[\7\nabla_\mu,\7\nabla_\nu]$,
$[\cD_\alpha,\7\nabla_\mu]$ and
$[\5\cD_\da,\7\nabla_\mu]$
involve structure functions
rather than structure constants. (\ref{alg}) may thus be
rightly called a generalization of a (graded) Lie algebra
and the cohomology of $\ext$ a generalization of (graded) Lie algebra
cohomology.

b) In order to avoid possible confusion,
I stress that $\cD_\alpha$ and $\5\cD_\da$ do {\em not} act in
a superspace but are algebraically defined on the
generalized tensor fields.
The $\cD_\alpha$-transformations
of $\7\lambda$, $\7F^{(+)}$,
$\phi$, $\7\psi$, $\7H$, $\varphi$, $\7\chi$ and their
complex conjugates are spelled out explicitly in appendix \ref{conv}.
{}From these one may derive the $\cD_\alpha$-transformations
of the first and higher order $\7\nabla$-derivatives of these fields
by means of the algebra (\ref{alg}) [e.g.,
$\cD_\alpha \7\nabla_\mu\hatlambda{}^i_\da=
([\cD_\alpha, \7\nabla_\mu]+\7\nabla_\mu \cD_\alpha ) \hatlambda{}^i_\da
=-\Ii(\sigma_\mu\hatlambda{}^j)_\alpha\f jki\hatlambda{}^k_\da$].
The $\5\cD_\da$-transformations can be obtained
from the $\cD_\alpha$-transformations by complex conjugation.

\mysection{Computation of the cohomology}\label{`vii'}\label{ALL}

\subsection{Cohomology for small ghost numbers}
\label{`vi'}\label{leq3}

As a first cohomological result we shall now derive
the cohomological groups $H^{g,4}(\ext|d)$ and $H^{g+4}(\ext,\Space)$
for $g<-1$. These groups can be straightforwardly obtained
from corresponding cohomological groups
of the so-called Koszul-Tate differential%
\footnote{We are dealing here with the standard
Koszul-Tate differential, trivially extended
to the constant ghosts. It must not be confused with the extended
Koszul-Tate differential introduced in \cite{Brandt:1997cz}
which acts also on ``global antifields'' conjugate to
the constant ghosts.} $\delta$
modulo $d$.
$\delta$ is part of $\ext$ 
and arises in the decomposition of
$\ext$ according to the antifield number ($\af$).
The antifield number
is defined according to standard conventions: 
$R^*_\Index$ has antifield number 3,
$C^*_i$ and $Q^{*\mu}_\Index$ have antifield number 2, all other
antifields have antifield number 1, the fields $\Phi^\FD$ and the
constant ghosts have antifield number 0.
The extended BRST differential decomposes into
three pieces with antifield numbers $-1$, 0 and 1, respectively,
with $\delta$ the piece with antifield number $-1$:
\bea
\ext=\delta+\gammaext+\soneext,\quad \af(\delta)=-1,\quad 
\af(\gammaext)=0,\quad
\af(\soneext)=1.
\label{deltagamma}
\eea
The $\delta$-transformations of the fields, constant ghosts and antifields
are:
\bea
&&
\delta\Phi^\FD=\delta c^\mu=\delta\xi^\alpha=\delta\5\xi^\da
=\delta c^{\mu\nu}=0,
\nonumber\\
&&
\delta\Phi^*_\FD=\frac{\7\6^RL}{\7\6\Phi^\FD}\quad\mbox{for}\quad
\Phi^*_\FD\in\{A^{*\mu}_i,\lambda^{*\alpha}_i,\5\lambda^{*\da}_i,
\phi^*_\Index,B^{*\mu\nu}_\Index,\psi^{*\alpha}_\Index,
\5\psi^{*\da}_\Index,
\varphi^*_\indec,\5\varphi^{*\indec},\chi^{*\alpha}_\indec,
\5\chi^{*\indec\da}\},
\nonumber\\
&&
\delta R^*_\Index=\Ii\,\6_\mu Q^{*\mu}_\Index\, ,\quad
\delta Q^{*\mu}_\Index=2\6_\nu B^{*\nu\mu}_\Index,\quad
\delta C^*_i=-\nabla_\mu A^{*\mu}_i
+\!\!\!\sum_{\Phi^\FD\not\in\{A_\mu^i\}}\!\!\!\Phi^*_\FD\delta_i\Phi^\FD.
\label{delta}
\eea
Notice that the constant ghosts are inert to both $\delta$ and $d$.
Therefore the cohomological groups $H(\delta|d)$ are the same as 
in the case of the non-extended BRST cohomology except that
the representatives can depend arbitrarily on the
constant ghosts.

\begin{pro}\label{prodelta|d}
Cohomology $H^4_k(\delta|d)$ for antifield numbers $k>1$:
\bea
&&\delta\omega^4_k+d\omega^3_{k-1}=0\ \then\ 
\omega^4_k\sim
\left\{\ba{ll}
0 & \mbox{for}\ k>3\\
k^\Index(c,\xi,\5\xi) R^*_\Index\, d^4x & \mbox{for}\ k=3\\
{}[k^{\ifree}(c,\xi,\5\xi)C^*_{\ifree}+
k^{[\Index\Jndex]}(c,\xi,\5\xi)f_{\Index\Jndex}]\,d^4x 
& \mbox{for}\ k=2,
\ea\right.
\nonumber\\
&&f_{\Index\Jndex}=Q^{*\mu}_{[\Index}H^{\phantom{*\mu}}_{\Jndex]\mu}-
\sfrac 12\vep_{\mu\nu\rho\sigma}B^{*\mu\nu}_\Index B^{*\rho\sigma}_\Jndex,
\label{prode1}
\eea
where $\sim$ is equivalence in $H(\delta|d)$
($\omega^4_k\sim \omega^4_k+\delta\omega^4_{k+1}+d\omega^3_{k}$),
$\ifree$ runs over those Abelian elements of $\lieg$ under which
all matter fields are uncharged, the $k(c,\xi,\5\xi)$'s are
arbitrary functions of the constant ghosts,
and we used the notation
$H_{\Index\mu}:=\delta_{\Index\Jndex}H^\Jndex_\mu$.
The 4-forms $R^*_\Index d^4x$, $C^*_{\ifree}d^4x$ and
$f_{\Index\Jndex}d^4x$, $ \Index<\Jndex$,
are nontrivial and inequivalent 
in $H(\delta|d)$:
\bea
&
k^\Index(c,\xi,\5\xi) R^*_\Index\sim 0\quad \then\quad k^\Index=0;
&
\label{prode2}\\
&
[k^{\ifree}(c,\xi,\5\xi)C^*_{\ifree}
+k^{[\Index\Jndex]}(c,\xi,\5\xi) f_{\Index\Jndex}]\,d^4x \sim 0
\ \then\
k^{\ifree}=k^{[\Index\Jndex]}=0.
&\quad
\label{prode3}
\eea
\end{pro}

Using lemma \ref{prodelta|d}, it is straightforward to prove the
following result:

\begin{pro}\label{pros|d<-1}
Cohomology $H^{g,4}(\ext|d)$ for ghost numbers $g<-1$:
\bea
&&\ext\omega^{g,4}+d\omega^{g-1,3}=0\ \then\
\omega^{g,4}\sim
\left\{\ba{ll}
0 & \mbox{for}\quad g<-3\\
k^\Index R^*_\Index\, d^4x & \mbox{for}\quad g=-3\\
{}(k^{\ifree}C^*_{\ifree}+k^{[\Index\Jndex]}f'_{\Index\Jndex})\,d^4x 
& \mbox{for}\quad g=-2,
\ea\right.\quad
\nonumber\\
&&
f'_{\Index\Jndex}=
Q^{*\mu}_{[\Index}(H^{}_{\Jndex]\mu}+\psi^*_{\Jndex]}\sigma_\mu\5\xi
-\xi\sigma_\mu\5\psi^*_{\Jndex]})-
\sfrac 12\vep_{\mu\nu\rho\sigma}B^{*\mu\nu}_\Index B^{*\rho\sigma}_\Jndex
+R^*_{[\Index}(\5\psi^{}_{\Jndex]}\5\xi-\xi\psi^{}_{\Jndex]}),
\label{s<3,1}
\eea
where $\sim$ is equivalence in $H(\ext|d)$
($\omega^{g,4}\sim\omega^{g,4}+\ext\omega^{g-1,4}+d\omega^{g,3}$), 
$\ifree$ runs over those Abelian elements of $\lieg$ under which
all matter fields are uncharged,
and
$k^\Index$, $k^{\ifree}$ and $k^{[\Index\Jndex]}$ are
complex numbers.
The 4-forms $R^*_\Index d^4x$, $C^*_{\ifree}d^4x$ and
$f'_{\Index\Jndex}d^4x $, $\Index<\Jndex$, are nontrivial and inequivalent 
in $H(\ext|d)$:
\bea
&
k^\Index R^*_\Index\, d^4x\sim 0\quad \then\quad
k^\Index=0;
&
\label{s<3,2}\\
&
(k^{\ifree}C^*_{\ifree}
+k^{[\Index\Jndex]}f'_{\Index\Jndex})\, d^4x\sim 0
\quad \then\quad
k^{\ifree}=k^{[\Index\Jndex]}=0.
\label{s<3,3}
\eea
\end{pro}

Using the relation between $H^{g,4}(\ext|d)$ and
$H^{g+4}(\ext,\Space)$ (see section \ref{DESCENT}) 
it is now immediate to
derive $H^{g}(\ext,\Space)$ for ghost numbers smaller than 3:

\begin{pro}\label{pros<3}
Cohomology $H^g(\ext,\Space)$ for ghost numbers
$g<3$:
\bea
&&\ext f^g=0\ \then\ 
f^g\sim
\left\{\ba{ll}
0 & \mbox{for}\quad g<0\\
k & \mbox{for}\quad g=0\\
k_\Index\4H^\Index & \mbox{for}\quad g=1\\
k_{\ifree}\4F^{\ifree}+\sfrac 12 k_{[\Index\Jndex]}\4H^\Index \4H^\Jndex
& \mbox{for}\quad g=2,\\
\ea\right.
\nonumber\\
&&\4H^\Index=\Ii(\hatpsi{}^\Index\5\xi-\xi\7\psi^\Index)
+\7c^\mu \7H_\mu^\Index,
\nonumber\\
&&\4F^{\ifree}=
\7c^\mu(\xi\sigma_\mu\hatlambda{}^{\ifree}+\7\lambda^{\ifree}\sigma_\mu\5\xi)
+\sfrac 14\7c^\mu\7c^\nu\vep_{\mu\nu\rho\sigma}\7F^{\ifree\rho\sigma},
\label{s<3}
\eea
where $\sim$ is equivalence in $H(\ext,\Space)$ ($f^g\sim f^g+\ext f^{g-1}$
with $f^g,f^{g-1}\in\Space$),
$\ifree$ runs over those Abelian elements of $\lieg$ under which
all matter fields are uncharged,
and $k$, $k_\Index$, $k_{\ifree}$ and $k_{[\Index\Jndex]}$ are
complex numbers.
The cocycles 1, $\4H^\Index$, $\4F^{\ifree}$ and $\4H^{\Index}\4H^{\Jndex}$,
$\Index<\Jndex$, are
nontrivial and inequivalent in $H(\ext,\Space)$:
\bea
&&
k\sim 0\quad\then\quad k=0;
\label{s<3,6}\\
&&
k_\Index\4H^\Index\sim 0\quad\then\quad k_\Index=0;
\label{s<3,7}\\
&&
k_{\ifree}\4F^{\ifree}+\sfrac 12 k_{[\Index\Jndex]}\4H^\Index \4H^\Jndex
\sim 0\ \then\ k_{\ifree}=k_{[\Index\Jndex]}=0.\quad
\label{s<3,8}
\eea
\end{pro}

\comment
In $n$-dimensional theories,
the representatives of $H^n_{n-p}(\delta|d)$ are related 
through descent equations for $\delta$ and $d$ to 
conserved local $p$-forms (i.e., $p$-forms which do not depend on
antifields and satisfy
$d\omega^p\approx 0$) representing
the so-called characteristic cohomology of the field
equations \cite{BBH1}. Therefore one can conclude from 
lemmas  \ref{prodelta|d} and \ref{pros|d<-1} that
$H^{-3,4}(\ext|d)$ and $H^{-2,4}(\ext|d)$
are isomorphic to the characteristic cohomology
in form-degrees $p=1$ and $p=2$, respectively, and
that the latter is represented by
the 1-forms $\star dB^\Index$ and the 2-forms
$\star dA^{\ifree}$ and $(\star dB^\Index)(\star dB^\Jndex)$
where $\star$ denotes Hodge-dualization and
$A^{\ifree}=dx^\mu A^{\ifree}_\mu$ and 
$B^\Index=(1/2)dx^\mu dx^\nu B^\Index_{\mu\nu}$.
We also observe
that $H^0(\ext,\Space)$, $H^1(\ext,\Space)$ and $H^2(\ext,\Space)$
are isomorphic to the characteristic cohomology
in form-degrees $p=0$, $p=1$ and $p=2$, respectively.
We shall see that a
similar result does not hold for $p=3$, 
cf.\ section \ref{negative}.

\subsection{Elimination of trivial pairs}\label{`vii.1'}\label{pairs}

To compute $H^g(\ext,\Space)$ for
ghost numbers $g\geq 3$ we use
the jet-coordinates $u^\ell,v^\ell,w^\Windex$
given in section \ref{CALC}.
Thanks to Eq.\ (\ref{susw}), the jet-variables 
$u^\ell$ and $v^\ell$ form trivial pairs in the
terminology of \cite{ten} and drop from (the nontrivial part of)
$H(\ext,\Space)$:

\begin{pro}\label{propairs}
$H(\ext,\Space)$ is isomorphic to the cohomology of $\ext$ in the
space $\Wspace$ of local functions of the variables $w^\Windex$
listed in Eq.\ (\ref{w}):
\[
H(\ext,\Space)\simeq H(\ext,\Wspace),\quad \Wspace=\{f(w)\}.
\]
\end{pro}

\subsection{Decomposition of the cohomological problem}
\label{`vii.2'}\label{decomps}

To compute $H(\ext,\Wspace)$, we use the counting operator
$N$ for the variables $\7C^i$, $c^{\mu\nu}$ and $\7R^\Index$
as a filtration,
\[
N=\7C^i\,\frac{\6}{\6\7C^i}
+\sfrac 12 c^{\mu\nu}\, \frac{\6}{\6c^{\mu\nu}}
+\7R^\Index\,\frac{\6}{\6\7R^\Index}\ .
\]
$\ext$ decomposes in $\Wspace$ into three parts with
$N$-degrees 1, 0 and $-1$ which we denote
by $\lie$, $\susy$ and $\curv$, respectively:
\bea
&&\ext f(w)=(\lie+\susy+\curv)f(w)
\nonumber\\
&&[N,\lie]=\lie,\quad [N,\susy]=0,\quad [N,\curv]=-\curv
\nonumber\\
&&\lie=\sfrac 12\f kji\7C^j\7C^k\,\frac{\6}{\6\7C^i}
       -c_\nu{}^\rho c_\rho{}^\mu\,\frac{\6}{\6 c_\nu{}^\mu}
\nonumber\\
&&\phantom{\lie=}
       +c_\nu{}^\mu \7c^\nu\,\frac{\6}{\6 \7c^\mu}
       +\sfrac 12c^{\mu\nu}(\xi\sigma_{\mu\nu})^\alpha\,
       \frac{\6}{\6 \xi^\alpha}
       -\sfrac 12c^{\mu\nu}(\5\sigma_{\mu\nu}\5\xi)^\da\,
       \frac{\6}{\6 \5\xi^\da}
\nonumber\\
&&\phantom{\lie=}
       +(\7C^i\delta_i+\sfrac 12 c^{\mu\nu}l_{\mu\nu})\7T^\Tindex\,
       \frac{\6}{\6 \7T^\Tindex}
\nonumber\\
&&\susy=2\Ii\xi\sigma^\mu\5\xi\,\frac{\6}{\6 \7c^\mu}
       +(\7c^\mu\7\nabla_\mu+\xi^\alpha\cD_\alpha
         +\5\xi^\da\5\cD_\da)\7T^\Tindex\,
       \frac{\6}{\6 \7T^\Tindex}
\nonumber\\
&&\curv=\cF^i\,\frac{\6}{\6\7C^i}+\cH^\Index\,\frac{\6}{\6\7R^\Index}
\label{decomp}
\eea
where we introduced
\bea
&&\cF^i=\Ii\7c^\mu(\xi\sigma_\mu\hatlambda{}^i-\7\lambda^i\sigma_\mu\5\xi)
  +\sfrac 12\7c^\mu\7c^\nu\7F^i_{\mu\nu}
\label{cF}\\
&&\cH^\Index=-2\7c^\mu\xi\sigma_\mu\5\xi\phi^\Index
  -\Ii\7c^\mu\7c^\nu(\xi\sigma_{\mu\nu}\7\psi^\Index
  -\hatpsi{}^\Index\5\sigma_{\mu\nu}\5\xi)
  -\sfrac{\Ii}6\7c^\mu\7c^\nu\7c^\rho\vep_{\mu\nu\rho\sigma}\7H^{\Index\sigma}.
\label{cH}
\eea
$\ext^2=0$ decomposes under the $N$-degree according to:
\bea
\lie^2=\curv^2=\{\lie,\susy\}=\{\curv,\susy\}=0,\quad
\{\lie,\curv\}+\susy^2=0.
\label{N2}
\eea
Let us denote by $f_m$ the piece with $N$-degree $m$ of a
function $f\in\Wspace$, and by
$\om$ and $\um$ the highest and lowest $N$-degrees
contained in $f$, respectively\footnote{We can 
always assume that $\om$ is finite because
it is bounded by the ghost number of $f$ (recall that there are no
$w$'s with negative ghost number).}.
The cocycle condition $\ext f(w)=0$ in $H(\ext,\Wspace)$ decomposes into:
\bea
&\lie f_\om=0&
\label{co1}\\
&\susy f_\om+\lie f_{\om-1}=0&
\label{co2}\\
&\curv f_\om+\susy f_{\om-1}+\lie f_{\om-2}=0&
\label{co3}\\
&\vdots&
\nonumber\\
&\curv f_{\um}=0.&
\label{co4}
\eea

\subsection{Lie algebra cohomology}\label{`vii.3'}\label{secLAC}

(\ref{co1}) shows that $f_\om$ is a cocycle of $\lie$.
We can assume that it is not a coboundary of $\lie$ because
otherwise we could remove it from the $\ext$-cocycle $f(w)$
without changing the cohomology class of the latter
(if $f_\om=\lie h_{\om-1}$, one replaces $f$ with 
$f-\ext h_{\om-1}$ which is equivalent
to $f$ in $H(\ext,\Wspace)$). Hence, $f_\om$ can be assumed to
be a nontrivial representative of 
the cohomology of $\lie$ in $\Wspace$ which
we denote by $H(\lie,\Wspace)$. This cohomology is well-known: it is
the Lie algebra cohomology
of $\liefull=\lieg+{\mathfrak so}(1,3)$, with $\liefull$
represented on the local functions of
$\7c^\mu$, $\xi^\alpha$, $\5\xi^\da$ and $\7T^\Tindex$. 
This cohomology is generated by so-called primitive
elements $\theta_\Lindex$ constructed of the
$\7C^i$ and $c^{\mu\nu}$,
$\liefull$-invariant functions of the
$\7c^\mu$, $\xi^\alpha$, $\5\xi^\da$ and $\7T^\Tindex$,
and linearly independent polynomials in the $\7R^\Index$
(notice that the $\7R^\Index$ are inert to $\lie$).
The $\theta$'s correspond one-to-one to the independent
Casimir operators of $\liefull$. 
The index $\Lindex$ of the $\theta$'s runs thus
from 1 to ${\mathrm rank}(\liefull)=\rankg+2$.
The $\theta$'s of $\lieg$ can be constructed by means of
suitable matrix representations $\{T_i^{(\Lindex)}\}$ of $\lieg$
(the superscript $(\Lindex)$ of $T_i^{(\Lindex)}$ indicates
that the respective representation may depend on the
value of $\Lindex$):
\bea
\theta_\Lindex=(-)^{m(\Lindex)-1}
\frac{m(\Lindex)!(m(\Lindex)-1)!}{(2m(\Lindex)-1)!}\,
\TR(\7C^{2m(\Lindex)-1}),\ \7C=\7C^i T_i^{(\Lindex)},\
\Lindex=1,\dots,\rankg,
\label{theta}
\eea
where $m(\Lindex)$ is the order of the corresponding Casimir operator of
$\lieg$.  
The $\theta$'s with $m(\Lindex)=1$ can be taken to
coincide with the Abelian $\7C$'s
by choosing $T_i^{(\Lindex)}=1$ for one of the Abelian elements of $\lieg$
and  $T_i^{(\Lindex)}=0$ for all other elements. We denote
the Abelian $\7C$'s by $\7C^{i_A}$:
\bea
\{\theta_\Lindex:\ m(\Lindex)=1\}=\{\7C^{i_A}\}=
\{\mbox{Abelian $\7C$'s}\}.
\label{abeliantheta}
\eea
${\mathfrak so}(1,3)$ contributes two additional $\theta$'s
with $m(\Lindex)=2$
(``Lorentz-$\theta$'s''):
\bea
\theta_{\rankg+1}=c_\mu{}^\nu c_\nu{}^\rho c_\rho{}^\mu,
\quad
\theta_{\rankg+2}=\vep_{\mu\nu\rho\sigma}
c^{\mu\nu}c^{\rho\lambda}c_\lambda{}^\sigma.
\label{Lorentztheta}
\eea
We denote the space of $\liefull$-invariant local 
functions of the $\7c^\mu$, $\xi^\alpha$, $\5\xi^\da$ and $\7T^\Tindex$
by $\inv$,
\bea
\inv=\{f(\7c,\xi,\5\xi,\7T):\ \delta_i f(\7c,\xi,\5\xi,\7T)=0,\ 
l_{\mu\nu}f(\7c,\xi,\5\xi,\7T)=0\}.
\eea
$H(\lie,\Wspace)$ can now be described as follows:
\bea
&&\lie f(w)=0\quad \LRA\quad f(w)=\lie h(w)+f^\Pindex P_\Pindex(\theta,\7R),
\quad f^\Pindex\in\inv;
\label{LAC1}\\
&&f^\Pindex P_\Pindex(\theta,\7R)=\lie h(w),\quad f^\Pindex\in\inv\quad
\then\quad f^\Pindex=0,
\label{LAC2}\eea
where $\{P_\Pindex(\theta,\7R)\}$ is a basis of the monomials
in the $\theta_\Lindex$ and $\7R^\Index$.
As mentioned above, this expresses the Lie algebra
cohomology of $\liefull$, with representation space given by
the local functions of the 
$\7c^\mu$, $\xi^\alpha$, $\5\xi^\da$ and $\7T^\Tindex$.
It is a well-known result, see, e.g., section 8 of \cite{report},
and implies directly the following: 

\begin{pro}\label{prohigh}
The piece with highest $N$-degree of a
representative $f$ of $H(\ext,\Wspace)$ can be assumed to be of the form
\bea
f_\om=f^{\Pindex_\om} P_{\Pindex_\om}(\theta,\7R),\quad 
\susy f^{\Pindex_\om}=0,\quad f^{\Pindex_\om}\neq\susy g^{\Pindex_\om},\quad
f^{\Pindex_\om},g^{\Pindex_\om}\in\inv,
\label{high}
\eea
where $\{P_{\Pindex_m}(\theta,\7R)\}=
\{P_{\Pindex}(\theta,\7R):N P_{\Pindex}(\theta,\7R)=
m P_{\Pindex}(\theta,\7R)\}$.
\end{pro}

\subsection{Supersymmetry algebra cohomology}\label{`vii.4'}\label{SUSY}

(\ref{high}) shows that the functions $f^{\Pindex_\om}$ 
are representatives of the cohomology of $\susy$
in the space $\inv$. We denote this cohomology by
$H(\susy,\inv)$. It is indeed well-defined because
$\susy$ squares to zero on all functions in $\inv$:
according to (\ref{N2}) one has $\susy^2 f=
-\{\lie,\curv\}f$ which vanishes for $f\in\inv$
because one has $\lie f=0$ and $\curv f=0$ for
$f\in\inv$ by definition of $\inv$. In particular one has 
\bea
f\in\inv\quad\then\quad \susy f=\ext f,
\label{susyinW}
\eea 
and thus 
\bea
H(\susy,\inv)=H(\ext,\inv).
\eea
The following lemma describes this cohomology
and is a key result of the computation.

\begin{pro}\label{prosusy}
Cohomology $H(\susy,\inv)$:

(i) The general solution of the cocycle condition in
$H^g(\susy,\inv)$ for the various ghost numbers $g$ is:
\bea
&&\susy f^g=0,\ f^g\in\inv\ \LRA\
f^g\sim\left\{\ba{ll}
k & \mbox{for}\quad g=0\\
k_\Index\4H^\Index & \mbox{for}\quad g=1\\
k_{\ifree}\4F^{\ifree}+\sfrac 12 k_{[\Index\Jndex]}\4H^\Index \4H^\Jndex
+k_{i_A}\cF^{i_A}
& \mbox{for}\quad g=2\\
\cO R(\7T)& \mbox{for}\quad g=3\\
\cP \Omega_1+\5\cP \5\Omega_2 & \mbox{for}\quad g=4\\
0& \mbox{for}\quad g\geq 5\\
\ea\right.
\nonumber\\
&&\cO=4\7c^\mu\xi\sigma_\mu\5\xi
-4\,\Xi_{\mu\nu}(\xi\sigma^{\mu\nu}\cD+\5\xi\5\sigma^{\mu\nu}\5\cD)
-\sfrac{\Ii}{2}\Xi_\mu
\sigma^\mu_{\alpha\da}[\cD^\alpha,\5\cD^\da]
\nonumber\\
&&
\cP=
-16\,\Xi_{\mu\nu}\,\5\xi\5\sigma^{\mu\nu}\5\xi
-4\Ii\,\Xi_\mu\5\xi\5\sigma^\mu\cD
+\Xi\,\cD^2
\nonumber\\
&&
\Xi_{\mu\nu}=-\sfrac 14\vep_{\mu\nu\rho\sigma}\7c^\rho\7c^\sigma,\quad
\Xi_\mu=-\sfrac 16\vep_{\mu\nu\rho\sigma}\7c^\nu\7c^\rho\7c^\sigma,\quad
\Xi=-\sfrac 1{24}\vep_{\mu\nu\rho\sigma}\7c^\mu\7c^\nu\7c^\rho\7c^\sigma
\nonumber\\
&&R(\7T)\in\inv,\quad \cD^2R(\7T)=\5\cD^2R(\7T)=0
\nonumber\\
&&\Omega_i=A_i(\varphi,\7\lambda)+\5\cD^2B_i(\7T),\quad
A_i(\varphi,\7\lambda),B_i(\7T)\in\inv\quad (i=1,2),
\label{cocsusy}
\eea
where $\sim$ is equivalence in $H(\susy,\inv)$ ($f^g\sim f^g+\susy f^{g-1}$
with $f^g,f^{g-1}\in\inv$), 
$i_A$ runs over all Abelian elements of $\lieg$ and 
$\cF^{i_A}$ are the Abelian $\cF$'s (\ref{cF}),
$k$, $k_\Index$, $k_{\ifree}$, $k_{\Index\Jndex}$ and $k_{i_A}$ are complex
numbers, we used the notation $\cD^2=\cD^\alpha\cD_\alpha$,
$\5\cD^2=\5\cD_\da\5\cD^\da$, $[\cD^\alpha,\5\cD^\da]=
\cD^\alpha\5\cD^\da-\5\cD^\da\cD^\alpha$ and other notation
as in lemma \ref{pros<3}.
The functions
$A_i(\varphi,\7\lambda)$ depend only
on the undifferentiated $\varphi^\indec$ and $\7\lambda^i$ but
not on any (generalized covariant) derivatives thereof.

(ii) The cocycles 1, $\4H^\Index$, $\4F^{\ifree}$,
$\4H^{\Index} \4H^{\Jndex}$, $\Index< \Jndex$,
and $\cF^{i_A}$ are nontrivial
and inequivalent in $H(\susy,\inv)$,
a cocycle $\cO R(\7T)$
is trivial in $H(\susy,\inv)$ iff
$R(\7T)=\cD\Omega_1+\5\cD\5\Omega_2$ for some
functions 
$\Omega_i^\alpha=A^\alpha_i(\varphi,\7\lambda)+\5\cD^2B^\alpha_i(\7T)$
($i=1,2$),
and a cocycle $\cP \Omega_1+\5\cP \5\Omega_2$
is trivial in $H(\susy,\inv)$ iff both $\Omega_1=\5\cD^2X(\7T)$ and
$\5\Omega_2=-\cD^2X(\7T)$ for some (the same!) function $X(\7T)\in\inv$:\,%
\footnote{
$\Omega_1=\5\cD^2X\wedge\5\Omega_2=-\cD^2X$ is equivalent to
$A_1=0\wedge A_2=0\wedge\5\cD^2B_1=\5\cD^2X\wedge\cD^2B_2=-\cD^2X$
because 
no function $A(\varphi,\7\lambda)$ is of the form
$\5\cD^2(\dots)$.
}
\bea
&
k\sim 0\quad\LRA\quad k=0;
&
\nonumber\\
&
k_\Index\4H^\Index\sim 0\quad\LRA\quad k_\Index=0;
&
\nonumber\\
&
k_{\ifree}\4F^{\ifree}+
\sfrac 12 k_{[\Index\Jndex]}\4H^\Index \4H^\Jndex+
k_{i_A}\cF^{i_A}\sim 0\quad\LRA\quad
k_{\ifree}=k_{[\Index\Jndex]}=k_{i_A}=0;
&
\nonumber\\
&
\cO R(\7T)\sim 0\quad\LRA\quad
R(\7T)=\cD\Omega_1+\5\cD\5\Omega_2,\quad
\Omega^\alpha_i=
A^\alpha_i(\varphi,\7\lambda)+\5\cD^2B^\alpha_i(\7T);
&
\nonumber\\
&
\cP \Omega_1+\5\cP \5\Omega_2\sim 0\quad\LRA\quad
[\Omega_1=\5\cD^2X(\7T)\ \wedge\ 
\5\Omega_2=-\cD^2X(\7T)].
&
\label{cobsusy}\eea
\end{pro}

\comments
a) In section \ref{leq3} we found that the cohomology groups
$H^0(\ext,\Space)$, $H^1(\ext,\Space)$ and $H^2(\ext,\Space)$
are isomorphic to the characteristic cohomology in 
form-degrees 0, 1 and 2, respectively (see comments at the end
of that section).
A similar result holds for 
$H^0(\susy,\inv)$, $H^1(\susy,\inv)$ and $H^2(\susy,\inv)$:
they are isomorphic to the gauge invariant characteristic cohomology
(characteristic cohomology in the space of gauge invariant
local forms)
in form-degrees 0, 1 and 2, respectively. 
Indeed one can show as in \cite{BBH2,HKS} that the
latter
is in form-degrees 0, 1 and 2 represented by:
\bea
p=0:&\quad&\omega^0=k;
\label{char0}\\
p=1:&&\omega^1=k_\Index \star dB^\Index;
\label{char1'}\\
p=2:&&\omega^2=k_{\ifree}\star dA^{\ifree}+\sfrac 12 k_{[\Index\Jndex]}
(\star dB^\Index)(\star dB^\Jndex)+k_{i_A}dA^{i_A}.
\label{char2'}
\eea

b) Notice that the lemma 
characterizes the cohomology in ghost number 3
through functions $R(\7T)\in\inv$ which satisfy
$\cD^2R(\7T)=\5\cD^2R(\7T)=0$ and are determined up to contributions
of the form
$\cD\Omega_1+\5\cD\5\Omega_2$ with
$\Omega^\alpha_i=
A^\alpha_i(\varphi,\7\lambda)+\5\cD^2B^\alpha_i(\7T)$
[such contributions can be dropped because of part (ii) of the lemma].
I have not determined the general solution
to these conditions (which appears to be
a rather involved problem) but would like to add the
following comments concerning this result.

The simplest nontrivial functions $R(\7T)$ are 
complex numbers, i.e., $R(\7T)=k\in{\mathbb C}$ with $k\neq 0$.
They yield field independent representatives with ghost number 3
given by
\bea
4k\,\7c^\mu\xi\sigma_\mu\5\xi.
\label{4theta}
\eea
All other representatives $\cO R(\7T)$
contain gauge invariant 
conserved currents $j^\mu$ given by 
the antifield independent parts of
$(1/2)\5\sigma^{\mu}_{\alpha\da}[\cD^\alpha,\5\cD^\da]R(\7T)$,
\bea
j^\mu=
\Big[\sfrac 12 \5\sigma^{\mu}_{\alpha\da}
[\cD^\alpha,\5\cD^\da]R(\7T)\Big]_{\Phi^*=0}.
\label{current1}
\eea
That these currents are indeed conserved can be
directly verified by means of
the algebra (\ref{alg}) which implies, for all
$\delta_i$-invariant functions $f(\7T)$:
\bea
\delta_i f(\7T)=0\quad \then\quad 
[\cD^2,\5\cD^2]f(\7T)=
-4\Ii\7\nabla_{\alpha\da}[\cD^\alpha,\5\cD^\da]f(\7T).
\eea
Specializing this formula to $R(\7T)$,
it yields $0=\7\nabla_{\alpha\da}[\cD^\alpha,\5\cD^\da]R(\7T)$
because of $\cD^2R(\7T)=\5\cD^2R(\7T)=0$.
Owing to (\ref{nablaformula}) and the gauge invariance of
$j^\mu$ (which follows
from $R(\7T)\in\inv$), this implies
that
$j^\mu$ is indeed conserved:
\bea
0\approx\nabla_\mu j^\mu=\6_\mu  j^\mu.
\label{current2}
\eea
$j^\mu$ gives thus rise to a cocycle 
$\omega^3=(1/6) dx^\mu dx^\nu dx^\rho\vep_{\mu\nu\rho\sigma}j^\sigma$ of the
gauge invariant characteristic cohomology
in form-degree 3. One can show that this 3-form is trivial in
the gauge invariant characteristic cohomology iff
$\cO R(\7T)$ is in $H(\susy,\inv)$ equivalent to
a function (\ref{4theta}).%
\footnote{Using part (ii) of lemma \ref{prosusy},
the algebra (\ref{alg}), and Eq.\ (\ref{nablaformula}),
it is straightforward to verify that
$\cO[R(\7T)-k]\sim 0$
implies $j^\mu\approx \6_\nu S^{\nu\mu}$
with
$S^{\nu\mu}=4\Ii(\cD\sigma^{\nu\mu}\Omega_1
+\5\cD\5\sigma^{\mu\nu}\5\Omega_2)$
for some gauge invariant $\Omega^\alpha_i=
A^\alpha_i(\varphi,\7\lambda)+\5\cD^2B^\alpha_i(\7T)$.
Hence, $\cO R(\7T)\sim 4k\,\7c^\mu\xi\sigma_\mu\5\xi$ implies
the triviality of $\omega^3$ in the
gauge invariant characteristic cohomology
($\omega^3\approx d\omega^2$ with
$\omega^2=\sfrac 14 dx^\mu dx^\nu\vep_{\mu\nu\rho\sigma}S^{\rho\sigma}$).
The proof of the reversed implication is more involved.
Let me just note
that one can prove it using the descent equations,
by showing that the triviality of $\omega^3$ in the
gauge invariant characteristic cohomology implies that
the cocycle of $(\ext+d)$ which arises from $\cO R(\7T)$
(see section \ref{DESCENT}) is trivial, up to a function (\ref{4theta}), 
in the cohomology of $(\ext+d)$ on
gauge invariant functions that do not depend on the ghost fields
(treating the antifields as gauge covariant quantities).
}
Hence, except for the field independent representatives
(\ref{4theta}), nontrivial functions $\cO R(\7T)$
correspond to representatives of the 
gauge invariant characteristic cohomology in form-degree 3.
However, this
correspondence is not one-to-one as it is not
surjective:
there
are representatives of the
gauge invariant characteristic cohomology in form-degree 3
which do not have a counterpart in 
$H^3(\susy,\inv)$.
In particular the
Noether currents of supersymmetry and Poincar\'e symmetry
(and also those of other conformal symmetries)
do not correspond to representatives of
$H^3(\susy,\inv)$ as one can already deduce from the
fact that these Noether currents
are not contravariant Lorentz-vector fields (note
that $j^\mu$ in (\ref{current1}) is a
contravariant Lorentz-vector field since
$R(\7T)$ is Lorentz-invariant owing to $R(\7T)\in\inv$).

The ``generic'' representatives of the
gauge invariant characteristic cohomology in form-degree 3
involve gauge invariant Noether currents, i.e.,
they correspond to
nontrivial global symmetries of the Lagrangian
by Noethers first theorem \cite{Noether}.
In addition there are representatives which are
trivial in the 
characteristic cohomology but nevertheless
nontrivial in the
gauge invariant characteristic cohomology [accordingly
the corresponding functions $\cO R(\7T)$
are nontrivial in $H^3(\ext,\inv)$ but
trivial in $H^3(\ext,\Space)$].
These are exhausted by
the 3-forms $dB^\Index$
and $(dA^{i_A})(\star dB^\Index)$, as can be shown
as analogous results in \cite{HKS,report},
and do have counterparts in $H^3(\susy,\inv)$:
$dB^\Index$ and $(dA^{i_A})(\star dB^\Index)$ correspond to
functions
$R(\7T)$ given by $\phi^\Index$ and
the imaginary part
of $\7\lambda^{i_A}\7\psi^\Index+(1/2) \phi^\Index\cD\7\lambda^{i_A}$,
respectively:
\bea
&&
R(\7T)=-\sfrac 12k_\Index\phi^\Index\quad\then\quad
\cO R(\7T)=k_\Index\cH^\Index,\quad
j^\mu= k_\Index H^{\Index\mu},
\label{R1}
\\
&&
R(\7T)=k_{i_A\Index}
(-\Ii\7\lambda^{i_A}\7\psi^\Index
+\Ii\hatlambda{}^{i_A}\hatpsi{}^\Index
-\Ii\phi^\Index\cD\7\lambda^{i_A})
\quad\then\quad
\nonumber\\
&&
j^\mu\approx k_{i_A\Index}
[-2\vep^{\mu\nu\rho\sigma}F_{\nu\rho}^{i_A}H^\Index_\sigma
+4\6_\nu(F^{i_A\nu\mu}\phi^\Index
+\lambda^{i_A}\sigma^{\mu\nu}\psi^\Index
+\5\lambda^{i_A}\5\sigma^{\mu\nu}\5\psi^\Index
)].\quad
\label{R2}\eea
Examples of representatives containing
nontrivial Noether currents arise from
functions $R(\7T)$ that are linear
combinations of the real parts
of $\7\lambda^{\ifree}\7\psi^\Index$:
\bea
&&
R(\7T)=\sfrac 12 k_{\ifree\Index}(\7\lambda^{\ifree}\7\psi^\Index
+\hatlambda{}^{\ifree}\hatpsi{}^\Index)
\quad\then
\nonumber\\
&&
j^\mu\approx k_{\ifree\Index}[
2F^{\ifree\mu\nu}H_\nu^\Index
+\6_\nu(\vep^{\nu\mu\rho\sigma}F^{i_A}_{\rho\sigma}\phi^\Index
+2\Ii\lambda^{i_A}\sigma^{\mu\nu}\psi^\Index
-2\Ii\5\lambda^{i_A}\5\sigma^{\mu\nu}\5\psi^\Index
)].\quad\quad
\label{R3}
\eea
This contains indeed the Noether currents
$k_{\ifree\Index}F^{\ifree\mu\nu}H_\nu^\Index$ of
nontrivial global symmetries generated by
\bea
\Delta A_\mu^\ifree=k_{\ifree\Index}H^\Index_\mu,\quad
\Delta B_{\mu\nu}^\Index=\sfrac 12k_{\ifree\Index}\,
\vep_{\mu\nu\rho\sigma}F^{\ifree\rho\sigma},\quad
\Delta (\mbox{other fields})=0.\
\label{hidden}
\eea

c) 
Almost all representatives of
$H(\susy,\inv)$ depend on
antifields because the
generalized tensor fields $\7T^\Tindex$ involve
antifields, see section \ref{CALC}.
Exceptions are the field independent representatives
(these are the constants $k$ for $g=0$, the representatives
(\ref{4theta}) for $g=3$, and the representatives
$\cP \Omega_1+\5\cP \5\Omega_2$
with $\Omega_i=k_i\in{\mathbb C}$ for $g=4$), and
the representatives $k_\Index\cH^\Index$, see (\ref{R1})
($\cH^\Index$ does not depend on antifields).
In addition there are a few 
representatives $\cP \Omega_1+\5\cP \5\Omega_2$
from which the antifield dependence
can be removed by subtracting trivial terms, see
equation (\ref{8.4}). The
antifield dependence of all other representatives
cannot be removed as a consequence of the
fact that the commutator algebra of supersymmetry
and gauge transformations
closes only on-shell. However, this
changes when one uses the formulation with 
the standard auxiliary fields in which the commutator algebra
closes off-shell (cf.\ section \ref{otherform}).
In that formulation,
every representative
$\cP \Omega_1+\5\cP \5\Omega_2$ can be brought in a form
that does not depend on antifields.
The reason is that, when one uses the auxiliary fields,
the algebra (\ref{alg}) has an
off-shell counterpart which is
realized on standard gauge covariant tensor
fields (rather than on the generalized tensor
fields $\7T^\Tindex$) and
arises from (\ref{alg}) by substituting $\nabla_\mu$, $\lambda^i_\alpha$, 
$\5\lambda^i_\da$ and $F^i_{\mu\nu}$ for
their hatted counterparts. In the formulation
with auxiliary fields, $H^4(\susy,\inv)$ can thus
be represented by functions
$\cP \Omega_1(T)+\5\cP \5\Omega_2(T)$ with
$\Omega_i(T)=A_i(\varphi,\lambda)+\5\cD^2B_i(T)\in\inv$ ($i=1,2$)
where $\cD_\alpha$ and $\5\cD_\da$ are now realized
on the ordinary tensor fields $T^\Tindex$. The antifield dependence
of other representatives of $H(\susy,\inv)$ (with $g<4$) can not be
removed by the use of auxiliary fields because of their
relation to the characteristic cohomology of the field equations
which is not affected by the use of auxiliary fields.

d) 
The results for $g\geq 3$ in lemma \ref{prosusy}
follow from the ``QDS-structure'' of the supersymmetry multiplets
formed by the $\7T^\Tindex$,
see proof of the lemma. These results are thus not restricted to
the theories studied here but apply analogously
to models with more general Lagrangian, or even with different
field content, as long as the models have QDS-structure.

\subsection{Completion and result of the computation}
\label{`vii.5'}\label{RESULT}

So far we have determined the part
with highest $N$-degree of the possible representatives
of $H(\ext,\Space)$ by
analysing equations (\ref{co1}) and (\ref{co2}) which
involve only $\lie$ and $\susy$. 
We have found that this part can be assumed to be of the form
$f^\Pindex P_\Pindex(\theta,\7R)$ where the
$f^\Pindex$ are representatives of $H(\susy,\inv)$
given by lemma \ref{prosusy}.
To complete the computation of $H(\ext,\Space)$ we have to
examine which functions $f^\Pindex P_\Pindex(\theta,\7R)$ can be completed to
(inequivalent) solutions of the remaining equations
(\ref{co3}) through (\ref{co4})
involving $\curv$ in addition to $\lie$ and $\susy$.
For this purpose the following result is helpful:

\begin{pro}[\cite{cqg}]\label{prouseful}
(i) There is no nonvanishing $\ext$-closed function with degree
4 in the $\7c^\mu$.

(ii) Let $f=f_3+f_4$
be an $\ext$-cocycle where $f_3$ and $f_4$ have
degree 3 and 4 in the $\7c^\mu$, respectively.
Then $f$ is the $\ext$-variation of a function $\eta_4$
with degree 4 in the $\7c^\mu$; $\eta_4$ is unique%
\footnote{This does not exclude that there are
functions $g$ which are not entirely of degree
4 in the $\7c^\mu$ and satisfy $f=\ext g$.} 
and
the solution of $\delta_-\eta_4=f_3$ with
$\delta_-$ as in (\ref{lastlabel}).
\end{pro}

Furthermore it is extremely useful to 
complete the $\theta_\Lindex$ of the semisimple
part of the Yang-Mills group
to corresponding 
``super-Chern-Simons functions'' $\7q_\Lindex$ given by:
\bea
\7q_\Lindex&=&m(\Lindex)\int_0^1 dt\ \TR(\7C\cF_t^{m(\Lindex)-1})\quad
\mbox{if}\quad m(\Lindex)> 3,
\label{m>3}\\
\7q_\Lindex&=&\TR[\7C\cF^2-\sfrac 12\7C^3\cF+\sfrac 1{10}\7C^5
+3\Ii\,\Xi\,(\xi\7\lambda\hatlambda\hatlambda
+\5\xi\hatlambda\7\lambda\7\lambda)]\quad
\mbox{if}\quad m(\Lindex)=3,
\label{m=3}\\[4pt]
\7q_\Lindex&=&\TR(\7C\cF-\sfrac 13\7C^3)\quad
\mbox{if}\quad m(\Lindex)=2\quad\mbox{and}\quad \Lindex\leq\rankg,
\label{m=2}
\eea
with $\7C$ and $\{T_i^{(\Lindex)}\}$ 
as in (\ref{theta}), and
\bea
&&
\7\lambda_\alpha=\7\lambda^i_\alpha T_i^{(\Lindex)},\quad
\hatlambda_\da=\hatlambda{}^i_\da T_i^{(\Lindex)},\quad
\cF=\cF^i T_i^{(\Lindex)},\quad\cF_t=t\cF+(t^2-t)\7C^2.
\label{lambdaF}
\eea
The $\7q$'s satisfy
$\7q_\Lindex=\theta_\Lindex+\dots$ and
the $\7q$'s in
(\ref{m>3}) and (\ref{m=3}) are $\ext$-cocycles (in contrast
to (\ref{m=2})):

\begin{pro}\label{proq}
The super-Chern-Simons functions
(\ref{m>3}) and (\ref{m=3})
are $\ext$-closed:
\bea
\ext\7q_\Lindex=0\quad\mbox{for}\quad m(\Lindex)\geq 3.
\eea
\end{pro}

Using lemmas \ref{prosusy}, \ref{prouseful} and \ref{proq} and results
on the standard (non-extended) BRST-cohomology one derives the
following lemma containing the result of the cohomological
analysis:


\begin{pro}\label{pros} 
The representatives of $H(\ext,\Space)$ can be grouped into seven types:
\bea
\ext\,f &=& 0,\quad
f\in\Space\quad\LRA\quad f=\sum_{(i)=(1)}^{(7)}f^{(i)}+\ext\,h,\quad
h\in\Space,
   \nonumber\\
f^{(1)} &=& P^{(1)}(\7q,\Lth)\quad{\mathrm where}\quad
    \frac{\6P^{(1)}(\7q,\Lth)}{\6\7q_\Lindex}=0\quad
    {\mathrm if}\quad m(\Lindex)=2,
  \label{om1}\\
f^{(2)} &=& \Big\{\4H^\Index 
    +\!\!\!\sum_{\Lindex:m(\Lindex)=2}\!\!\! X^\Index_\Lindex\,
    \frac{\6}{\6\7q_\Lindex}\Big\} P^{(2)}_\Index(\7q,\Lth) ,
  \nonumber\\
 &&\hspace{-4ex}
 X^\Index_\Lindex=
    (\sfrac 18\phi^\Index\cP-\Ii\,\Xi_\mu \5\xi\5\sigma^\mu\7\psi^\Index
    +\sfrac 12\Xi\,\7\psi^\Index\cD)
    \,\TR(\7\lambda\7\lambda) 
    -(\Ii\,\Xi_\mu\xi\7\psi^\Index+\sfrac 12\Xi\,\7H_\mu^\Index)\,
    \TR(\7\lambda\sigma^\mu\hatlambda)
    +\CC,
  \label{om2}\\
f^{(3)} &=& \Big\{\4H^{\Index}\7R^{\Jndex} 
    +X^{[\Index\Jndex]}
    +\!\!\!\sum_{\Lindex:m(\Lindex)=2}\!\!\! 
    X^{[\Index}_\Lindex\7R^{\Jndex]}\,
    \frac{\6}{\6\7q_\Lindex}
    \Big\}
    P^{(3)}_{[\Index\Jndex]}(\7q,\Lth) ,
  \nonumber\\
 &&\hspace{-4ex}   
 X^{[\Index\Jndex]}=
    2\Ii\,\Xi_{\mu\nu}(\xi\sigma^{\mu\nu}\7\psi
    -\5\xi\5\sigma^{\mu\nu}\hatpsi)^{[\Index}\phi^{\Jndex]}
    +\Xi^\mu(\Ii\phi^{[\Index}\7\nabla_\mu\phi^{\Jndex]}
    -\7\psi^{[\Index}\sigma_\mu\hatpsi{}^{\Jndex]}),
  \label{om3}\\
f^{(4)} &=& 
    \Big\{
    \4F^\ifree
    +X^{[\ifree\jfree]}\,
    \frac{\6}{\6\7C^\jfree}
    +X^{\ifree\Index} \frac{\6}{\6\7R^\Index}
    +\!\!\!\sum_{\Lindex:m(\Lindex)=2}\!\!\! 
    X^\ifree_\Lindex\frac{\6}{\6\7q_\Lindex}    
    \Big\}
    \frac{\6P^{(4)}(\7q,\7C_\free,\Lth,\7R)}{\6\7C^\ifree}\, ,
  \nonumber\\       
 &&\hspace{-4ex}    
 X^{[\ifree\jfree]}=
    -\Ii\,\Xi_\mu\7\lambda^{[\ifree}
    \sigma^\mu\hatlambda{}^{\jfree]},
  \quad 
 X^{\ifree\Index}=
    \Xi_\mu \7\lambda^\ifree\sigma^\mu\5\xi\phi^\Index
    +\Ii\,\Xi\,\7\lambda^\ifree\psi^\Index-\CC ,
  \nonumber\\ 
  &&\hspace{-4ex}    
 X^\ifree_\Lindex=
    \Xi\,\5\xi\hatlambda{}^\ifree\TR(\7\lambda\7\lambda)
    +2\,\Xi\,\7\lambda^{\ifree}\,\TR(\7\lambda\hatlambda\5\xi)
    +\CC ,
  \label{om4}\\
f^{(5)} &=& \Big\{
    \4H^\Index\4H^\Jndex 
    +\!\!\!\sum_{\Lindex:m(\Lindex)=2}\!\!\!
    X^{[\Index\Jndex]}_\Lindex\,\frac{\6}{\6\7q_\Lindex}
    +X^{[\Index\Jndex]\Kdex}\,\frac{\6}{\6\7R^\Kdex}
    +X^{[\Index\Jndex](\Kdex\Ldex)}\,\frac{\6^2}{\6\7R^\Ldex\6\7R^\Kdex}
    \Big\}
    P^{(5)}_{[\Index\Jndex]}(\7q,\Lth,\7R) ,
  \nonumber\\
 &&\hspace{-4ex}    
 X^{[\Index\Jndex]}_\Lindex=
    \{
    \sfrac{\Ii}8\phi^{[\Index}\hatpsi{}^{\Jndex]}\5\xi\,\cP
    +\sfrac 12\Xi_\mu\,
    \5\xi\5\sigma^\mu\7\psi^{[\Index}
    (\hatpsi\5\xi-\xi\7\psi)^{\Jndex]}
    +\Xi_\mu\,\5\xi\5\sigma^{\mu\nu}\5\xi\,
    \7\psi^{[\Index}(H+\Ii\7\nabla\phi)_\nu^{\Jndex]}
  \nonumber\\
    &&\hspace{-4ex}
    \phantom{X^{[\Index\Jndex]}_\Lindex=}
    +\sfrac {\Ii}4\Xi\,
    [2\7H_\mu^{[\Index}\7\psi^{\Jndex]}\sigma^\mu\5\xi
    +(\xi\7\psi-3\hatpsi\5\xi)^{[\Index}\7\psi^{\Jndex]}\cD
    +\phi^{[\Index}(\7H+\Ii\7\nabla\phi)_\mu^{\Jndex]}
    \5\xi\5\sigma^\mu\cD
    ]
    \}
    \,\TR(\7\lambda\7\lambda)
  \nonumber\\
    &&\hspace{-4ex}
    \phantom{X^{[\Index\Jndex]}_\Lindex=}
    +
    \{
    \Xi_\mu\,\xi\7\psi^{[\Index}(\xi\7\psi-\hatpsi\5\xi)^{\Jndex]}
    +2\Ii\,\Xi\,\xi\7\psi^{[\Index}\7H_\mu^{\Jndex]}
    \}
    \,\TR(\hatlambda\5\sigma^\mu\7\lambda)
    +\CC,
  \nonumber\\
 &&\hspace{-4ex}    
  X^{[\Index\Jndex]\Kdex}=
    \Xi_{\mu\nu}\{
    2\xi\sigma^{\mu\nu}\7\psi^{[\Index}\, \xi\7\psi^{\Jndex]}\phi^\Kdex
    +\xi\sigma^\mu\hatpsi{}^{[\Index}
    \7\psi^{\Jndex]}\sigma^\nu\5\xi\phi^\Kdex
    -2\xi\sigma^{\mu\nu}\xi\,\phi^{[\Index}\7\psi^{\Jndex]}\7\psi^\Kdex
    -2\xi\sigma^\nu\5\xi\phi^{[\Index}
    \hatpsi{}^{\Jndex]}\5\sigma^\mu\7\psi^\Kdex
    \}
  \nonumber\\
    &&\hspace{-4ex}
    \phantom{X^{[\Index\Jndex]\Kdex}=}
    +\Xi_\mu\{
    -2\Ii\xi\sigma^{\mu\nu}\7\psi^{[\Index}\7H_\nu^{\Jndex]}\phi^\Kdex
    +\xi\7\psi^{[\Index}\7\nabla^\mu\phi^{\Jndex]}\phi^\Kdex
    +\sfrac{\Ii}2\xi\sigma^\mu
    \5\cD(\phi^{[\Index}\7\psi^{\Jndex]}\7\psi^\Kdex)
  \nonumber\\
    &&\hspace{-4ex}
    \phantom{X^{[\Index\Jndex]\Kdex}=} 
    +\sfrac{\Ii}2\phi^{[\Index}(\xi\cD+\5\xi\5\cD)
    (\hatpsi{}^{\Jndex]}\5\sigma^\mu\7\psi^\Kdex) 
    +\sfrac{\Ii}2
    \xi\sigma^\mu\hatpsi{}^{[\Index}\7\psi^{\Jndex]}\7\psi^\Kdex
    \}
  \nonumber\\
    &&\hspace{-4ex}
    \phantom{X^{[\Index\Jndex]\Kdex}=}  
    +\Xi\,\{
    \sfrac 12\7\psi^{[\Index}\sigma^\mu\hatpsi{}^{\Jndex]}\7H_\mu^\Kdex
    -\7\psi^\Kdex\sigma^\mu\hatpsi{}^{[\Index}\7H_\mu^{\Jndex]}
  \nonumber\\
    &&\hspace{-4ex}
    \phantom{X^{[\Index\Jndex]\Kdex}=}  
    +\sfrac{\Ii}2(
    \7H^{\mu[\Index}\phi^{\Jndex]}\7\nabla_\mu\phi^\Kdex
    -\7H^{\mu\Kdex}\phi^{[\Index}\7\nabla_\mu\phi^{\Jndex]}
    -\phi^\Kdex\7H^{\mu[\Index}\7\nabla_\mu\phi^{\Jndex]}
    )
    \}-\CC,
  \nonumber\\
 &&\hspace{-4ex}    
  X^{[\Index\Jndex](\Kdex\Ldex)}=
    2\Ii\,\Xi\, (\hatpsi{}^\Index\5\xi\,\hatpsi{}^\Jndex\5\xi\,
    \phi^\Kdex\phi^\Ldex
    -\phi^{[\Index}\hatpsi^{\Jndex]}\5\xi\,\xi\7\psi^{(\Kdex}\phi^{\Ldex)})
    -\CC,
  \label{om5}\\
f^{(6)} &=& \Big\{
    (\cO R^\Pindex(\7T)) + X^{\Pindex i_A}\,\frac{\6}{\6\7C^{i_A}}
    +\!\!\!\sum_{\Lindex:m(\Lindex)=2}\!\!\!
    X^\Pindex_\Lindex\,\frac{\6}{\6\7q_\Lindex}
    +X^{\Pindex\Index}\,\frac{\6}{\6\7R^\Index}
    \Big\}
    P_\Pindex(\7q,\7C_\Abel,\Lth,\7R),
  \nonumber\\
 &&\hspace{-4ex}    
  X^{\Pindex i_A}=
    \{-2\Ii\,\Xi_\mu \,\7\lambda^{i_A}\sigma^\mu\5\xi
    +\sfrac 12 \Xi\,(\cD\7\lambda^{i_A})
    +2\,\Xi\,\7\lambda^{i_A}\cD-\CC\}R^\Pindex (\7T),
  \nonumber\\
 &&\hspace{-4ex}    
  X^\Pindex _\Lindex=
    8\,\Xi\,R^\Pindex (\7T)\,\TR(\xi\lambda\,\hatlambda\5\xi),
  \quad   
  X^{\Pindex \Index}=
    2\Ii\,\Xi\,(\5\psi^\Index\5\xi-\xi\psi^\Index
    +\phi^\Index\xi\cD+\phi^\Index\5\xi\5\cD)R^\Pindex (\7T),
  \label{om6}\\
f^{(7)} &=& \Big\{
    (\cP\Omega^\Pindex_1+\5\cP\5\Omega^\Pindex_2)
    +8\,\Xi\,(\hatlambda{}^{i_A}\5\xi\,\Omega^\Pindex_1
    +\xi\7\lambda^{i_A}\5\Omega^\Pindex_2)\,
    \frac{\6}{\6\7C^{i_A}}
    \Big\}
    P_\Pindex(\7q,\7C_\Abel,\Lth,\7R),
  \label{om7}
\eea
where $\Lth$, $\7C_\Abel$ and $\7C_\free$ denote
collectively the Lorentz-$\theta$'s (\ref{Lorentztheta}),
the Abelian $\7C$'s, and the $\7C^\ifree$, respectively,
$\CC$ denotes 
complex conjugation when $\{T_i^{(\Lindex)}\}$ is antihermitian%
\footnote{If
$\{T_i^{(\Lindex)}\}$ is not antihermitian,
then $+\CC$ denotes the addition of terms that
were the complex conjugation
if $\{T_i^{(\Lindex)}\}$ was antihermitian
(i.e., using $-T_i^{(\Lindex)}$ in place of $T_i^{(\Lindex)\dagger}$).
},
and other notation is as in lemmas \ref{pros<3} and \ref{prosusy}.
In particular, $R^\Pindex(\7T)$, $\Omega_1^\Pindex$ and
$\5\Omega_2^\Pindex$ are functions as
$R(\7T)$, $\Omega_1$ and
$\5\Omega_2$ in (\ref{cocsusy}), respectively, and can be
assumed not to be of the trivial form given in (\ref{cobsusy}).
\end{pro}

\mysection{Consistent deformations, counterterms and anomalies}
\label{`viii'}\label{DEFOS}

We shall now discuss the results for the cohomological
groups $H^{0,4}(\ext|d)$ and $H^{1,4}(\ext|d)$ 
because of their relevance to algebraic renormalization
and consistent deformations. In 
algebraic renormalization, $H^{0,4}(\ext|d)$
yields the possible counterterms that
are Poincar\'e invariant, gauge invariant and
N=1 supersymmetric on-shell, and $H^{1,4}(\ext|d)$ provides
the Poincar\'e invariant candidate
gauge and supersymmetry anomalies to lowest nontrivial order,
cf.\ \cite{Piguet:1995er,Weinberg:1996kr,report}.

The other applications concern
the Poincar\'e invariant and
N=1 supersymmetric consistent deformations
of the classical theories.
In this context a
deformation is called consistent if it is 
a continuous
deformation%
\footnote{A deformation is called continuous if it is
a formal power series in deformation parameters.} 
of the Lagrangian and its gauge symmetries
such that the deformed Lagrangian is invariant under
the deformed gauge transformations up to a total divergence.
A deformation is considered trivial when it
can be removed through field redefinitions.
Such deformations of $n$-dimensional gauge theories
are completely controlled
by  $H^{0,n}(\gauge|d)$ and $H^{1,n}(\gauge|d)$ where
$n$ is the spacetime dimension and
$\gauge$ is the standard (non-extended) BRST differential 
\cite{BH,Henneaux:1998bm}.
In particular, 
$H^{0,n}(\gauge|d)$ provides the nontrivial
deformations to first order in the deformation parameters,
while $H^{1,n}(\gauge|d)$ 
yields all possible restrictions or obstructions
to the extendability of the deformations
to second and all higher orders.
Analogously the cohomology groups
$H^{0,n}(\ext|d)$ and $H^{1,n}(\ext|d)$ of an extended
BRST differential for gauge and global symmetries 
control those consistent deformations
that are invariant under 
the global symmetry transformations contained in $\ext$,
where these transformations may get nontrivially deformed
but their commutator algebra does not change
on-shell modulo gauge transformations.
The latter statement on the algebra of the global symmetries holds because
this algebra
is encoded in the extended BRST-transformations of the constant ghosts.
Representatives of $H^{0,n}(\ext|d)$ do not contain terms that
would modify the extended BRST-transformations of the constant ghosts
because they do not involve ``global antifields''
conjugate to the constant ghosts\footnote{See \cite{Brandt:1997cz} for the
concept of such antifields,
and \cite{FB} for a discussion of deformations of
the global symmetry algebra
in the framework of the extended antifield formalism.}. 
Hence, $H^{0,n}(\ext|d)$
yields only deformations that preserve
the algebra of the global symmetries contained in $\ext$
(in contrast, the algebra of the gauge transformations may
get nontrivially deformed). The
consistent deformations which arise from our
results on $H^{0,4}(\ext|d)$ are thus precisely those 
which are Poincar\'e invariant and N=1 supersymmetric,
where the supersymmetry transformations
may get nontrivially deformed but their
algebra is still the standard N=1 supersymmetry
algebra (on-shell, modulo gauge transformations).
In view of the results derived in \cite{Haag:1975qh} 
it is unlikely that there are more general physically reasonable
deformations (which change nontrivially 
the N=1 supersymmetry algebra) 
but this question is
beyond the scope of this work.

We shall now spell out explicitly the 
antifield independent parts of the representatives
of $H^{0,4}(\ext|d)$. These give the
first order deformations of the Lagrangian,
and the possible counterterms to the Lagrangian that are
invariant, at least on-shell, under
the gauge, Lorentz and supersymmetry transformations
up to total divergences (the parts with antifield numbers 
1 and 2 of the representatives
of $H^{0,4}(\ext|d)$ yield the corresponding
first order deformations of, and counterterms to, the
symmetry transformations and their commutator algebra,
respectively).
As explained in section  \ref{DESCENT},
the representatives of $H^{0,4}(\ext|d)$
are obtained from those of $H^4(\ext,\Space)$ 
by substituting $c^\mu+dx^\mu$ for $c^\mu$
and then picking the 4-form of the resultant expression.
The representatives of $H^4(\ext,\Space)$ are obtained
from lemma \ref{pros} by selecting from among the
functions $f^{(i)}$ those with ghost number 4.%
\footnote{
There are no functions $f^{(1)}$ or $f^{(3)}$
with ghost number 4. The functions $f^{(2)}$
with ghost number 4 arise from polynomials
$P^{(2)}_\Index$
that are linear combinations $k^\Lindex_\Index\7q_\Lindex$
of the $\7q_\Lindex$ with $m(\Lindex)=2$ and of the
Lorentz-$\theta$'s where, however, 
only $k^\Lindex_\Index\7q_\Lindex$ yields deformations
and counterterms
of the Lagrangian, symmetry transformations and their algebra
(the representatives with Lorentz-$\theta$'s have antifield
number 3).
The functions $f^{(4)}$, $f^{(5)}$,
$f^{(6)}$, and $f^{(7)}$ with ghost number 4 arise from 
$P^{(4)}=(1/6)k_{[\ifree\jfree\kfree]}
\7C^\ifree\7C^\jfree\7C^\kfree
-\Ii k_{\ifree\Index}\7C^\ifree\7R^\Index$,
$P^{(5)}_{[\Index\Jndex]}=\Ii k_{[\Index\Jndex]\Kdex}\7R^\Kdex$,
$P_\Pindex=-\Ii\7C^{i_A}$, and $P_\Pindex=1$
respectively,
where we introduced
cosmetic factors $1/6$ and $\pm\Ii$.
}
Up to trivial terms, the antifield
independent part of the
general representative of $H^{0,4}(\ext|d)$ is:
\bea
&&
\omega^{0,4}|_{\Phi^*=0}=
d^4x\,\Big(\stackrel{(1)}{L}_{{\mathrm CM}}+
\stackrel{(1)}{L}_{{\mathrm SYM}}+
\stackrel{(1)}{L}_{{{\mathrm St}}}+
\stackrel{(1)}{L}_{{\mathrm FT}}+
\stackrel{(1)}{L}_{{\mathrm Noe,CM',CS,FI}}+
\stackrel{(1)}{L}_{{\mathrm generic}}\Big)
\nonumber\\[4pt]
&&
\stackrel{(1)}{L}_{{\mathrm CM}}=
\!\!\!\sum_{\Lindex:m(\Lindex)=2}\!\!\!
k_\Index^\Lindex
\Big\{
H_\mu^\Index\,\TR[
\vep^{\mu\nu\rho\sigma}(A_\nu\6_\rho A_\sigma
+\sfrac 23 A_\nu A_\rho A_\sigma)
-\lambda\sigma^\mu\5\lambda
]
\nonumber\\[-14pt]
&&
\phantom{
\stackrel{(1)}{L}_{{\mathrm CM}}=
\!\!\!\sum_{\Lindex:m(\Lindex)=2}\!\!\!
k_\Index^\Lindex
\Big\{
}
+\phi^\Index\,\TR[
-\sfrac 12 F_{\mu\nu}F^{\mu\nu}
+\Ii\nabla_\mu\lambda\sigma^\mu\5\lambda
-\Ii\lambda\sigma^\mu\nabla_\mu\5\lambda
-\sfrac 14 (\cD\lambda)^2
]
\nonumber\\[-14pt]
&&
\phantom{
\stackrel{(1)}{L}_{{\mathrm CM}}=
\!\!\!\sum_{\Lindex:m(\Lindex)=2}\!\!\!
k_\Index^\Lindex
\Big\{
}
-\TR[
(\psi^\Index\sigma^{\mu\nu}\lambda
+\5\psi^\Index\5\sigma^{\mu\nu}\5\lambda)F_{\mu\nu}
-\sfrac{\Ii}{2}(\psi^\Index\lambda+\5\psi^\Index\5\lambda)\cD\lambda
]
\Big\}
\label{CM}\\[-8pt]
&&
\stackrel{(1)}{L}_{{\mathrm SYM}}=
k_{[\ifree\jfree\kfree]}
(-\sfrac 12 A_\mu^\ifree A_\nu^\jfree F^{\kfree \mu\nu}
+\Ii \lambda^\ifree\sigma^\mu\5\lambda^\jfree A_\mu^\kfree)
\label{YM}\\[4pt]
&&
\stackrel{(1)}{L}_{{\mathrm St}}=
k_{\ifree\Index}(\sfrac 12 F^{\ifree \mu\nu} B_{\mu\nu}^\Index
+\lambda^\ifree \psi^\Index+\5\lambda^\ifree \5\psi^\Index)
\label{St}\\[4pt]
&&
\stackrel{(1)}{L}_{{\mathrm FT}}=
k_{[\Index\Jndex]\Kdex}\Big[
\sfrac 12\vep^{\mu\nu\rho\sigma}
H_\mu^\Index H_\nu^\Jndex B_{\rho\sigma}^\Kdex
-H^{\mu\Index}\phi^{\Jndex}\6_\mu\phi^\Kdex
+H^{\mu\Kdex}\phi^{\Index}\6_\mu\phi^{\Jndex}
+\phi^\Kdex H^{\mu\Index}\6_\mu\phi^{\Jndex}
\nonumber\\[-3pt]
&&
\phantom{
\stackrel{(1)}{L}_{{\mathrm FT}}=
k_{[\Index\Jndex]\Kdex}\Big[
}
+\Ii\psi^{\Index}\sigma^\mu\5\psi^{\Jndex} H_\mu^\Kdex
+\Ii(\psi^{\Index}\sigma^\mu\5\psi^\Kdex
-\psi^\Kdex\sigma^\mu\5\psi^{\Index}) H_\mu^{\Jndex}
\Big]
\label{FT}\\
&&
\stackrel{(1)}{L}_{{\mathrm Noe,CM',CS,FI}}=
(\sfrac 12 A_\mu^{i_A}\5\sigma^{\mu\alpha\da}
[\cD_\alpha,\5\cD_\da]
-\Ii(\cD\lambda^{i_A})-2\Ii\lambda^{i_A}\cD+2\Ii\5\lambda^{i_A}\5\cD)R_{i_A}(T)
\label{Noether}\\[4pt]
&&
\stackrel{(1)}{L}_{{\mathrm generic}}=
\cD^2[A_1(\varphi,\lambda)+\5\cD^2B_1(T)]
+\5\cD^2[\5A_2(\5\varphi,\5\lambda)+\cD^2\5B_2(T)]
\label{generic}
\eea
where
the $k$'s are complex numbers, 
$R_{i_A}(\7T)$, $A_i(\varphi,\7\lambda)$ and $B_i(\7T)$ are functions as 
in lemma \ref{prosusy}, 
$T^\Tindex$, $\cD_\alpha T^\Tindex$, $\5\cD_\da T^\Tindex$
denote the antifield independent parts of
$\7T^\Tindex$, $\cD_\alpha \7T^\Tindex$, $\5\cD_\da \7T^\Tindex$,
respectively, 
$[\cD_\alpha,\5\cD_\da]$ is the commutator of $\cD_\alpha$ and $\5\cD_\da$,
and $A_\mu$, $\lambda$, $\5\lambda$ are matrices
constructed from the gauge fields and gauginos:
\bea
&
T^\Tindex=\7T^\Tindex|_{\Phi^*=0} ,\quad
\cD_\alpha T^\Tindex=(\cD_\alpha \7T^\Tindex)|_{\Phi^*=0} ,\quad
\5\cD_\da T^\Tindex=(\5\cD_\da \7T^\Tindex)|_{\Phi^*=0} ,\quad
[\cD_\alpha,\5\cD_\da]=\cD_\alpha\5\cD_\da-\5\cD_\da\cD_\alpha,
&
\nonumber\\
&
A_\mu=A_\mu^iT^{(\Lindex)}_i,\quad
\lambda_\alpha=\lambda^i_\alpha T^{(\Lindex)}_i,\quad
\5\lambda_\da=\5\lambda^i_\da T^{(\Lindex)}_i.
&
\label{defsnot}\eea

Analogously one obtains
the representatives of $H^{1,4}(\ext|d)$ 
from  the
functions $f^{(i)}$  with ghost number 5 given in lemma \ref{pros}.
The antifield independent parts of these representatives are 
(the superscripts indicate
from which $f^{(i)}$ they derive):
\bea 
&&\cA^{(1)}=
\!\!\!\sum_{\Lindex:m(\Lindex)=3}\!\!\! k^\Lindex\,
\TR\left\{ Cd(AdA+\sfrac 12A^3)
+\Ii(\xi\sigma\,\5\lambda+\lambda\,\sigma\5\xi)
(AdA+(dA)A+\sfrac 32A^3)
+3\Ii\, d^4x\, (\5\xi\5\lambda\, \lambda\lambda
+\xi\lambda\, \5\lambda\5\lambda)
\right\}
              \label{anchir}\\
&&\cA^{(4a)}=
k_{[\ifree\jfree\kfree\lfree]}\left\{
\sfrac 12(\star F^\ifree)A^\jfree A^\kfree C^\lfree
+\sfrac 16(\xi\sigma\5\lambda^\ifree-\lambda^\ifree\sigma\5\xi)
A^\jfree A^\kfree A^\lfree
+\Ii\, d^4x\,\lambda^\ifree\sigma^\mu\5\lambda^\jfree A_\mu^\kfree C^\lfree
\right\}
\label{an4a}\\[4pt]
&&\cA^{(4b)}=
\!\!\!\sum_{\Lindex:m(\Lindex)=2}\!\!\!
k^\Lindex_\ifree\left\{
(\star F^\ifree)\,
\TR(CdA+\Ii A\xi\sigma\5\lambda+\Ii A\lambda\sigma\5\xi)
+(\xi\sigma\5\lambda^\ifree-\lambda^\ifree\sigma\5\xi)\,
\TR(AdA+\sfrac 23A^3)
\right.
\nonumber\\[-12pt]
&&\phantom{
\cA^{(4b)}=
\!\!\!\sum_{\Lindex:m(\Lindex)=2}\!\!\!k^\Lindex_\ifree\Big\{
}
\left.
+d^4x\left[\5\xi\5\lambda^\ifree\,\TR(\lambda\lambda)
+2\lambda^\ifree\,\TR(\lambda\5\lambda\5\xi)+\CC\right]\right\}
\label{an4b}\\[-6pt]
&&\cA^{(4c)}=
k_{[\ifree\jfree]\Index}\left\{
(\star F^\ifree)(A^\jfree Q^\Index+C^\jfree B^\Index)
+(\xi\sigma\5\lambda^\ifree-\lambda^\ifree\sigma\5\xi)A^\jfree B^\Index
\right.
\nonumber\\
&&\phantom{
\cA^{(4c)}=k_{[\ifree\jfree]\Index}\Big\{
}
\left.
+d^4x\,(\sfrac{\Ii}{2}
\lambda^\ifree\sigma^\mu\5\lambda^\jfree Q^\Index_\mu
+\Ii \lambda^\ifree\sigma^\mu\5\xi A^\jfree_\mu
-\lambda^\ifree\psi^\Index C^\jfree +\CC)
\right\}
\label{an4c}\\
&&\cA^{(5)}=
\!\!\!\sum_{\Lindex:m(\Lindex)=2}\!\!\!k^\Lindex_{[\Index\Jndex]}\left\{
(\star dB^\Index)(\star dB^\Jndex)\,
\TR(CdA+\Ii A\xi\sigma\5\lambda+\Ii A\lambda\sigma\5\xi)
+2\Ii (\star dB^\Index)(\5\psi^\Jndex\5\xi-\xi\psi^\Jndex)\,
\TR(AdA+\sfrac 23A^3)
\right.
\nonumber\\[-12pt]
&&
\phantom{
\cA^{(5)}=
\!\!\!\sum_{\Lindex:m(\Lindex)=2}\!\!\!k^\Lindex_{[\Index\Jndex]}\Big\{
}
\left.
+d^4x\left[
2\Ii\, \5\psi^\Index\5\xi\, H_\mu^\Jndex\,
\TR(\lambda\sigma^\mu\5\lambda)
+\cP^{\Index\Jndex}\,\TR(\lambda\lambda)
+\CC
\right]\right\}
\nonumber\\[-6pt]
&&\phantom{
\cA^{(5)}=
}
\mbox{with}\quad 
\cP^{\Index\Jndex}=\sfrac{\Ii}{8}\phi^\Index\,\5\psi^\Jndex\5\xi\,\cD^2
+\sfrac{\Ii}{4}\phi^\Index(H_\mu+\Ii\6_\mu\phi)^\Jndex\,\5\xi\5\sigma^\mu\cD
+\sfrac{\Ii}{4}(\xi\psi^\Index-3\5\psi^\Index\5\xi)\,\psi^\Jndex\cD
+\sfrac{\Ii}{2}H_\mu^\Index\,\psi^\Jndex\sigma^\mu\5\xi
\label{an5}\\[4pt]
&&\cA^{(6a)}=
d^4x\left\{
\sfrac 14 C^{i_A}A_{\alpha\da}^{j_A}\,[\cD^\alpha,\5\cD^\da]
+2\Ii A_\mu^{i_A}A_\nu^{j_A}\,\xi\sigma^{\mu\nu}\cD
-2\Ii C^{i_A}\lambda^{j_A}\cD
\right.
\nonumber\\
&&
\phantom{
\cA^{(6a)}=
d^4x\,\Big\{
}
\left.
-\sfrac{\Ii}{2}C^{i_A}(\cD\lambda^{j_A})
+2\xi\sigma^\mu\5\lambda^{i_A}A_\mu^{j_A}
+\CC
\right\}R_{[i_Aj_A]}(T)
\label{an6a}\\
&&\cA^{(6b)}=
d^4x\left\{
\sfrac 14 Q^\Index_{\alpha\da}\,[\cD^\alpha,\5\cD^\da]
+4\Ii B_{\mu\nu}^\Index\,\xi\sigma^{\mu\nu}\cD
-2\Ii\,\xi\psi^\Index
+2\Ii \phi^\Index\,\xi\cD+\CC\right\}R_\Index(T)\quad
\label{an6b}\\[6pt]
&&\cA^{(7)}=
d^4x\left(C^{i_A}\cD^2-4\Ii A_\mu^{i_A}\,\5\xi\5\sigma^\mu\cD
+8\5\lambda^{i_A}\5\xi\right)
\left[A_{i_A}(\varphi,\lambda)+\5\cD^2B_{i_A}(T)\right]
\nonumber\\[4pt]
&&
\phantom{\cA^{(7)}=}
+d^4x\left(C^{i_A}\5\cD^2+4\Ii A_\mu^{i_A}\,\xi\sigma^\mu\5\cD
+8\xi\lambda^{i_A}\right)\left[A'_{i_A}(\5\varphi,\5\lambda)+\cD^2B'_{i_A}(T)
\right]
\label{an7}
\eea
where
$R_{[i_Aj_A]}(\7T)$ and $R_\Index(\7T)$ are
functions as $R(\7T)$ in
lemma \ref{prosusy},
$A_{i_A}(\varphi,\lambda)$, $B_{i_A}(T)$,
$A'_{i_A}(\5\varphi,\5\lambda)$, $B'_{i_A}(T)$ are
gauge invariant and Lorentz invariant functions,
$\CC$ is used as in lemma \ref{pros},
and
\bea 
&
C=C^iT^{(\Lindex)}_i,\quad 
Q^\Index=dx^\mu Q_\mu^\Index,\quad
\sigma_{\alpha\da}=dx^\mu\sigma_{\mu\, \alpha\da},\quad
A^i=dx^\mu A^i_\mu,\quad
A=A^i T^{(\Lindex)}_i,
&
\\
&
\star F^i=\sfrac 14 dx^\mu dx^\nu\vep_{\mu\nu\rho\sigma}F^{i\rho\sigma},\quad
B^\Index=\sfrac 12 dx^\mu dx^\nu B_{\mu\nu}^\Index,\quad
\star dB^\Index= dx^\mu H_\mu^\Index.
&
\label{AL}\eea
We leave it to the reader to work out the
antifield dependent terms of the representatives
of $H^{0,4}(\ext|d)$ and $H^{1,4}(\ext|d)$ 
and add the following comments:

a)
$\stackrel{(1)}{L}_{{\mathrm CM}}$ contains
``Chapline-Manton'' couplings between the 2-form gauge potentials
and non-Abelian Chern-Simons 3-forms of the type
frequently encountered in supergravity models, see, e.g.,
\cite{NT,Bergshoeff:1982um,CM,Green:1984sg}.

b)
$\stackrel{(1)}{L}_{{\mathrm SYM}}$ contains
the cubic interaction vertices of standard super-Yang-Mills theories.
In particular it gives rise to deformations of free
supersymmetric gauge theories to standard 
non-Abelian super-Yang-Mills theories,
with the coefficients $k_{[\ifree\jfree\kfree]}$
becoming the structure constants of the
non-Abelian gauge group (the Jacobi identity
for the structure constants arises at second order
of the deformation, cf.\ \cite{BHT}; see also comment h) below).

c)
$\stackrel{(1)}{L}_{{\mathrm St}}$ 
gives rise to the supersymmetric version of the Stueckelberg mechanism
\cite{Bizdadea:1998jp} for 2-form gauge potentials.

d)
$\stackrel{(1)}{L}_{{\mathrm FT}}$ contains
the trilinear vertices $\vep^{\mu\nu\rho\sigma}
H_\mu^\Index H_\nu^\Jndex B_{\rho\sigma}^\Kdex$
of ``Freedman-Townsend models'' \cite{OP,FT}.
In particular it yields deformations of
gauge theories for free linear multiplets
to supersymmetric Freedman-Townsend models
as derived in \cite{Lindstrom:1983rt,Clark:1989gx,Brandt:1998pp}.

e)
One may distinguish four different types of functions
$\stackrel{(1)}{L}_{{\mathrm Noe,CM',CS,FI}}$, depending
on the functions $R_{i_A}(T)$ they involve.
The simplest choice is $R_{i_A}(T)=k_{i_A}\in{\mathbb C}$. It
yields Fayet-Iliopoulos terms \cite{FI}:%
\footnote{In the formulation with auxiliary fields
$\cD\lambda^{i_A}$ is proportional to the
auxiliary field $D^{i_A}$, cf.\ section \ref{otherform}.}
\bea
\stackrel{(1)}{L}_{{\mathrm FI}}=-\Ii k_{i_A}\cD\lambda^{i_A}.
\label{FI}
\eea
All other choices lead to terms containing couplings
$A_\mu^{i_A}j_{i_A}^\mu$
of Abelian gauge fields to conserved currents
$j_{i_A}^\mu=(1/2) \5\sigma^{\mu}_{\alpha\da}
[\cD^\alpha,\5\cD^\da]R_{i_A}(T)$, cf.\ comment b)
in section \ref{SUSY}.
Generically these conserved currents are nontrivial
Noether currents (up to trivial currents). 
The only exceptions are the currents which arise from
$R_{i_A}(T)=-\sfrac 12k_{i_A\Index}\phi^\Index+k_{i_Aj_A\Index}
(-\Ii\lambda^{j_A}\psi^\Index+\Ii\5\lambda{}^{j_A}\5\psi{}^\Index
-\Ii\phi^\Index\cD\lambda^{j_A})$,
see (\ref{R1}) and the text before that equation.
These $R_{i_A}$'s yield
Chern-Simons type couplings between Abelian gauge fields 
and the 2-form gauge potentials, and
Chapline-Manton couplings of the 2-form gauge potentials 
to Abelian Chern-Simons
3-forms:
\bea
&&
\stackrel{(1)}{L}_{{\mathrm CS}}=
k_{i_A\Index}A_\mu^{i_A}H^{\Index\mu}+\dots
=\sfrac 12k_{i_A\Index}\vep^{\mu\nu\rho\sigma}
A_\mu^{i_A}\6_\nu B_{\rho\sigma}^\Index+\dots
\label{CS'}
\\
&&
\stackrel{(1)}{L}_{{\mathrm CM'}}=-2k_{i_Aj_A\Index}
\vep^{\mu\nu\rho\sigma}A_\mu^{i_A}F_{\nu\rho}^{j_A}H^\Index_\sigma
+\dots
\label{CM'}
\eea
An example for a deformation involving a nontrivial
Noether current arises from $R_{i_A}(T)=
\sfrac 12 k_{i_A\ifree\Index}(\lambda^{\ifree}\psi^\Index
+\5\lambda{}^{\ifree}\5\psi{}^\Index)$ which
yields
\bea
\stackrel{(1)}{L}_{{\mathrm Noe}}=
2k_{i_A\ifree\Index}A_\mu^{i_A}F^{\ifree\mu\nu}H_\nu^\Index+\dots
\label{superHK}
\eea
Supersymmetric models with these interaction vertices
were constructed in \cite{Brandt:1998pp}.

f)
The invariants (\ref{generic}) have been termed ``generic''
because there are infinitely many of them,
with arbitrarily high mass dimensions. For instance, an invariant
with mass dimension 8 arises from a constribution
$\TR(\lambda\lambda\5\lambda\5\lambda)$ to $B_1$,
\bea
\cD^2\5\cD^2\TR(\lambda\lambda\5\lambda\5\lambda)
=\TR(16F_{\mu\nu}F_{\rho\sigma}F^{\mu\rho}F^{\nu\sigma}-
4F_{\mu\nu}F_{\rho\sigma}F^{\mu\nu}F^{\rho\sigma}+\dots).
\label{mass8}
\eea
In the formulation with 
the standard auxiliary fields (see section \ref{otherform}),
all invariants (\ref{generic})
can be written as
off-shell invariants, cf. comment c) in section \ref{SUSY}. 
Hence, in that formulation
deformations (\ref{generic}) preserve the form of the
supersymmetry and gauge transformations.
Notice also that they can be written as standard superspace
integrals in a superfield formulation
($\cD^2$ and $\5\cD^2$
then turn into superspace integrals $\int d^2\theta$ and
$\int d^2\5\theta$).

g) (\ref{anchir}) is a supersymmetric generalization of the
non-Abelian chiral anomalies. Remarkably it is not accompanied
by antifield dependent terms, i.e., it is already
a complete representative of $H^{1,4}(\ext|d)$, 
whether or not one uses auxiliary fields
(alternative forms and discussions of supersymmetric 
non-Abelian chiral anomalies
can be found in
\cite{Piguet:1984aa,Clark:1984wn,Nielsen:1984nk,%
Garreis:1985ef,%
Girardi:1985hf,Bonora:1985ib,Guadagnini:1985ar,%
Ferrara:1985me,Bonora:1985vn,Itoyama:1985qi,Harada:1985wa,%
Piguet:1986ug,Pernici:1986nb,%
McArthur:1986xd,Guadagnini:1986ea,Itoyama:1986ni,Krivoshchekov:1987ep,%
kaiser,Marinkovic:1991ny,%
Maggiore:1996gr,Ohshima:1999jg,Gates:2000dq,Gates:2000gu}).
I note that (\ref{anchir})
has a direct
generalization in supergravity \cite{cqg,sugra}.

h) (\ref{an4a}), (\ref{an4b}) and (\ref{an4c})
are present only when the Yang-Mills gauge group
contains Abelian gauge symmetries
under which all matter fields are uncharged.
They are unlikely to represent anomalies
of quantum theories but give important
restrictions to consistent deformations at higher
orders in the deformation parameters.
In particular (\ref{an4a}) enforces at second order
that the coefficients $k_{[\ifree\jfree\kfree]}$
in (\ref{YM}) satisfy
the Jacobi identity for structure constants of a Lie algebra,
cf. \cite{BHT}.

i) (\ref{an5}) may be 
viewed as an analogue with ghost number
1 of non-Abelian Chapline-Manton type couplings (\ref{CM}).

j) (\ref{an6a}) is the analogue  with ghost number 1 of the
couplings (\ref{Noether}). Owing to the antisymmetry
of $R_{[i_A j_A]}(T)$ these representatives exist only if the
Yang-Mills gauge group contains at least two Abelian factors.
As in the case of the couplings (\ref{Noether}) one may distinguish
between different types of representatives (\ref{an6a}), 
depending on the choice
of $R_{[i_A j_A]}(T)$. The simplest choice is
$R_{[i_A j_A]}(T)=k_{[i_A j_A]}\in{\mathbb C}$ and
gives the analogue  with ghost number 1 of the
Fayet-Iliopoulos terms (\ref{FI}):
\bea
k_{[i_A j_A]}\left[
-\Ii C^{i_A}\cD\lambda^{j_A}
+2(\xi\sigma^\mu\5\lambda^{i_A}+\lambda^{i_A}\sigma^\mu\5\xi)A_\mu^{j_A}
\right] d^4x.
\label{anFI}
\eea
All other choices yield representatives containing
conserved currents given by
(\ref{R1}), (\ref{R2}) or Noether currents such as in (\ref{R3}).  
The representatives arising from (\ref{R1})
do not contain antifields and
are somewhat reminiscent of chiral anomalies because
they read
\bea
k_{[i_A j_A]\Index}C^{i_A}A^{j_A}dB^\Index+\dots
\label{anCS}
\eea
The representatives arising from (\ref{R2}) may
be viewed as an analogue with ghost number
1 of Abelian Chapline-Manton type couplings (\ref{CM'}). They
read:
\bea
k_{[i_A j_A]k_A\Index}C^{i_A}A^{j_A}(dA^{k_A})(\star dB^\Index)+\dots
\eea
The representatives containing the  currents (\ref{R3}) read
\bea
k_{[i_A j_A]\ifree \Index}C^{i_A}A^{j_A}
(\star dA^\ifree)(\star dB^\Index)+\dots
\eea

k) Similarly there are several types of representatives (\ref{an6b}).
The simplest arise from
$R_{\Index}(T)=k_{\Index}\in{\mathbb C}$ and read
\bea
2\Ii \, k_{\Index}(\5\psi^\Index\5\xi-\xi\psi^\Index)\, d^4x.
\label{Q1}
\eea
The other representatives (\ref{an6b}) involve conserved currents
and are of the form
\bea
d^4x\,(Q_\mu^\Index j^\mu_\Index+\dots),\quad
j^\mu_\Index=\sfrac 12\sigma^{\mu}_{\alpha\da}
[\cD^\alpha,\5\cD^\da]R_\Index(T).
\label{Q2}
\eea
Again, one may distinguish between representatives containing
the currents (\ref{R1}), (\ref{R2}), or Noether currents
such as in (\ref{R3}). Those with the currents (\ref{R1}) 
are not accompanied by antifields and explicitly given by
(one can assume $k_{\Index\Jndex}=k_{[\Index\Jndex]}$
because the part with $k_{(\Index\Jndex)}$ is trivial):
\bea
k_{[\Index\Jndex]}\left[Q^\Index dB^\Jndex-\Ii 
B_{\mu\nu}^\Index(\xi\sigma^{\mu\nu}\psi^\Jndex
+\5\xi\5\sigma^{\mu\nu}\5\psi^\Jndex)d^4x 
+2\Ii \phi^\Index(\xi\psi^\Jndex-\5\psi^\Jndex\5\xi)d^4x  \right].
\label{Q3}
\eea

l) The representatives  (\ref{an7}) are the counterparts with
ghost number 1 of the invariants (\ref{generic}).
The simplest nontrivial representatives  (\ref{an7}) arise
from $A_{i_A}=k_{i_A}\in{\mathbb C}$,
$A'_{i_A}=k'_{i_A}\in{\mathbb C}$, $B_{i_A}=B'_{i_A}=0$. They read
\bea
8(k_{i_A}\xi\lambda^{i_A}+k'_{i_A}\5\lambda^{i_A}\5\xi)\,d^4x.
\eea
Significant representatives  (\ref{an7}) are in particular
the supersymmetric generalizations of
Abelian chiral anomalies. They
arise from
$A_{i_A}(\varphi,\lambda)=-\Ii k^\Lindex_{i_A}\TR(\lambda\lambda)
-\Ii k_{(i_A j_A k_A)}\lambda^{j_A}\lambda^{k_A}$,
$A'_{i_A}(\5\varphi,\5\lambda)=\Ii k^\Lindex_{i_A}\TR(\5\lambda\5\lambda)
+\Ii k_{(i_A j_A k_A)}\5\lambda^{j_A}\5\lambda^{k_A}$, $B_{i_A}=B'_{i_A}=0$.
This yields supersymmetrized Abelian chiral anomalies
in a form as in Eq. (5.14) of \cite{White:1992ai}.
By adding a coboundary of $H^{1,4}(\ext|d)$ to these
anomalies, they
can be brought to
a form analogous to (\ref{anchir}) 
as can be inferred from (\ref{8.4}).

m) 
Notice:
when the Yang-Mills gauge group
is semisimple and no linear multiplets are present,
(\ref{generic}) and (\ref{anchir}) exhaust the nontrivial representatives
of $H^{0,4}(\ext|d)$ and $H^{1,4}(\ext|d)$, respectively.
Indeed, all other representatives require that
the Yang-Mills gauge group contains Abelian factors or 
that linear multiplets are present.
Recall that the representatives (\ref{generic}) can
be written as off-shell invariants when one uses
the auxiliary fields, see item f). 
Hence, these off-shell invariants provide
all Lorentz-invariant
and N=1 supersymmetric
deformations of standard super-Yang-Mills theories
with semisimple gauge group (to all orders!).

\mysection{Remarks on the cohomology in negative ghost numbers}\label{`ix0'}
\label{negative}

In \cite{BBH1} it was shown that the 
standard (non-extended) BRST cohomological groups 
$H^{p-n,n}(\gauge|d)$ in $n$-dimensional theories
for negative ghost numbers
are isomorphic to
the characteristic cohomology in form-degree $p$ ($0<p<n$)
whose representatives are conserved local $p$-forms
($d\omega^p\approx 0$). 
The cohomological groups $H^{p-4,4}(\ext|d)$, $0<p<4$,
have a similar interpretation: they correspond to
conserved local $p$-forms that are invariant under N=1 supersymmetry
and Poincar\'e transformations up to trivially conserved
$p$-forms. To show this we use that
each representative of $H^{p-4,4}(\ext|d)$ is related
through
the descent equations for $\ext$ and $d$ to a local $p$-form
$\omega^{0,p}$ with ghost number 0 which satisfies 
$\ext\omega^{-1,p+1}+d\omega^{0,p}=0$ and
$\ext\omega^{0,p}+d\omega^{1,p-1}=0$ (see proof of lemma
\ref{proDescent}).
The parts of these equations with antifield number 0 read
(see section \ref{leq3} for the notation):
\bea
\delta\omega^{-1,p+1}_1+d\omega^{0,p}_0=0&\quad\LRA\quad&
d\omega^{0,p}_0\approx 0
\label{cons1}\\
\gammaext\omega^{0,p}_0+\delta\omega^{0,p}_1+d\omega^{1,p-1}_0=0
&\quad\LRA\quad& \gammaext\omega^{0,p}_0\approx -d\omega^{1,p-1}_0,
\label{cons2}
\eea
where subscripts denote the antifield numbers
($\omega^{g,p}_k$ is the part with antifield number $k$ contained
in $\omega^{g,p}$). Notice that 
$\omega^{0,p}_0$ has vanishing ghost and antifield number and thus
does not depend on ghost fields, constant ghosts or antifields.
(\ref{cons1}) shows that it is conserved. Furthermore
it is gauge invariant, N=1 supersymmetric and Poincar\'e
invariant up to 
trivially conserved $p$-forms.
This is seen from (\ref{cons2}) because
$\gammaext$ contains the gauge transformations (with ghost fields
in place of gauge parameters), the N=1 supersymmetry transformations
and the Poincar\'e transformations (multiplied by the
corresponding constant ghosts, respectively).

Moreover $\omega^{0,p}_0$ is trivial in the characteristic
cohomology ($\omega^{0,p}_0\approx d\omega^{0,p-1}_0$) iff
the corresponding representative of $H^{p-4,4}(\ext|d)$ is
trivial in the cohomology of $\ext$ modulo $d$.
For $p<3$ we had seen this already in section \ref{leq3}
where we found that $H^{-3,4}(\ext|d)$ and $H^{-2,4}(\ext|d)$
are isomorphic to the characteristic
cohomology in form-degrees 1 and 2, respectively.
For $p=3$ the assertion can be proved by the arguments  used in
the proof of lemma \ref{pros|d<-1}: 
one finds
that the part $\omega^{-1,4}_1$ of a nontrivial
representative of $H^{-1,4}(\ext|d)$ is a 
nontrivial representative of $H^{4}_1(\delta|d)$%
\footnote{
Applying the arguments in the proof of lemma \ref{pros|d<-1}
to the case $g=-1$
one finds that the nontrivial representatives have $\unk=1$
(but not $\unk=2$ or $\unk=3$ because there are no Lorentz-invariant
homogeneous polynomials of degree 1 or 2 in the constant ghosts).
}.
As shown in \cite{BBH1}, it thus
corresponds to a nontrivial global symmetry
and, via descent
equations for $\delta$ and $d$, to a nontrivial
conserved 3-form containing the Noether current
of this global symmetry%
\footnote{In this context a global symmetry is called trivial if
it equals a gauge transformation 
(in general with field dependent parameters) on-shell.
A conserved current $j^\mu$ is called trivial if 
$j^\mu\approx\6_\nu S^{\nu\mu}$ for some local functions
$S^{\nu\mu}=-S^{\mu\nu}$.}.
Hence, all nontrivial representatives of $H^{-1,4}(\ext|d)$ 
correspond to nontrivial global symmetries of the action
and nontrivial conserved currents. 
These currents are N=1 supersymmetric and Poincar\'e
invariant up to 
trivial conserved currents (see discussion above).
Accordingly the global symmetries corresponding to
representatives of $H^{-1,4}(\ext|d)$ 
commute on-shell with the
N=1 supersymmetry transformations and the Poincar\'e transformations
up to gauge transformations%
\footnote{Let $\Delta_A$ be
(infinitesimal) global symmetry transformations 
and $j_A^\mu$ the corresponding 
conserved currents. One has $\Delta_A j_B^\mu\sim j_{[A,B]}^\mu$
where $j_{[A,B]}^\mu$ is the Noether current
of $[\Delta_A,\Delta_B]$ and $\sim$ is equality up to
trivial currents \cite{Dickey:1991xa,Barnich:1996mr}. 
Furthermore a current is trivial iff the corresponding
global symmetry is trivial \cite{BBH1}.}.

$H^{-3,4}(\ext|d)$ and $H^{-2,4}(\ext|d)$ were already given
in lemma \ref{pros|d<-1}. $H^{-1,4}(\ext|d)$ can be
obtained from $H^3(\ext,\Space)$ which has
three types of representatives as one infers from lemma \ref{pros}:
$f^{(3)}$ with
$P^{(3)}_{[\Index\Jndex]}=k_{[\Index\Jndex]}\in\mathbb{C}$
which correspond to the global symmetries of the action under
rotations acting on the
indices $\Index$ of the linear multiplets,
$f^{(4)}$ with $P^{(4)}=(1/2)k_{\ifree\jfree}\7C^\ifree\7C^\jfree$
($k_{\ifree\jfree}\in\mathbb{C}$) which correspond to the global 
symmetries under
rotations of the indices $\ifree$,
and $f^{(6)}=\cO R(\7T)$ where, however,
(\ref{4theta}), (\ref{R1}) and (\ref{R2}) do not
yield nontrivial representatives of $H^{-1,4}(\ext|d)$
[(\ref{4theta}) does not correspond to any
 representatives of $H^{-1,4}(\ext|d)$ because it
is field independent, while
(\ref{R1}) and (\ref{R2}) are trivial in $H^3(\ext,\Space)$
though nontrivial in $H^3(\ext,\inv)$].

\comments
a) We just observed that the representatives of 
$H^{-1,4}(\ext|d)$ correspond to nontrivial
conserved currents that are N=1 supersymmetric
and Poincar\'e invariant up to trivially conserved currents,
and to the corresponding global symmetries which commute
on-shell with the
N=1 supersymmetry transformations and the Poincar\'e transformations
up to gauge transformations.
However, the correspondence is not one-to-one because
there are conserved currents and global symmetries
of this type which have no counterpart
in $H^{-1,4}(\ext|d)$. Indeed, consider the
currents $j^{\mu\Index}=\6^\mu\phi^\Index$.
They are the Noether currents of the
global symmetries under constant shifts of the $\phi^\Index$.
These shift symmetries are evidently nontrivial. Furthermore they
commute with all N=1 supersymmetry and Poincar\'e
transformations. Accordingly their currents are conserved, nontrivial,
and both N=1 supersymmetric and Poincar\'e invariant
up to trivial currents, respectively.
Nevertheless they do not give rise to
representatives of 
$H^{-1,4}(\ext|d)$. This can be seen in various ways.
One way is the following: the corresponding 
representatives of
$H^3(\ext,\Space)$ would have
dimension $-1$ and be linear in the fields of the
linear multiplets (this follows from the structure of the
shift symmetries and their currents), and would thus
inevitably have to be
proportional to $\cO \phi^\Index$; however, $\cO \phi^\Index$
is trivial in  $H^3(\ext,\Space)$
for one has
$\cO \phi^\Index=-2\cH^\Index=-2\ext \7R^\Index$.
Another way is based on the
algebraic structure which is associated to the local BRST cohomology
and described in section 3 of \cite{Brandt:1997cz}.
In the present case this algebra links the 
representatives of $H^{-1,4}(s|d)$ corresponding to
super-Poincar\'e
symmetries and the symmetries under constant shifts of the $\phi$'s
to the representatives $R^*_\Index d^4x$ of
$H^{-3,4}(s|d)$
in a nontrivial way (owing to the presence of $\phi$-dependent
gauge transformations in the commutator of two supersymmetry
transformations on $B_{\mu\nu}$)
which obstructs the 
existence of representatives of 
$H^{-1,4}(\ext|d)$ corresponding to the shift symmetries.

b) The $p$-forms $\omega^{0,p}$ mentioned above, which are related
through the descent equations to the representatives of $H^{p-4,4}(\ext|d)$
for $0<p<4$,
make up a complete set of field
dependent representatives of $H^{0,p}(\ext|d)$ for $p<4$.
This can be inferred from comment c) in section \ref{DESCENT} using
the result that $H^0(\ext,\Space)$ is
represented just by a number (the latter result follows from lemma \ref{pros},
a number being a solution $f^{(1)}$).

\mysection{Other formulations of supersymmetry}\label{`ix'}\label{otherform}

In this section it is shown that our cohomological
results do not depend on
the chosen formulation of supersymmetry. 
Alternative well-known formulations with auxiliary fields or
linearly realized supersymmetry (in particular, the
standard superspace formulation)
lead to exactly the same results.
The reason is that the additional fields and antifields
which occur in these formulations give only rise to trivial
pairs and thus do not contribute
nontrivially to the cohomology, cf.\ section \ref{pairs}.

\subsection{Formulation with auxiliary fields}\label{`ix.1'}\label{AUX}

Let us first discuss the formulation with
the standard auxiliary fields so that the commutator algebra of the
Poincar\'e, supersymmetry and gauge transformations closes off-shell.
These auxiliary fields are real fields $D^i$
for the super-Yang-Mills multiplets and
complex fields $F^\indec$ for the chiral multiplets.
The formulation with these fields differs from
the one used here through the following changes as compared
to section \ref{BRST}:
\ben
\item
Field content: one adds the fields $D^i$, $F^\indec$, $\5F_\indec$
and their antifields $D^*_i$, $F^*_\indec$, $\5F^{*\indec}$.

\item
Lagrangian: 
\begin{align}
L=&
    -\sfrac 14 \Kill_{ij}F^i_{\mu\nu} F^{j\mu\nu}
    +\sfrac{\Ii}2 \Kill_{ij}(\nabla_\mu\5\lambda^i\5\sigma^\mu\lambda^j
                         -\5\lambda^i\5\sigma^\mu \nabla_\mu\lambda^j)
    +\sfrac 12\Kill_{ij} D^i D^j
\nonumber\\
   &+\sfrac 12\6_\mu\phi^t\6^\mu\phi
    -\sfrac 12H_\mu^t H^\mu
     +\sfrac{\Ii}{2}(\6_\mu\5\psi^t\5\sigma^\mu\psi
                    -\5\psi^t\5\sigma^\mu\6_\mu\psi)
\nonumber\\
   &+\nabla_\mu\5\varphi \nabla^\mu\varphi
     +\sfrac{\Ii}{4}(\nabla_\mu\5\chi\5\sigma^\mu\chi
                     -\5\chi\5\sigma^\mu \nabla_\mu\chi)
     +\sfrac 14\5F F
\nonumber\\
   &+\Ii D^i\5\varphi T_i \varphi+\5\varphi T_i\chi\lambda^i
    -\5\lambda^i\5\chi T_i\varphi.
\label{LWithAux}\end{align}

\item 
Extended BRST transformations of the fields: the only changes are
in the transformations of $\lambda_\alpha$, $\chi_\alpha$ and
their complex conjugates, and the addition of the
extended BRST-transformations of the auxiliary fields: 
\begin{align}
\ext\lambda^i_\alpha=&-\f jki C^j\lambda^k_\alpha
  +\7c^\mu\6_\mu\lambda^i_\alpha
  -\sfrac 12 c^{\mu\nu}(\sigma_{\mu\nu}\lambda^i)_\alpha
  -\Ii\xi_\alpha D^i+(\sigma^{\mu\nu}\xi)_\alpha F^i_{\mu\nu}
\nonumber\\
\ext D^i=&-\f jki C^jD^k+\7c^\mu\6_\mu D^i
  +\xi\sigma^\mu \nabla_\mu\5\lambda^i+\nabla_\mu\lambda^i\sigma^\mu\5\xi
\nonumber\\
\ext\chi_\alpha=&-C^iT_i\chi_\alpha
  +\7c^\mu\6_\mu \chi_\alpha
  -\sfrac 12 c^{\mu\nu}(\sigma_{\mu\nu}\chi)_\alpha
  +\xi_\alpha F-2\Ii(\sigma^\mu\5\xi)_\alpha \nabla_\mu\varphi
\nonumber\\
\ext F=&-C^iT_iF+\7c^\mu\6_\mu F
  -2\Ii \nabla_\mu\chi\sigma^\mu\5\xi
  +4\5\lambda^i\5\xi T_i\varphi.
\label{sWithAux}
\end{align}

\item 
The extended BRST-transformations of the antifields
are according to (\ref{sAF}) obtained from
an extended Lagrangian which is now given by
\begin{equation}
L_{{\mathrm ext}}=L-(\ext\Phi^N)\Phi^*_N\, .
\label{sAFWithAux}
\end{equation}
\een
The passage from the formulation with auxiliary
fields to the formulation without auxiliary
fields is done by elimination of the 
auxiliary fields using the solution of their
``extended equations of motion'' derived from $L_{{\mathrm ext}}$
and
setting the antifields $D^*_i$, $F^*_\indec$, $\5F^*_\indec$
to zero.
This amounts to
the following identifications:
\bea
&D^i \equiv -\Ii \Kill^{ij}
(\5\varphi T_j\varphi+\lambda^*_j\xi
+\5\xi\5\lambda^*_j),\quad D^*_i\equiv 0,&
\nonumber\\
&F\equiv -4\5\xi\5\chi^{*},\quad
\5F\equiv 4\chi^*\xi,\quad
F^*\equiv 0\equiv \5F^*.
\label{aux}
\eea
These identifications reproduce the formulae given in
section \ref{BRST} und turn representatives of
the cohomologies
$H(\ext,\Space)$ and $H(\ext|d)$ in the formulation with
the auxiliary fields into representatives of
the same cohomologies in the formulation without
auxiliary fields (see, e.g., section 15 of \cite{BBH1}).
From the cohomological perspective, this
reflects that the auxiliary fields and their
antifields give only rise to trivial pairs
(see section \ref{pairs}). These trivial pairs 
are $(D^*_i,\ext D^*_i)$, $(F^*_\indec,\ext F^*_\indec)$,
$(\5F^*_\indec,\ext \5F^*_\indec)$ and all their derivatives, for one has:
\begin{align}
\ext D^*_i&=\Kill_{ij}D^j
+\Ii\,(\5\varphi T_i\varphi+\lambda^*_i\xi+\5\xi\5\lambda^*_i)
+C^j\f jik D^*_k+\7c^\mu\6_\mu D^*_i
\nonumber\\
\ext F^*&=\sfrac 14 \5F-\chi^*\xi+C^i F^*T_i+\7c^\mu\6_\mu F^*
\nonumber\\
\ext\5F^{*}&=\sfrac 14 F+\5\xi\5\chi^{*}-C^i T_i\5F^{*}
+\7c^\mu\6_\mu \5F^{*}.
\end{align}

\subsection{Formulation with linearly realized supersymmetry}
\label{`ix.2'}\label{LINEAR}

The standard formulation with linearly realized
supersymmetry (often formulated in superspace)
uses complete real vector multiplets
in place of
only the gauge potentials and the gauginos,
and a higher gauge symmetry which gives rise to
additional ghost fields.
The linearized additional gauge transformations shift
the additional fields of the vector multiplets
by additional gauge parameters.
As a consequence,
the extended BRST transformations act on these fields
as nonlinearly extended shift transformations
involving the additional ghost fields.
The extended BRST transformations of the
antifields of the additional ghost fields
are nonlinearly extended shift transformations
involving the antifields of the additional fields 
of the vector multiplets. 
Hence, all additional
fields, antifields and their derivatives
give only rise to trivial pairs. 
Elimination of these trivial pairs reproduces
the familiar Wess-Zumino gauged models described in section \ref{AUX}.
If one also removes the auxiliary fields as outlined there,
one ends up precisely with the formulation as in section \ref{BRST}.
Let us show
this explicitly for the Abelian case with one
vector multiplet and one chiral matter multiplet
(linear multiplets need not be considered here because
supersymmetry is already linearly realized on them).
We denote the fields of the vector multiplet
by $V$, $\zeta_\alpha$, $Z$, $A'_\mu$, $\lambda'_\alpha$
and $D'$ and those of the chiral multiplet by
$\varphi'$, $\chi'_\alpha$ and $F'$, where 
$V$, $A'_\mu$ and $D'$ are real fields, while
the other fields are complex. The ghost fields
are denoted by $\Lambda$, $\Gamma_\alpha$ and $\Sigma$ and
are all complex (they form a chiral supersymmetry multiplet).
The extended BRST transformations of these fields are
\begin{align}
\ext V&=\Ii(\Lambda-\5\Lambda)+\xi\zeta+\5\zeta\5\xi+\7c^\mu\6_\mu V
\nonumber\\
\ext \zeta_\alpha&=\Gamma_\alpha
+\7c^\mu\6_\mu \zeta_\alpha
-\sfrac 12 c^{\mu\nu}(\sigma_{\mu\nu}\zeta)_\alpha
-(\sigma^\mu\5\xi)_\alpha (A'_\mu+\Ii\6_\mu V)
-\xi_\alpha Z
\nonumber\\
\ext Z &=\Sigma
+\7c^\mu\6_\mu Z
+2\Ii \6_\mu\zeta\sigma^\mu\5\xi+2\Ii\5\lambda'\5\xi
\nonumber\\
\ext  A'_\mu&=-\6_\mu(\Lambda+\5\Lambda)
+\7c^\nu\6_\nu A'_\mu
-c_\mu{}^\nu A'_\nu
-\Ii\,\xi\sigma_\mu\5\lambda'+\Ii\,\lambda'\sigma_\mu\5\xi
+\Ii\6_\mu(\xi\zeta-\5\zeta\5\xi)
\nonumber\\
\ext \lambda'_\alpha&=\7c^\mu\6_\mu\lambda'_\alpha
-\sfrac 12 c^{\mu\nu}(\sigma_{\mu\nu}\lambda')_\alpha
-\Ii\xi_\alpha D'+2(\sigma^{\mu\nu}\xi)_\alpha \6_\mu A'_\nu
\nonumber\\
\ext D'&=\7c^\mu\6_\mu D'
+\xi\sigma^\mu\6_\mu\5\lambda'
+\6_\mu\lambda'\sigma^\mu\5\xi.
\nonumber\\
\ext \varphi'&=-2\Ii\Lambda\varphi'
+\7c^\mu\6_\mu \varphi'+\xi\chi'
\nonumber\\
\ext \chi'_\alpha&=-2\Ii\Lambda\chi'_\alpha-2\Gamma_\alpha \varphi'
+\7c^\mu\6_\mu \chi'_\alpha
-\sfrac 12 c^{\mu\nu}(\sigma_{\mu\nu}\chi')_\alpha
-2\Ii(\sigma^\mu\5\xi)_\alpha\6_\mu \varphi'+\xi_\alpha F'
\nonumber\\
\ext F'&=-2\Ii\Lambda F'+2\Gamma\chi'+2\Sigma\varphi'
+\7c^\mu\6_\mu F'
-2\Ii\6_\mu\chi'\sigma^\mu\5\xi
\nonumber\\
\ext \Lambda&=\7c^\mu\6_\mu\Lambda+\Ii\xi\Gamma
\nonumber\\
\ext \Gamma_\alpha&=\7c^\mu\6_\mu\Gamma_\alpha
-\sfrac 12 c^{\mu\nu}(\sigma_{\mu\nu}\Gamma)_\alpha
+\xi_\alpha\Sigma-2(\sigma^\mu\5\xi)_\alpha\6_\mu\Lambda
\nonumber\\
\ext \Sigma&=\7c^\mu\6_\mu\Sigma-2\Ii\6_\mu\Gamma\sigma^\mu\5\xi.
\label{vmult}
\end{align}
The presence of the shift terms $\Ii(\Lambda-\5\Lambda)$,
$\Gamma_\alpha$ and $\Sigma$ in $\ext V$, $\ext \zeta_\alpha$ and $\ext Z$
implies that $(V,\ext V)$, $(\zeta_\alpha,\ext \zeta_\alpha)$, $(Z,\ext Z)$
and their derivatives
form indeed trivial pairs. The same holds of course for
$(\5\zeta_\da,\ext \5\zeta_\da)$, $(\5Z,\ext \5Z)$ and their derivatives.
It is straightforward to verify that the following identifications
reproduce the $\ext$-transformations 
of $A_\mu$, $\varphi$, $\lambda$, $D$, $\chi$, $F$ and $C$ given in
(\ref{sA}), (\ref{svarphi}), (\ref{sWithAux}) and (\ref{sC})
for the Abelian case (with $C^iT_i\varphi\equiv -\Ii C\varphi$ etc):
\bea
&&A_\mu= A'_\mu,\quad 
\lambda_\alpha=\lambda'_\alpha,\quad
D= D',\quad
C= -\Lambda-\5\Lambda+\Ii\xi\zeta-\Ii\5\zeta\5\xi,
\nonumber\\
&&\varphi={\mathrm e}^V\varphi',\quad
\chi_\alpha={\mathrm e}^V(\chi'_\alpha+2\zeta_\alpha\varphi'),\quad
F={\mathrm e}^V(F'-2Z\varphi'-2\zeta\chi'-2\zeta\zeta\varphi').\quad
\eea
The antifields of the additional fields give also only
rise to trivial pairs, as one has
\begin{align}
\ext\Lambda^{*}&=
\Ii V^*+\6_\mu A^{*\mu\prime}+2 \6_\mu\Gamma^*\sigma^\mu\5\xi
-2\Ii(\varphi^{*\prime}\varphi'+\chi^{*\prime}\chi'+F^{*\prime}F')
\nonumber\\
\ext\Gamma^{*\alpha}&=-\zeta^{*\alpha}-\Ii\xi^\alpha\Lambda^*
-2\Ii\6_\mu\Sigma^*\sigma^\mu\5\xi+2\varphi'\chi^{*\alpha\prime}
+2\chi^{\alpha\prime}F^{*\prime}
\nonumber\\
\ext\Sigma^*&=Z^*+\Gamma^*\xi+2\varphi'F^{*\prime}.
\end{align}
The non-Abelian case is analogous except that
the extended BRST transformations
of the fields receive further nonlinear contributions,
and, as a consequence, the relation of primed
and unprimed fields becomes more complicated (in particular
$A_\mu$, $\lambda$ and $D$ do not coincide anymore with their
primed counterparts).

\mysection{Other Lagrangians}\label{`x'}\label{otheraction}

Even though our analysis was performed for Lagrangians
(\ref{L}), it is not restricted to these.
In particular it applies analogously to more general
Lagrangians which are of the form (\ref{generic})
in the
formulation with auxiliary fields
(with $\cD_\alpha$ and $\5\cD_\da$ realized off-shell
on ordinary tensor fields,
cf.\ comment c) in section \ref{SUSY}), such as
\bea
L=\cD^2[G_{ij}(\varphi)\lambda^i\lambda^j+P(\varphi)
+\5\cD^2 K(\varphi,\5\varphi,\phi)]+\CC
\label{L2}
\eea
where $G_{ij}(\varphi)\lambda^i\lambda^j$,
$P(\varphi)$ and $K(\varphi,\5\varphi,\phi)$
are gauge invariant functions.
(\ref{L2}) is a generalization of the Lagrangian (\ref{L}) which
is still at most quadratic in derivatives but
in general not power counting renormalizable as it
contains terms such as $(G_{ij}(\varphi)+\CC)F_{\mu\nu}^iF^{j\mu\nu}$.
The special choice $G_{ij}(\varphi)=\Kill_{ij}/16$,
$P(\varphi)=0$ and $K(\varphi,\5\varphi,\phi)=
(\5\varphi\varphi-\phi^t\phi)/32$ reproduces
the Lagrangians (\ref{L}) after elimination of the auxiliary fields,
up to a total divergence.
Let us briefly indicate how one can apply
our analysis and results to such models or even
more general Lagrangians
(\ref{generic}) containing terms with more than two derivatives
such as (\ref{mass8}), assuming that
supersymmetry is not spontaneously broken (see comment
below) and that the field equations satisfy
the standard
regularity conditions (see \cite{report}):

\blist
\item
The extended BRST transformations of
the fields and constant ghosts are as in section \ref{AUX}
when one uses the formulation with auxiliary fields
(recall that the deformations (\ref{generic}) do not change the
gauge or supersymmetry transformations when one
uses the formulation with the auxiliary fields, cf.\ comment f) in section
\ref{DEFOS}).

\item
The extended BRST transformations of the antifields
contain the 
Euler-Lagrange derivatives of the Lagrangian with respect
to the corresponding fields and are
thus more involved than in the simple models (\ref{L}).
As a consequence the explicit
form of the $w$-variables (see section \ref{CALC}) changes. A subset of these
variables even may fail to be local in the strict sense of 
section \ref{DESCENT} (because the
algorithm \cite{jet} may not terminate anymore).
This failure of locality does not happen for
Lagrangians (\ref{L2})%
\footnote{Lemma \ref{proW} still holds for these Lagrangians but
the proof of the lemma is more involved than for the simple
Lagrangians (\ref{L}). Without going into detail I note
that one may use the derivative order of the variables
(rather than their dimension) to prove that the
algorithm \cite{jet} terminates for
Lagrangians (\ref{L2}).
}
but it will generically happen for more general Lagrangians.
In order to apply our analysis in the latter case
one must relax the definition of local functions
and forms accordingly. 
In particular this is necessary and natural when dealing with effective
Lagrangians containing terms such as (\ref{mass8})
multiplied by free parameters (coupling constants).
One may then use the definition that local functions and forms
are formal power series in the free parameters,
with each term of the
series local in the strict sense, cf.\ the remarks on
effective theories in \cite{report}.

\item
The descent equations and lemma \ref{proDescent} hold also in
the generalized models.         

\item
The gauge covariant algebra (\ref{alg}) does not change
because it reflects the extended BRST transformations of
the ghosts which do not change.

\item
The structure of the cohomological groups presented in
section \ref{leq3} does not change because 
$\ext R^*_\Index$
and $\ext C^*_\ifree$ 
do not change
(this follows from the fact that the gauge transformations
do not change). However, $f_{\Index\Jndex}$, $f'_{\Index\Jndex}$,
$\4H^\Index$ and $\4F^\ifree$ can receive
additional terms because of changes of the
field equations.

\item
Subsections \ref{pairs}, \ref{decomps} and \ref{secLAC} 
evidently apply also to the generalized models.

\item
$H(\susy,\inv)$ is still given
by lemma \ref{prosusy}, except   
for the possible modifications of $\4H^\Index$ and $\4F^\ifree$
mentioned above. Indeed, the
results for $g\leq 2$ derive as before
from the results in section \ref{leq3}
and from Eq.\ (\ref{g=012}) which holds also in the generalized models
(the proof of that equation applies also
to the generalized models).
The results for $g\geq 3$ are solely based
on the algebra (\ref{alg}) and the structure of
the supersymmetry multiplets which do not change.

\item 
Lemmas \ref{prouseful} and \ref{proq} apply
without modifications also to the generalized models.

\item
The possible 
changes of lemma \ref{pros}
are induced by the modifications of $\4H^\Index$ and $\4F^\ifree$
mentioned above and
concern the representatives $f^{(2)}$, $f^{(3)}$, $f^{(4)}$ and
$f^{(5)}$. 
$f^{(2)}$ and $f^{(5)}$ 
have counterparts
also in the generalized models but their explicit form
changes when $\4H^\Index$ receives additional
terms (the additional terms cause changes
of the $X$-functions occurring in $f^{(2)}$ and $f^{(5)}$;
the existence of these representatives is still 
guaranteed by the vanishing of $H^g(\susy,\inv)$ for $g>4$).
The existence and precise form
of representatives
$f^{(3)}$ and $f^{(4)}$ depends on the modifications
of $\4H^\Index$ and $\4F^\ifree$ and varies from
case to case. However, the modifications of $f^{(3)}$
do not concern consistent deformations, counterterms
or anomalies because there are no representatives 
$f^{(3)}$ with ghost numbers 4 or 5.
The representatives $f^{(4)}$ are mainly of interest
for the deformation of free theories 
with Lagrangians of the form (\ref{L}).

\elist

\comment
In models with spontaneously broken supersymmetry
the goldstino fields and their extended BRST transformations
form trivial pairs (the extended BRST transformations
of the goldstino fields substitute for the
constant supersymmetry ghosts in the new 
jet coordinates $\{u^\ell,v^\ell,w^\Windex\}$).
As a consequence the structure of the
extended BRST cohomology is
essentially the same as its non-supersymmetric
counterpart, provided
one relaxes the definition of local functions and forms
suitably, if necessary (see remarks above).
The representatives of the cohomology are
supersymmetrizations of their counterparts
in the non-supersymmetric cohomology, similarly to the
supersymmetrized actions
constructed in
\cite{Uematsu:1982rj,Samuel:1983uh,Clark:1996aw}.

\mysection{Conclusion}\label{`xi'}\label{conclusion}

The major advances of our analysis
as compared to previous work are: (i) we have computed
the cohomology in the space of all local forms rather
than only in the restricted space of forms with
bounded power-counting dimension;
(ii) we have included linear multiplets
in addition to super Yang-Mills multiplets and chiral multiplets.
Furthermore we have computed the cohomology for all ghost numbers
even though the results for $H^{g,4}(\ext|d)$, $g>1$
are currently only of mathematical interest as no
physical interpretation of these cohomological groups
is known to date.
Let us briefly summarize the results for the cohomological
groups
$H^{0,4}(\ext|d)$ and $H^{1,4}(\ext|d)$ which
are most important for algebraic renormalization,
candidate anomalies and supersymmetric
consistent deformations of the models under study.

The results are particularly simple when
the Yang-Mills gauge group is semisimple
and no linear multiplets are present:
then all representatives of
$H^{0,4}(\ext|d)$ can be written
in the form (\ref{generic}) (times the volume element, and
up to antifield dependent terms in the formulation without
auxiliary fields),
and the representatives of $H^{1,4}(\ext|d)$ 
are exhausted by (\ref{anchir}) (up to cohomologically trivial terms,
respectively). Hence, for semisimple
gauge group and in absence of linear multiplets, (i) all 
Poincar\'e invariant and N=1 supersymmetric
consistent deformations of the action which preserve
the N=1 supersymmetry algebra on-shell modulo gauge transformations
can be constructed from standard superspace integrals
and preserve the form of the gauge transformations and
N=1 supersymmetry transformations when one
uses the auxiliary fields
(accordingly in the formulation without
auxiliary fields only the supersymmetry transformations of the
fermion fields get deformed as one sees by elimination of
the auxiliary fields);
(ii) all counterterms that are gauge invariant,
Poincar\'e invariant and N=1 supersymmetric at least on-shell
can be written even as off-shell invariants by means of
the auxiliary fields
and are constructible from standard superspace integrals;
(iii) the consistent Poincar\'e invariant
candidate gauge and supersymmetry anomalies
are exhausted by supersymmetric generalizations of the
well-known non-Abelian chiral anomalies and these
can be written in the universal form (\ref{anchir})
whether or not one uses the auxiliary fields.
These results are not restricted to Lagrangians
(\ref{L}) but apply also to a general class of
Lagrangians
and in particular to effective super-Yang-Mills
theories
with semisimple gauge group, see section \ref{otheraction}.

When the Yang-Mills gauge group contains Abelian factors
or when linear multiplets are present, there
are a number of additional cohomology classes 
of $H^{0,4}(\ext|d)$ and $H^{1,4}(\ext|d)$ whose
representatives cannot be written as in
(\ref{generic}) or (\ref{anchir}).
We have computed them explicitly
for the simple Lagrangians (\ref{L}) because they are particularly
relevant to 
the general classification of supersymmetric
consistent interactions which naturally starts off from free models
with Lagrangians (\ref{L}).
The antifield independent parts of the
additional representatives
are given in
equations (\ref{CM})--(\ref{Noether}) and
(\ref{an4a})--(\ref{an7}).
{}From among these only the candidate anomalies (\ref{an7}) and a few 
special solutions are off-shell invariants
or can be written as off-shell invariants by means of the
auxiliary fields. The special off-shell solutions
are particular 
representatives (\ref{Noether}) given by the
Fayet-Iliopoulos terms (\ref{FI}) and the Chern-Simons type
representatives (\ref{CS'}),
analogous representatives (\ref{an6a}) 
with ghost number 1 given in (\ref{anFI}) and
(\ref{anCS}), and particular representatives (\ref{an6b}) 
given in (\ref{Q1}) and (\ref{Q3}).
All other functions (\ref{CM})--(\ref{Noether}) and forms
(\ref{an4a})--(\ref{an6b}) are
accompanied by antifield dependent terms
and may look somewhat different when
one uses a more general Lagrangian  (some of them
might even disappear), cf.\ section \ref{otheraction}.
The interaction terms contained in (\ref{CM})--(\ref{Noether})
include, among other things, Yang-Mills, Freedman-Townsend and  
Chapline-Manton vertices, and
Noether couplings of gauge fields to gauge invariant conserved
currents that are supersymmetric up to trivial currents.

Finally we remark that the results derived in this paper
are characteristic of theories
with a particular supersymmetry multiplet structure 
on tensor fields which we have
termed ``QDS structure'' (see \cite{glusy} and proof of
lemma \ref{prosusy}). Indeed, when one reviews the
derivation of the results one observes that, essentially, they can be put
down to three central ingredients: the characteristic
cohomology of the field equations in form-degrees smaller than 3
(see section \ref{leq3}),
standard Lie algebra cohomology (see section \ref{secLAC}), 
and what we have termed supersymmetry
algebra cohomology (see section \ref{SUSY}). 
While the characteristic cohomology
and Lie algebra cohomology are not affected by the
supersymmetry multiplet structure,
this structure is decisive for the supersymmetry
algebra cohomology. The QDS structure underlies
the results for ghost numbers larger than 2
in lemma \ref{prosusy} and these results
hold analogously for all models with QDS structure as can be inferred
from the proof of the lemma.

\appendix

\mysection{Conventions and useful formulae}\label{`A'}\label{conv}

\noindent
Minkowski metric, $\vep$-tensors:
\beann & &\eta_{\mu\nu}={\mathrm{diag}}(1,-1,-1,-1),\quad
\vep^{\mu\nu\rho\sigma}=\vep^{[\mu\nu\rho\sigma]},\quad \vep^{0123}=1,\\
& &\vep^{\alpha\beta}=-\vep^{\beta\alpha},\quad
\vep^{\da\dbe}=-\vep^{\dbe\da},\quad
\vep^{12}=\vep^{\dot 1\dot 2}=1,\\
& &\vep_{\alpha\gamma}\vep^{\gamma\beta}=\delta_\alpha^\beta=
{\mathrm{diag}}(1,1),\quad
\vep_{\da\dg}\vep^{\dg\dbe}=\delta_\da^\dbe={\mathrm{diag}}(1,1)            
\eeann
$\sigma$-matrices:
\beann
\sigma^0=\left( \begin{array}{rr}1&0\\ 0&1\end{array}\right),\quad
\sigma^1=\left( \begin{array}{rr}0&1\\ 1&0\end{array}\right),\quad
\sigma^2=\left( \begin{array}{rr}0&-\Ii\\ \Ii&0\end{array}\right),\quad
\sigma^3=\left( \begin{array}{rr}1&0\\ 0&-1\end{array}\right) 
\\
\5\sigma^{\mu\, \da\alpha}=
\vep^{\da\dbe}\vep^{\alpha\beta}\sigma^\mu_{\beta\dbe},\quad 
\sigma^{\mu\nu}=
\sfrac 14(\sigma^\mu\5\sigma^\nu-\sigma^\nu\5\sigma^\mu),\quad
\5\sigma^{\mu\nu}
=\sfrac 14(\5\sigma^\mu\sigma^\nu-\5\sigma^\nu\sigma^\mu) \eeann
Raising, lowering, contraction  of spinor indices:
\beann \psi_\alpha=\vep_{\alpha\beta}\psi^\beta,\quad
\psi^\alpha=\vep^{\alpha\beta}\psi_\beta,\quad
\5\psi_\da=\vep_{\da\dbe}\5\psi^\dbe,\quad
\5\psi^\da=\vep^{\da\dbe}\5\psi_\dbe,\quad
\psi\chi=\psi^\alpha\chi_\alpha,\quad
\5\psi\5\chi=\5\psi_\da\5\chi^\da
\eeann
Lorentz vector indices in spinor notation:
\beann V_{\alpha\da}=\sigma^\mu_{\alpha\da} V_\mu \eeann
Grassmann parity:
\beann |X_{\alpha_1\ldots\alpha_n}^{\da_1\ldots\da_m} |=
m+n+\gh(X)+{\mbox{form-degree}}(X)\quad ({\mathrm{mod}}\ 2)
\eeann                       
Complex conjugation:
\beann \overline{XY}=(-)^{|X|\, |Y|}\5X\ \5Y \eeann
Hodge dual:
\beann
\star (dx^{\mu_1}\dots dx^{\mu_p})=
\sfrac 1{(n-p)!}dx^{\nu_1}\dots dx^{\nu_{n-p}}
\vep_{\nu_1\dots\nu_{n-p}}{}^{\mu_1\dots\mu_p}
\eeann
Symmetrization and antisymmetrization of indices:
\beann T_{(a_1\ldots a_n)}=\frac 1{n!}\sum_{\pi\in S_n}
T_{a_{\pi(1)}\ldots a_{\pi(n)}},\quad
T_{[a_1\ldots a_n]}=\frac 1{n!}\sum_{\pi\in S_n}(-)^{sign(\pi)}
T_{a_{\pi(1)}\ldots a_{\pi(n)}}                              
\eeann
Frequently used functions and operators:
\beann
\Xi_{\mu\nu}&=&-\sfrac 14\vep_{\mu\nu\rho\sigma}\7c^\rho\7c^\sigma,\quad
\Xi_\mu=-\sfrac 16\vep_{\mu\nu\rho\sigma}\7c^\nu\7c^\rho\7c^\sigma,\quad
\Xi=-\sfrac 1{24}\vep_{\mu\nu\rho\sigma}\7c^\mu\7c^\nu\7c^\rho\7c^\sigma
\\
\Theta&=&\7c^\mu\xi\sigma_\mu\5\xi,\quad
\vartheta^\alpha=\7c^\mu(\5\xi\5\sigma_\mu)^\alpha,\quad
\5\vartheta^\da=\7c^\mu(\5\sigma_\mu\xi)^\da
\\
\7C^i&=&C^i+A^i_\mu \7c^\mu
\\
\7R&=&R+\Ii Q_\mu \7c^\mu+\sfrac{\Ii}2 B_{\mu\nu} \7c^\nu\7c^\mu
\\
\4H^\Index&=&\Ii(\hatpsi{}^\Index\5\xi-\xi\7\psi^\Index)
+\7c^\mu \7H_\mu^\Index
\\
&=&\Ii(\5\psi{}^\Index\5\xi-\xi\psi^\Index)
+\7c^\mu H_\mu^\Index+\delta^{\Index\Jndex}(\vartheta\psi^*_\Jndex
-\5\psi^*_\Jndex\5\vartheta+2\,\Xi_{\mu\nu}B^{*\mu\nu}_\Jndex
-\Xi_\mu Q^{*\mu}_\Jndex-\Ii\,\Xi\,R^*_\Jndex)
\nonumber\\
\4F^{\ifree}&=&
\7c^\mu(\xi\sigma_\mu\hatlambda{}^{\ifree}+\7\lambda^{\ifree}\sigma_\mu\5\xi)
+\sfrac 14\7c^\mu\7c^\nu\vep_{\mu\nu\rho\sigma}\7F^{\ifree\rho\sigma}
\\
&=&\vartheta\lambda^\ifree-\5\lambda^\ifree\5\vartheta
-\Xi_{\mu\nu}F^{\ifree\mu\nu}
+\Kill^{\ifree\jfree}(-2\,\Xi_{\mu\nu}\,\xi\sigma^{\mu\nu}\lambda^*_\jfree
-2\,\Xi_{\mu\nu}\,\5\lambda^*_\jfree\5\sigma^{\mu\nu}\5\xi
+\Xi_\mu A^{*\mu}_\jfree+\Xi\,C^*_\jfree)
\\
\cF^i&=&\Ii\7c^\mu(\xi\sigma_\mu\hatlambda{}^i-\7\lambda^i\sigma_\mu\5\xi)
  +\sfrac 12\7c^\mu\7c^\nu\7F^i_{\mu\nu}
\\
&=&-\Ii\vartheta\lambda^i-\Ii\5\vartheta\5\lambda^i
  +\sfrac 12\7c^\mu\7c^\nu F^i_{\mu\nu}
\\
\cH^\Index&=&-2\7c^\mu\xi\sigma_\mu\5\xi\phi^\Index
  -\Ii\7c^\mu\7c^\nu(\xi\sigma_{\mu\nu}\7\psi^\Index
  -\hatpsi{}^\Index\5\sigma_{\mu\nu}\5\xi)
  -\sfrac{\Ii}6\7c^\mu\7c^\nu\7c^\rho\vep_{\mu\nu\rho\sigma}\7H^{\Index\sigma}
\\
&=&-2\Theta\phi^\Index+2\,\Xi_{\mu\nu}\,
  (\xi\sigma^{\mu\nu}\psi^\Index
  +\5\psi{}^\Index\5\sigma^{\mu\nu}\5\xi)
  -\Ii\,\Xi_\mu H^{\Index\mu}
\\
\cO&=&4\Theta
-4\,\Xi_{\mu\nu}(\xi\sigma^{\mu\nu}\cD+\5\xi\5\sigma^{\mu\nu}\5\cD)
-\sfrac{\Ii}{2}\Xi_\mu
\sigma^\mu_{\alpha\da}[\cD^\alpha,\5\cD^\da]
\\
\cP&=&
-16\,\Xi_{\mu\nu}\,\5\xi\5\sigma^{\mu\nu}\5\xi
-4\Ii\,\Xi_\mu\,\5\xi\5\sigma^\mu\cD
+\Xi\,\cD^2
=4\Ii\,\vartheta\vartheta
-4\Ii\,\Xi_\mu\5\xi\5\sigma^\mu\cD
+\Xi\,\cD^2
\eeann
$\cD_\alpha$-transformations:
\beann
&&\cD_\alpha\7\lambda^i_\beta=-\7F^{i(+)}_{\alpha\beta}
  + \vep_{\alpha\beta}\,\Kill^{ij}\,\5\varphi T_j\varphi,\quad
  \cD_\alpha\hatlambda{}^i_\dbe=0,\quad
  \cD_\alpha\7F^{i(+)}_{\beta\gamma}=
  -2\Kill^{ij}\,\vep_{\alpha(\beta}\,\5\varphi T_j\7\chi_{\gamma)},\quad
  \cD_\alpha\7F^{i(-)}_{\dbe\dg}
  =-2\Ii \7\nabla_{\alpha(\dbe}\hatlambda{}^i_{\dg)}
\\
&&\cD_\alpha\phi=\7\psi_\alpha,\quad 
  \cD_\alpha\7\psi_\beta=0,\quad 
  \cD_\alpha\hatpsi_\dbe=-\7H_{\alpha\dbe}-\Ii \7\nabla_{\alpha\dbe}\phi,\quad
  \cD_\alpha\7H_{\beta}{}^\dbe=-\Ii\7\nabla_{(\alpha}{}^\dbe\psi_{\beta)}
\\
&&\cD_\alpha\varphi=\7\chi_\alpha,\quad 
  \cD_\alpha\5\varphi=0,\quad 
  \cD_\alpha\7\chi_\beta=0,\quad 
  \cD_\alpha\hatchi_\dbe=-2\Ii\7\nabla_{\alpha\dbe}\5\varphi
\eeann

\mysection{Proofs of the lemmas}\label{`D'}\label{proofs}

\proof{proDescent}
First we show that
(\ref{p1}) implies descent equations
of the standard form:
\bea
\ext\omega^{g,4}+d\omega^{g+1,3}=0,\quad
\ext\omega^{g+1,3}+d\omega^{g+2,2}=0,\quad\dots\quad
\ext\omega^{g+4-m,m}=0,
\label{des}
\eea
for some local forms $\omega^{g+4-p,p}$ and some
form-degree $m$ which is not a priori known
but turns out to be necessarily 0 whenever $\omega^{g,4}$ is nontrivial
(see below). The descent equations follow from the
so-called algebraic Poincar\'e lemma (cf.
section 4.5 of \cite{report}) which
describes the cohomology $H(d)$ (locally) in the jet space associated
with the fields, antifields and the constant ghosts.
It states that $H^p(d)$
vanishes locally%
\footnote{Global aspects are left out of consideration
in this work.
They may be studied along the lines of 
\cite{viallet,BBHgrav}.} 
in form-degrees $p=1,2,3$ and that $H^0(d)$ is represented
by constants which means in the present case that
$H^0(d)$ is given by $\CONST$, the 
vector space of polynomials in the constant ghosts.
One derives (\ref{des}) 
in the usual manner:
acting with $\ext$ on an equation $\ext\omega^{g+4-p,p}+d\omega^{g+5-p,p-1}=0$
gives $d(\ext\omega^{g+5-p,p-1})=0$. If $p-1> 0$, the 
algebraic Poincar\'e lemma guarantees that there is
a local form $\omega^{g+6-p,p-2}$ such that
$\ext\omega^{g+5-p,p-1}+d\omega^{g+6-p,p-2}=0$. If $p-1=0$, then the
algebraic Poincar\'e lemma alone only allows one to
conclude $\ext\omega^{g+4,0}=M$
for some $M\in\CONST$.
Nevertheless, one can assume $M=0$ without loss of generality, 
for the following reason:
an $\ext$-exact function which depends only on the constant ghosts 
is necessarily the $\ext$-transformation of a function
which also depends only on the constant ghosts because the
$\ext$-transformations of the fields and antifields do not
contain parts depending only on the constant ghosts;
hence, $\ext\omega^{g+4,0}=M$ implies
$M=\ext N$ for some $N\in\CONST$.
The descent
equations (\ref{des}) then hold with $\omega^{g+4,0}-N$
in place of $\omega^{g+4,0}$ as one has
$\ext(\omega^{g+4,0}-N)=0$ and
$\ext\omega^{g+3,1}+d(\omega^{g+4,0}-N)=0$.

$H(\ext+d)$ enters in the usual manner by writing
(\ref{des}) into the compact form
\bea
(\ext+d)\4\omega=0,\quad \4\omega=\sum_{p=m}^4\omega^{g+4-p,p}.
\label{tildeomega}
\eea
Using standard arguments (see \cite{Dragon:1996md,ten,report})
one deduces that $H(\ext+d)$ is isomorphic
to $H^{*,4}(\ext|d)\oplus H(\ext,\CONST)$.

The isomorphism between $H(\ext+d)$ and
$H(\ext,\Space)$ rests on the fact that $(\ext+d)$ arises from
$\ext$ on all variables except for
$x^\mu$ simply through the
substitution $c^\mu\rightarrow c^\mu+dx^\mu$.%
\footnote{This is a direct consequence of the presence of
spacetime translations in $\ext$:
$\ext$ contains the translational
piece $c^\mu\6_\mu$ (except on $x^\mu$) and thus
$(\ext+d)$ contains $(c^\mu+dx^\mu)\6_\mu$.}
Apart from this substitution, the only difference
between $(\ext+d)$ and $\ext$ is that one has
$(\ext+d)x^\mu=dx^\mu$ but $\ext x^\mu=0$.
However, this difference does not matter
because $x^\mu$ drops essentially from $H(\ext+d)$.
This is seen after changing variables from
$c^\mu$ to $\4c^\mu=c^\mu+dx^\mu-x^\nu {c_\nu}^\mu$
(with all other variables left unchanged).
Using the new variables, $(x^\mu,dx^\mu)$
become trivial pairs which drop
from $H(\ext+d)$ by the standard arguments
 (see section \ref{pairs}) because
the $x^\mu$ and $dx^\mu$ form ``$(\ext+d)$-doublets'' due to
$(\ext+d)x^\mu=dx^\mu$, and the $(\ext+d)$-transformations
of the other variables do not involve $x^\mu$ or
$dx^\mu$ when expressed in terms of the new variables 
(one has $\ext+d=\4c^\mu\6_\mu+\dots$, except on $x^\mu$).

Hence, one has $H(\ext+d)\simeq H(\ext+d,\4{\Space})$ where
$\4{\Space}$ is the analog of $\Space$, with $\4c^\mu$ in place
of $\7c^\mu$.
The isomorphism $H(\ext+d,\4{\Space})\simeq H(\ext,\Space)$ is now
evident, since
$(\ext+d)$ acts in $\4\cF$ exactly in the
same manner as $\ext$ acts in $\Space$ (modulo the substitution
$\4c^\mu\rightarrow \7c^\mu$).
\QED

\proof{proBasis}
$[\6_\mu,\6_\nu]=0$ implies the
following identities:
\bea
&&[\6_+^+,\6_+^-]Z_n^m=
[\6_+^+,\6_-^+]Z_n^m=
[\6_-^-,\6_+^-]Z_n^m=
[\6_-^-,\6_-^+]Z_n^m=0,
\nonumber\\
&&[\6_+^+,\6_-^-]Z_n^m=(m+n+2)\Box Z_n^m,
\nonumber\\
&&[\6_-^+,\6_+^-]Z_n^m=(m-n)\Box Z_n^m,
\nonumber\\
&&\6_+^- \6_-^+Z_n^m=\sfrac 12 n(m+2)\, \Box Z_n^m
+\6_+^+\6_-^-Z_n^m,
\label{bas5}\eea
These identities can be used to construct a
basis for the derivatives of a field or antifield: 
every $k$th order derivative can be 
expressed as a linear combination of the components of
polynomials of degree $k$ in the operations
$\6^+_+$, $\6^+_-$, $\6^-_+$, $\6_-^-$ applied to the field
or antifield;
using (\ref{bas5}) one
can construct an appropriate basis
of these polynomials; 
the $u$'s, linearized $v$'s, and $w_{(0)}$'s
just provide such a basis for the derivatives of all fields and antifields.
For instance,
consider a scalar field $\phi$ and its antifield $\phi^*$.
Using (\ref{bas5}) one verifies straightforwardly that
bases for all derivatives of $\phi$ and $\phi^*$ are
$\cup_{p,q}\Box^p(\6_+^+)^q\phi$ and $\cup_{p,q}\Box^p(\6_+^+)^q\phi^*$,
respectively. $\cup_{p,q}\Box^p(\6_+^+)^q\phi^*$
is contained in
$\{u^\ell\}$.
One has $\ext\phi^*=-\Box\phi+\dots$ and
thus $\ext\Box^p(\6_+^+)^q\phi^*=-\Box^{p+1}(\6_+^+)^q\phi+\dots$.
Hence the set of the linearized variables $v^\ell=\ext u^\ell$ contains
a basis for all derivatives of
$\phi$ except for the $(\6_+^+)^q\phi$. The latter are
contained in $\{w_{(0)}^\Windex\}$.
Analogous arguments apply to the other fields and
antifields, see \cite{sugra} for further details.
\QED

\proof{proW} 
We only need to prove that the algorithm in section 2 of \cite{jet}
produces $w$'s that are local functions (eq. (\ref{susw})
holds by the algorithm). 
This can be done along the lines
of section 3 of \cite{jet} using the following dimension assignments:
\bea
\ba{c|c|c|c|c|c|c}
X & R,\7c^\mu & \xi & C,Q_\mu,c^{\mu\nu} &
A_\mu,B_{\mu\nu},\phi,\varphi,\6_\mu & \lambda,\psi,\chi & \Phi^*_\FD
\\
\hline
{\mathrm dim}(X) & -1 & -1/2 & 0 & 1 & 3/2 & 4-{\mathrm dim}(\Phi^\FD).
\ea
\label{dimensions}\eea
With these assignments we have: 
(i) $\ext$ has dimension 0,
(ii) all $u$'s, $v$'s, $w_{(0)}$'s with
non-positive dimension have positive ghost numbers 
and dimensions $\geq -1$,
(iii) all $u$'s, $v$'s, $w_{(0)}$'s
with negative ghost numbers have dimensions $\geq 5/2$.
Because of (i) and since $\ext$ has ghost number 1, each 
$w$-variable constructed by means of the algorithm in \cite{jet}
has a definite dimension and a definite ghost number.
Properties (ii) and (iii) imply that
there are only finitely many monomials in the $u$'s, $v$'s and $w_{(0)}$'s
with a given dimension and a given ghost number.
Hence, each $w$-variable is a finite sum of
such monomials and thus a local function. \QED

\proof{prosW}
$\ext^2=0$ and (\ref{sf(T)}) imply
\begin{align}
0=\ext^2f(\7T)&=
(\ext\cC^\gindex)\Delta_\gindex f(\7T)
+(-)^{|\gindex|+1}\cC^\gindex\cC^\hindex\Delta_\hindex\Delta_\gindex
f(\7T)
\nonumber\\
&=(\ext\cC^\gindex)\Delta_\gindex f(\7T)
+\sfrac 12(-)^{|\gindex|+1}\cC^\gindex\cC^\hindex
[\Delta_\hindex,\Delta_\gindex\}f(\7T)
\label{prosW1}\end{align}
where $|\gindex|$ is the Grassmann parity of $\Delta_\gindex$
and $[\ ,\ \}$ is the graded commutator:
\beann
{}[\Delta_\hindex,\Delta_\gindex\}:=\Delta_\hindex\Delta_\gindex
-(-)^{|\hindex|\,|\gindex|}\Delta_\gindex\Delta_\hindex.
\eeann
The transformations of the $\cC$'s have the form
\bea
\ext\cC^\iindex=\sfrac 12(-)^{|\gindex|+1}\cC^\gindex\cC^\hindex
\cF_{\hindex\gindex}{}^\iindex(\7T)
\label{prosW2}\eea
for functions $\cF_{\hindex\gindex}{}^\iindex(\7T)$ which
can be read off from (\ref{s7C}), (\ref{s7c}), (\ref{scmunu}),
(\ref{sxi}) and (\ref{s5xi}).
Since (\ref{prosW1}) holds for all functions $f(\7T)$
and since the $\cC^\gindex$
are independent variables, we conclude
\bea
[\Delta_\hindex,\Delta_\gindex\}=
-\cF_{\hindex\gindex}{}^\iindex(\7T)\Delta_\iindex .
\label{prosW3}\eea
This yields (\ref{alg}) when spelled out explicitly. \QED

\proof{prodelta|d}
The lemma is proved exactly as the corresponding results in
\cite{BBH1} (see also \cite{HKS}).
We shall therefore only sketch the basic ideas
and refer to \cite{BBH1} for details.

The results for $k>3$ follow from the general
theorems 8.3, or 10.1 and 10.2 in \cite{BBH1} because
the models under study have Cauchy order 3 and
reducibility order 1 in the terminology used there
(the proofs of these theorems given in
\cite{BBH1} apply also in
presence of the constant ghosts because
the latter are inert to both $\delta$ and $d$).
To prove the results
for $k=3$ and $k=2$ we
first consider the linearized theory and derive
$H^4_k(\delta^{(0)}|d)$ for $k=3$ and $k=2$
where $\delta^{(0)}$ is the Koszul-Tate differential
of the linearized models (the $\delta^{(0)}$-transformations
arise from (\ref{delta}) by linearization).
The cocycle condition in $H^4_k(\delta^{(0)}|d)$ is
$\delta^{(0)}\omega^4_k+d\omega^3_{k-1}=0$ 
and
reads in dual notation 
\bea
\delta^{(0)} f_k+\6_\mu f_{k-1}^\mu=0,
\label{de0}
\eea 
where we used
$\omega^4_k=d^4x f_k$ and
$\omega^3_{k-1}=(1/6)dx^\mu dx^\nu dx^\rho 
\vep_{\mu\nu\rho\sigma}f_{k-1}^\sigma$, i.e.,
$f_k$ and $f_{k-1}^\sigma$ are local functions (rather than forms)
with antifield number $k$ and $k-1$, respectively.
One now considers the Euler-Lagrange derivatives of (\ref{de0})
with respect
to the various fields and antifields. In particular, the Euler-Lagrange derivatives
with respect
to $R^*_\Index$, $Q^{*\mu}_\Index$, $B^{*\mu\nu}_\Index$,
$B^\Index_{\mu\nu}$, $C^*_i$ and $A^{*\mu}_i$ yield
the following equations, respectively:
\begin{align}
&
\delta^{(0)}\,\frac{\7\6f_k}{\7\6R^*_\Index}=0
\label{de1}\\
&
\delta^{(0)}\,\frac{\7\6f_k}{\7\6Q^{*\mu}_\Index}=
\Ii\,\6_\mu\,\frac{\7\6f_k}{\7\6R^*_\Index}
\label{de2}\\
&
\delta^{(0)}\,\frac{\7\6f_k}{\7\6B^{*\mu\nu}_\Index}=
-\6_\mu\,\frac{\7\6f_k}{\7\6Q^{*\nu}_\Index}
+\6_\nu\,\frac{\7\6f_k}{\7\6Q^{*\mu}_\Index}
\label{de3}\\
&
\delta^{(0)}\,\frac{\7\6f_k}{\7\6B^\Index_{\mu\nu}}=
\sfrac 14\delta_{\Index\Jndex}\,
\vep^{\mu\nu\rho\sigma}\vep_\rho{}^{\lambda\kappa\tau}
\6_\sigma\6_\lambda\,\frac{\7\6f_k}{\7\6B^{*\kappa\tau}_\Jndex}
\label{de4}\\
&
\delta^{(0)}\,\frac{\7\6f_k}{\7\6C^*_i}=0
\label{de5}\\
&
\delta^{(0)}\,\frac{\7\6f_k}{\7\6A^{*\mu}_i}=
\6_\mu\,\frac{\7\6f_k}{\7\6C^*_i}\ .
\label{de6}
\end{align}
We discuss first the case
$k=3$. ${\7\6f_3}/{\7\6R^*_\Index}$ has antifield
number 0, i.e., it does not depend on antifields.
(\ref{de1}) is thus trivially satisfied
and imposes no condition in the case $k=3$.
${\7\6f_3}/{\7\6Q^{*\mu}_\Index}$ has
antifield number 1 and thus
$\delta^{(0)}{\7\6f_3}/{\7\6Q^{*\mu}_\Index}$ vanishes
on-shell in the linearized theory.
Hence (\ref{de2}) imposes
$\6_\mu({\7\6f_3}/{\7\6R^*_\Index})\approx^{(0)}0$
where $\approx^{(0)}$ is equality on-shell in the
linearized theory, i.e., ${\7\6f_3}/{\7\6R^*_\Index}$
are cocycles of the characteristic cohomology
(= cohomology of $d$ on-shell) of the linearized theory
in form-degree 0. According to theorems 8.1 and 8.2 of
\cite{BBH1} (or theorem 6.2 of \cite{report}), 
the vanishing of $H^4_4(\delta^{(0)}|d)$
implies
that the characteristic cohomology of the linearized theory
is in form-degree 0 represented
just by constants, which in our case are functions of the
constant ghosts. This gives
\bea
\frac{\7\6f_3}{\7\6R^*_\Index}=k^\Index(c,\xi,\5\xi)
+\delta^{(0)}g^\Index_1\ ,
\label{de7}
\eea
for some local functions $g^\Index_1$ with antifield number 1.
Using (\ref{de7}) in (\ref{de2}) and 
proceeding then as in section 9 of \cite{BBH1},
one eventually obtains
\bea
f_3=k^\Index(c,\xi,\5\xi)R^*_\Index+\delta^{(0)}g_4
+\6_\mu g^\mu_3\ ,
\label{de8}
\eea
for some local functions $g_4$ and $g^\mu_3$ with
antifield numbers 4 and 3, respectively.
Hence, $H^4_3(\delta^{(0)}|d)$ is represented
just by the 4-forms $k^\Index(c,\xi,\5\xi)R^*_\Index d^4x$.
Owing
to $\delta R^*_\Index=\delta^{(0)}R^*_\Index$,
this result for $H^4_3(\delta^{(0)}|d)$ 
extends straightforwardly to  $H^4_3(\delta|d)$ which
proves the assertion in (\ref{prode1}) for $k=3$.

The case $k=2$ can be treated analogously.
${\7\6f_2}/{\7\6R^*_\Index}$ vanishes ($f_2$
has antifield number 2 and thus
cannot depend on $R^*_\Index$) while
${\7\6f_2}/{\7\6Q^{*\mu}_\Index}$
and ${\7\6f_2}/{\7\6C^*_i}$ do not depend on
antifields. Hence (\ref{de1}), (\ref{de2}) and (\ref{de5})
are trivially satisfied and impose no condition for
$k=2$. 
(\ref{de6}) imposes that ${\7\6f_2}/{\7\6C^*_i}$
are cocycles of the characteristic cohomology of the linearized theory
in form-degree 0. Analogously to
(\ref{de7}) we conclude
\bea
\frac{\7\6f_2}{\7\6C^*_i}=
k^i(c,\xi,\5\xi)+\delta^{(0)}g^i_1\ ,
\label{de7a}
\eea
for some local functions $g^i_1$ with antifield number 1.
(\ref{de3}) imposes that
$dx^\mu {\7\6f_2}/{\7\6Q^{*\mu}_\Index}$ are
cocycles of the characteristic cohomology
of the linearized theory
in form-degree 1. The latter is
isomorphic to $H^4_3(\delta^{(0)}|d)$
and this isomorphism is established through
descent equations for $\delta^{(0)}$ and $d$
(again, see theorems 8.1 and 8.2 of
\cite{BBH1} or theorem 6.2 of \cite{report}, and the proofs
of these theorems). Using the result for $H^4_3(\delta^{(0)}|d)$
which we just derived and 
the fact that the descent equations for $\delta^{(0)}$ and $d$
relate $R^*_\Index d^4x$ to the 1-forms
$dx^\mu H_{\Index\mu}$, we conclude
\bea
\frac{\7\6f_2}{\7\6Q^{*\mu}_\Index}=
k^{\Index\Jndex}(c,\xi,\5\xi)H_{\Jndex\mu}
+\delta^{(0)}g^\Index_{\mu, 1}
+\6_\mu g^\Index_{0}\ ,
\label{de9}
\eea
for some local functions $g^\Index_{\mu, 1}$ and $g^\Index_{0}$
with antifield numbers 1 and 0, respectively.
Inserting (\ref{de9}) in (\ref{de3}) and using that
$\6_\mu H_{\Index\nu}-\6_\nu H_{\Index\mu}=
\vep_{\mu\nu\rho\sigma}\delta^{(0)}B^{*\rho\sigma}_\Index$, one
obtains from (\ref{de3})
(owing to the acyclicity of $\delta^{(0)}$ in 
positive antifield numbers, see, e.g., section 5 of \cite{report}):
\bea
\frac{\7\6f_2}{\7\6B^{*\mu\nu}_\Index}=
-k^{\Index\Jndex}(c,\xi,\5\xi)
\vep_{\mu\nu\rho\sigma}B^{*\rho\sigma}_\Jndex
-\6_\mu g^\Index_{\nu, 1}+\6_\nu g^\Index_{\mu, 1}
+\delta^{(0)}g^\Index_{\mu\nu, 2}\ ,
\label{de10}
\eea
for some local functions $g^\Index_{\mu\nu, 2}$ with
antifield number 2. Using (\ref{de10}) in (\ref{de4}),
the latter gives (again owing to the acyclicity of $\delta^{(0)}$ in 
positive antifield numbers):
\bea
\frac{\7\6f_2}{\7\6B_{\mu\nu}^\Index}=
\sfrac 12\delta_{\Index\Jndex}\,
k^{\Jndex\Kdex}\vep^{\mu\nu\rho\sigma}\6_\sigma Q^*_{\Kdex\rho}
-\sfrac 32\delta_{\Index\Jndex}\,
\6_\rho\6^{[\rho}g^{\mu\nu]\Jndex}{}_{,2}
+\delta^{(0)}g^{\mu\nu}_{\Index,3}\ ,
\label{de11}
\eea
for some local functions $g^{\mu\nu}_{\Index,3}$ with
antifield number 3.
Using (\ref{de7a}) through (\ref{de11}) and
proceeding then as in section 9 of \cite{BBH1},
one obtains
\bea
f_2&=&
k^i(c,\xi,\5\xi)C^*_i+
\sfrac 12[k^{\Index\Jndex}(c,\xi,\5\xi)-k^{\Jndex\Index}(c,\xi,\5\xi)]
Q^{*\mu}_\Index H_{\Jndex\mu}
\nonumber\\
&&-\sfrac 12k^{\Index\Jndex}(c,\xi,\5\xi)
\vep_{\mu\nu\rho\sigma}B^{*\mu\nu}_\Index B^{*\rho\sigma}_\Jndex
+\delta^{(0)}g_3+\6_\mu g^\mu_2\ ,
\label{de12}
\eea
for some local functions $g_3$ and $g^\mu_2$ with
antifield numbers 3 and 2, respectively.
Hence, $H^4_2(\delta^{(0)}|d)$ is represented
by the 4-forms $k^i(c,\xi,\5\xi)C^*_id^4x$
and $k^{[\Index\Jndex]}(c,\xi,\5\xi)f_{\Index\Jndex}d^4x$
with $f_{\Index\Jndex}$ as given in the lemma.
$k^{[\Index\Jndex]}(c,\xi,\5\xi)f_{\Index\Jndex}d^4x$
is a cocycle of $H^4_2(\delta|d)$ owing
to $\delta Q^{*\mu}_\Index=\delta^{(0)}Q^{*\mu}_\Index$ and
$\delta B^{*\mu\nu}_\Index=\delta^{(0)}B^{*\mu\nu}_\Index$.
In contrast, $\delta^{(0)}C^*_i$ and
$\delta C^*_i$ do not coincide except for $i=\ifree$.
The nonlinear terms in $\delta C^*_i$ for other $i$ obstruct
the completability of these terms to
cocycles of  $H^4_2(\delta|d)$, see section 13
of \cite{BBH1}. This
yields the result for $k=2$ asserted in (\ref{prode1}).

To prove (\ref{prode2}) we take the
Euler-Lagrange derivative of 
$ R^*_\Index k^\Index(c,\xi,\5\xi)=\delta g_4+\6_\mu g^\mu_3$
(which is equivalent to $R^*_\Index k^\Index(c,\xi,\5\xi) d^4x\sim 0$)
with respect to $R^*_\Index$. The result is
\bea
k^\Index(c,\xi,\5\xi)=- \delta\,\frac{\7\6g_4}{\7\6R^*_\Index}\ .
\label{de13}
\eea
The presence of $\delta$ implies that
the right hand side contains only terms that are at least
linear in fields or antifields, unless it vanishes.
Hence both sides of (\ref{de13}) must vanish which
gives $k^\Index(c,\xi,\5\xi)=0$.

(\ref{prode3}) can be proved in an analogous manner. The
Euler-Lagrange derivatives of
$C^*_{\ifree}k^{\ifree}(c,\xi,\5\xi)
+ f_{\Index\Jndex}k^{[\Index\Jndex]}(c,\xi,\5\xi)=
\delta g_3+\6_\mu g^\mu_2$ with respect to
$C^*_{\ifree}$ and $Q^{*\mu}_\Index$ read, respectively:
\bea
&&
k^{\ifree}(c,\xi,\5\xi)=\delta\,\frac{\7\6g_3}{\7\6C^*_{\ifree}}\ ,
\label{de14}\\
&&
H_{\Jndex\mu}k^{[\Index\Jndex]}(c,\xi,\5\xi)
=\delta\,\frac{\7\6g_3}{\7\6Q^{*\mu}_\Index}
-\Ii\,\6_\mu\,\frac{\7\6g_3}{\7\6R^*_\Index}\ .
\label{de15}
\eea
(\ref{de14}) implies $k^{\ifree}(c,\xi,\5\xi)=0$ by the
same arguments as in the text after (\ref{de13}).
(\ref{de15}) states that the 1-forms
$dx^\mu H_{\Jndex\mu}k^{[\Index\Jndex]}(c,\xi,\5\xi)$ are
trivial in the characteristic cohomology.
This is equivalent to the statement
that the 4-forms 
$d^4x R^*_\Jndex k^{[\Index\Jndex]}(c,\xi,\5\xi)$
are trivial in $H^4_3(\delta|d)$, see text before (\ref{de9}).
Using (\ref{prode2}), which we have already proved, we thus conclude from
(\ref{de15}) that the $k^{[\Index\Jndex]}(c,\xi,\5\xi)$ vanish.
\QED

\proof{pros|d<-1}
We decompose the
cocycle condition $\ext\omega^{g,4}+d\omega^{g-1,3}=0$
into pieces with different antifield numbers.
This gives
\bea
&&
\delta\omega^{g,4}_\unk+d\omega^{g-1,3}_{\unk-1}=0,\quad
\gammaext\omega^{g,4}_\unk
+\delta\omega^{g,4}_{\unk+1}
+d\omega^{g-1,3}_{\unk}=0,\ \dots
\label{s<3,4}
\eea
where $\omega^{g,4}_k$ and $\omega^{g-1,3}_k$ are
the pieces with antifield number $k$ contained
in $\omega^{g,4}$ and $\omega^{g-1,3}$, respectively,
and $\unk$ is the smallest antifield number which
occurs in the decomposition of  $\omega^{g,4}$.
The first equation in (\ref{s<3,4}) states that
$\omega^{g,4}_\unk$ is a cocycle of
$H^4_\unk(\delta|d)$. Without loss of generality
we can assume that it is nontrivial in
$H^4_\unk(\delta|d)$ because otherwise
we could remove it from $\omega^{g,4}$ by
subtracting a coboundary of $H^{g,4}(\ext|d)$ from it
(if $\omega^{g,4}_\unk=\delta\omega^{g-1,4}_{\unk+1}
+d\omega^{g,3}_{\unk}$, consider
$\omega^{g,4}-\ext\omega^{g-1,4}_{\unk+1}
-d\omega^{g,3}_{\unk}$).
Now, if $g<-3$, then $\unk>3$ (the antifield
number of a form cannot be smaller than minus
its ghost number). Since $H^4_\unk(\delta|d)$
vanishes for $\unk>3$ according to
lemma \ref{prodelta|d}, we conclude that 
$H^{g,4}(\ext|d)$ vanishes for $g<-3$. 
This yields the result for $g<-3$ in (\ref{s<3,1}).
If $g=-3$,
then $\unk\geq 3$ and lemma \ref{prodelta|d}
implies that 
we can assume that $\unk=3$ and that
$\omega^{-3,4}_{3}$ is a linear combination
of the 4-forms $R^*_\Index d^4x$ with
numerical coefficients (constant ghosts cannot occur in
these coefficients for $g=-3$ because $R^*_\Index$ has
ghost number $-3$).
Since $R^*_\Index d^4x$ is a cocycle not only of
$H(\delta|d)$ but also of $H(\ext|d)$ (owing to
$\ext R^*_\Index =(\delta+\7c^\mu\6_\mu) R^*_\Index
=\6_\mu(\Ii Q^{*\mu}_\Index+\7c^\mu R^*_\Index)$), this 
yields the result for $g=-3$ in (\ref{s<3,1}).
If $g=-2$, then $\unk\geq 2$. Now lemma \ref{prodelta|d}
leaves two possibilities, $\unk= 2$ or $\unk= 3$.
In the case $\unk= 2$ it implies that 
$\omega^{-2,4}_{2}$ is a linear combination
of the 4-forms $C^*_{\ifree}d^4x$ and
$f_{\Index\Jndex}d^4x$ with numerical coefficients
(as $C^*_{\ifree}$ and $f_{\Index\Jndex}$ have ghost number $-2$).
One may now verify by direct computation that
$f'_{\Index\Jndex}d^4x$ completes
$f_{\Index\Jndex}d^4x$ to a cocycle of $H(\ext|d)$.
$C^*_{\ifree}d^4x$ is already a cocycle of $H(\ext|d)$
(owing to $\ext C^*_{\ifree}=
(\delta+\7c^\mu\6_\mu) C^*_{\ifree}
=\6_\mu(-A^{*\mu}_{\ifree}+\7c^\mu C^*_{\ifree})$).
In the case $g=-2$, $\unk=3$ we conclude from
lemma \ref{prodelta|d} that 
$\omega^{-2,4}_{3}$ is a linear combination
of the  4-forms $R^*_\Index d^4x$ with coefficients
$k^\Index(c,\xi,\5\xi)$ that
are linear combinations of the constant ghosts.
Owing to $\gammaext R^*_\Index=\6_\mu(\7c^\mu R^*_\Index)$,
the second equation in (\ref{s<3,4}) imposes
$[\gammaext k^\Index(c,\xi,\5\xi)]R^*_\Index d^4x=\delta(\dots )+d(\dots )$,
i.e., the triviality of 
$[\gammaext k^\Index(c,\xi,\5\xi)]R^*_\Index d^4x$  in $H(\delta|d)$.
This implies $\gammaext k^\Index(c,\xi,\5\xi)=0$
according to (\ref{prode2}) because
$\gammaext k^\Index(c,\xi,\5\xi)$ is again a function
of the constant ghosts. 
$\gammaext k^\Index(c,\xi,\5\xi)=0$ implies $k^\Index(c,\xi,\5\xi)=0$
since no linear combination of the constant ghosts is
$\gammaext$-closed as can be easily verified directly [or
deduced from the Lie algebra cohomology
of the Lorentz group, as $\gammaext$ contains the Lorentz transformations;
see also section \ref{secLAC}].
This completes the proof of (\ref{s<3,1}).

To prove
(\ref{s<3,2}) we use that $k^\Index R^*_\Index d^4x\sim 0$ is
equivalent to $k^\Index R^*_\Index=\ext f+\6_\mu f^\mu$ (for some
$f$ and $f^\mu$)
and take the Euler-Lagrange derivative of the latter equation
with respect to $R^*_\Index$. This gives
\bea
k^\Index=-\ext\,\frac{\7\6f}{\7\6R^*_\Index}
+\7c^\mu\6_\mu\,\frac{\7\6f}{\7\6R^*_\Index}
-2\xi\sigma^\mu\5\xi\,\frac{\7\6f}{\7\6Q^{*\mu}_\Index}\ .
\label{s<3,5}
\eea
All terms on the right hand side depend on fields, antifields
or constant ghosts while
the left hand side is a pure number.
Hence both sides of (\ref{s<3,5}) must vanish and
we conclude $k^\Index=0$.

(\ref{s<3,3}) can be proved analogously by taking the
Euler-Lagrange derivatives of the equation
$k^\ifree C^*_\ifree+ k^{[\Index\Jndex]}f'_{\Index\Jndex}=
\ext f+\6_\mu f^\mu$ 
with respect to $C^*_\ifree$ and $Q^{*\mu}_\Index$, respectively.
This gives
\bea
&&
k^\ifree=\ext\,\frac{\7\6f}{\7\6C^*_\ifree}
-\7c^\mu\6_\mu\,\frac{\7\6f}{\7\6C^*_\ifree}
+2\Ii\,\xi\sigma^\mu\5\xi\,\frac{\7\6f}{\7\6 A^{*\mu}_\ifree}
\label{s<3,6a}\\
&&
k^{[\Index\Jndex]}H_{\mu\Jndex}=
\ext\,\frac{\7\6f}{\7\6Q^{*\mu}_\Index}
-\7c^\nu\6_\nu\,\frac{\7\6f}{\7\6Q^{*\mu}_\Index}
+c_\mu{}^\nu\,\frac{\7\6f}{\7\6Q^{*\nu}_\Index}
-\Ii\,\6_\mu\,\frac{\7\6f}{\7\6R^*_\Index}
+2\Ii\,\xi\sigma^\nu\5\xi\,\frac{\7\6f}{\7\6B^{*\mu\nu}_\Index}\ .\quad
\label{s<3,7a}
\eea
(\ref{s<3,6a}) implies $k^\ifree=0$ 
by the same arguments
as the text after (\ref{s<3,5}).
Now consider (\ref{s<3,7a}). The ghost-independent
part of that equation is of the form $k^{[\Index\Jndex]}H_{\mu\Jndex}
=\delta(\dots)+\6_\mu(\dots)$ which 
implies $k^{[\Index\Jndex]}=0$ by the same arguments
as in the text after (\ref{de15}).
\QED

\proof{pros<3}
We use that $H^g(\ext,\Space)$ is isomorphic to $H^{g-4,4}(\ext|d)\oplus H^g(\ext,\CONST)$,
where the representatives of $H^{g-4,4}(\ext|d)$ and $H^g(\ext,\Space)$
are related through descent
equations for $\ext$ and $d$, see lemma \ref{proDescent} and its proof.
According to lemma \ref{pros|d<-1},
$H^{g-4,4}(\ext|d)$ is represented for $g<3$ by 
linear combinations of $R^*_\Index d^4x$, $C^*_i d^4x$ and 
$f'_{\Index\Jndex} d^4x$. 
These 4-forms are related through
descent equations for $\ext$ and $d$ to 
the 0-forms $\Ii\4H^\Jndex\delta_{\Index\Jndex}$, 
$\4F^{\jfree}\Kill_{\ifree\jfree}$ and
$(1/2)\4H^\Kdex \4H^\Ldex\delta_{\Kdex\Index}\delta_{\Ldex\Jndex}$,
respectively,
as can be verified by direct computation.
The contributions of $H(\ext|d)$ to the cohomology groups $H^g(\ext,\Space)$,
$g<3$, are thus representatives
$k_\Index\4H^\Index$ (for $g=1$), and
$k_\ifree \4F^{\ifree}+
(1/2)k_{[\Index\Jndex]}\4H^\Index \4H^\Jndex$ (for $g=2$).
The contribution of $H(\ext,\CONST)$ is exhausted by complex numbers
for $g=0$. The reason is that $\ext$ contains the Lorentz transformations
which enforces that
representatives of 
$H(\ext,\CONST)$ be Lorentz-invariant (as
the Lie algebra cohomology of the Lorentz-algebra
enters here, see section \ref{secLAC} for details);
however it is impossible to build nonvanishing Lorentz-invariants
with ghost number 1 or 2 in $\CONST$ 
owing to the Grassmann gradings of the constant ghosts 
(in particular, $\7c^\mu\7c_\mu$ and $\xi^\alpha\xi_\alpha$ vanish).
This leaves us with
$H^0(\ext,\CONST)$ which is evidently represented
just by complex numbers because 
functions in $\CONST$ with ghost number 0 are pure numbers.
This ends the proof since
$\CONST$ does not contain polynomials with
negative ghost numbers.\QED

\proof{propairs} 
The proof is standard and uses a contracting
homotopy $\varrho$ for the $u$'s and $v$'s. $\varrho$ is defined in the
space of functions $f(u,v,w)$ according to
\[
\varrho f(u,v,w)=\int_0^1 \frac{dt}{t}\,
u^\ell\,\frac{\6f(tu,tv,w)}{\6v^\ell} .
\]
Owing to (\ref{susw}), one has
\[
\{\ext,\varrho\}f(u,v,w)=
\int_0^1 \frac{dt}{t}\Big[
u^\ell\,\frac{\6}{\6u^\ell}+v^\ell\,\frac{\6}{\6v^\ell}\Big]f(tu,tv,w)
=f(u,v,w)-f(0,0,w).
\]
This implies
\[
\ext f(u,v,w)=0\ \then\ f(u,v,w)=f(0,0,w)+\ext\varrho f(u,v,w),
\]
i.e., only the piece $f(0,0,w)$ contained in an $\ext$-cocycle
$f(u,v,w)$ can be nontrivial. 
Owing to $f(0,0,w)\in\Wspace$ and $\ext\Wspace\subset\Wspace$
(the latter follows from (\ref{susw})), one concludes
$H(\ext,\Space)\simeq H(\ext,\Wspace)$.
\QED

\proof{prohigh}
Since coboundaries of $\lie$ can be removed
from $f_\om$ (see text at the beginning of section \ref{secLAC}),
we conclude from (\ref{LAC1}) that we can assume
$f_\om=f^{\Pindex_\om} P_{\Pindex_\om}(\theta,\7R)$ for some
$f^{\Pindex_\om}\in\inv$. Using this in Eq.\ (\ref{co2})
we obtain $(\susy f^{\Pindex_\om}) P_{\Pindex_\om}(\theta,\7R)
+\lie f_{\om-1}=0$. The latter implies $\susy f^{\Pindex_\om}=0$
for all $f^{\Pindex_\om}$ because of (\ref{LAC2}) (since
$f^{\Pindex_\om}\in\inv$ implies $\susy f^{\Pindex_\om}\in\inv$).
Assume now that $f^{\Pindex_\om}=\susy g$ for one of the
$f^{\Pindex_\om}$ and some $g\in\inv$. Then this $f^{\Pindex_\om}$ can be removed from
$f_\om$ by the redefinition $f-\ext[g P_{\Pindex_\om}(\theta,\7R)]$
(this redefinition only removes the piece with $f^{\Pindex_\om}$
from $f_\om$ and modifies in addition $f_{\om-1}$).
Hence we can indeed assume without loss of generality that
$f^{\Pindex_\om}\neq\susy g^{\Pindex_\om}$ for all $f^{\Pindex_\om}$.
\QED

\proof{prosusy}
The proof is based on methods and results developed in \cite{glusy}
and \cite{sugra}. Therefore we shall only sketch the main line
of reasoning and refer for details to the respective sections and 
equations in these works.

(i) The results (\ref{cocsusy}) for $g=0,1,2$ arise as follows.
Owing to (\ref{susyinW}), nontrivial representatives
of $H^g(\ext,\Space)$ which are in $\inv$ are also nontrivial representatives
of $H^g(\susy,\inv)$. For $g=0,1,2$ these are given by
lemma \ref{pros<3}. In addition $H^g(\susy,\inv)$ contains
functions $f\in\inv$ which are trivial in $H^g(\ext,\Space)$
but nontrivial in $H^g(\susy,\inv)$: $f=\ext\beta$ for some $\beta$,
but not for any $\beta\in\inv$. For $g<3$,
the latter are exhausted
by (linear combinations of) the Abelian $\cF$'s, up to
trivial terms:
\bea
f=\ext\beta,\ f\in\inv,\ \beta\in\Space\quad\then\quad
f=\left\{\ba{ll}
0 & \mbox{if}\quad g=0\\
\ext\beta',\ \beta'\in\inv & \mbox{if}\quad g=1\\
k_{i_A}\cF^{i_A}+\ext\beta',\ \beta'\in\inv & \mbox{if}\quad g=2.
\ea\right.
\label{g=012}
\eea
This can be proved as an analogous result in appendix D of
\cite{sugra}. The basic ideas are as follows.
One can assume $\beta\in\Wspace$, see lemma \ref{propairs}.
The result (\ref{g=012}) for $g=0$ simply reflects that
$\Wspace$ contains no function with negative ghost number, i.e.,
there is no $\beta$ with ghost number $-1$. The
result (\ref{g=012}) for $g=1$ holds because functions in $\Wspace$ 
with ghost number 0 can only depend
on the $\7T$'s which implies $\beta=\beta(\7T)\in\inv$.
The result (\ref{g=012}) for $g=2$ is obtained by differentiation of 
$f=\ext\beta$ with respect to
$\7C^i$ or $c^{\mu\nu}$ which yields 
$\ext(\6\beta/\6\7C^i)=0$ and $\ext(\6\beta/\6c^{\mu\nu})=0$,
without loss of generality (see \cite{sugra}).
Using now lemma \ref{pros<3} and
that $\6\beta/\6\7C^i$ and $\6\beta/\6c^{\mu\nu}$ have ghost number
0 in the case $g=2$, one concludes that these expressions are
complex numbers, and thus $\beta=k'_i\7C^i+k'_{\mu\nu}c^{\mu\nu}
+\beta'(\7c,\xi,\5\xi,\7T)$. $f=\ext\beta\in\inv$
implies then $k'_i=0$ unless $\delta_i$ is Abelian,
$k'_{\mu\nu}=0$, and $\beta'\in\inv$.

The results (\ref{cocsusy}) for $g\geq 3$ are derived
using a decomposition of the cocycle condition $\susy f=0$
with respect to the degree in the translation ghosts $\7c^\mu$
(``$\7c$-degree'').
$\susy$ decomposes into pieces $\delta_-$, $\delta_0$,
$\delta_+$ with $\7c$-degrees $-1$, 0, 1, respectively,
where $\delta_0$ consists of two differentials $b$ and $\5b$
that involve $\xi$ and $\5\xi$, respectively:
\bea
&&\susy=\delta_-+\delta_0+\delta_+,\quad \delta_0=b+\5b,
\nonumber\\
&&\delta_-=2\Ii\xi\sigma^\mu\5\xi\,\frac{\6}{\6 \7c^\mu}\, ,\quad
b=\xi^\alpha\cD_\alpha\, ,\quad \5b=\5\xi^\da\5\cD_\da\, ,\quad 
\delta_+=\7c^\mu\7\nabla_\mu\, .
\label{lastlabel}
\eea
Here $\cD_\alpha$, $\5\cD_\da$ and $\7\nabla_\mu$ act nontrivially
only on the $\7T$'s.
To prove the results (\ref{cocsusy}) for $g\geq 3$
one only needs the cohomology of $\delta_-$,
and the cohomologies of
$b$ and $\5b$ in certain spaces specified below.
The cohomology of $\delta_-$ 
was given in eq.\ (6.5) of \cite{glusy} (see also \cite{Dixon:1995jt});
it reads:
\bea
&\delta_- f(\7c,\xi,\5\xi)=0\ \LRA\
f(\7c,\xi,\5\xi)=P(\5\vartheta,\xi)+
Q(\vartheta,\5\xi)+\Theta M+\delta_- g(\7c,\xi,\5\xi),&
\nonumber\\
&\Theta=\7c^\mu\xi\sigma_\mu\5\xi,\quad
\vartheta^\alpha=\7c^\mu(\5\xi\5\sigma_\mu)^\alpha,\quad
\5\vartheta^\da=\7c^\mu(\5\sigma_\mu\xi)^\da,&
\label{delta-}
\eea
where nonvanishing $P(\5\vartheta,\xi)$,
$Q(\vartheta,\5\xi)$ and $\Theta M$ are nontrivial in
$H(\delta_-)$.
This result holds in the space of all polynomials in the $\7c^\mu$,
$\xi^\alpha$, $\5\xi^\da$, irrespectively of whether or not
they are Lorentz-invariant. Of course, it holds analogously
in the space of polynomials that can also depend on the $\7T$'s
because the latter are inert to $\delta_-$ (in that space
(\ref{delta-}) thus holds with $f$, $P$, $Q$, $M$, $g$ depending
also on the $\7T$'s).
The cohomology of $\5b$ is needed 
in the space of local functions of the $\5\xi^\da$ and $\7T^\Tindex$
which are invariant
under ${\mathfrak sl}(2,{\mathbb C})$-transformations 
of the dotted spinor indices\footnote{$\5b$ is indeed a differential 
because of $\5b^2=(1/2)\5\xi^\da\5\xi^\dbe\{\5\cD_\da,\5\cD_\dbe\}=0$ which
follows from (\ref{alg}).}. This cohomology is given by
\bea
\5b f(\5\xi,\7T)=\5l_{\da\dbe}f(\5\xi,\7T)=0&\quad  \LRA\quad &
f(\5\xi,\7T)=A(\varphi,\7\lambda)+\5\cD^2B(\7T)+\5b g(\5\xi,\7T),
\nonumber\\
&&\5l_{\da\dbe}B(\7T)=\5l_{\da\dbe}g(\5\xi,\7T)=0,
\label{bcohom}
\eea
where $\5l_{\da\dbe}=
-\sfrac 12\5\sigma^{\mu\nu}_{\da\dbe}l_{\mu\nu}$ generates the
${\mathfrak sl}(2,{\mathbb C})$-transformations 
of the dotted spinor indices 
($\5l_{\da\dbe}\5\psi_\dg=-\vep_{\dg(\da}\5\psi_{\dbe)}$ etc.).
(\ref{bcohom}) can be derived from its linearized version which
concerns the cohomology of the operator $\5b_0=\5\xi^\da\5D_\da$ involving
only the linear part 
$\5D_\da$ of $\5\cD_\da$ (i.e., $\5D_\da\7T^\Tindex$
is the part of $\5\cD_\da\7T^\Tindex$ which is linear in the
$\7T$'s; e.g., one has $\5\cD_\da\hatlambda{}^i_\dbe=
-\7F_{\da\dbe}^{i(-)}+\vep_{\da\dbe}\Kill^{ij}\5\varphi T_j\varphi$
and thus $\5D_\da\hatlambda{}^i_\dbe=
-\7F_{\da\dbe}^{i(-)}$) and is given by:
\bea
\5b_0 f(\5\xi,\7T)=\5l_{\da\dbe}f(\5\xi,\7T)=0&\quad \LRA\quad&
f(\5\xi,\7T)=A(\varphi,\7\lambda)+\5D^2B(\7T)+\5b_0 g(\5\xi,\7T),
\nonumber\\
&&\5l_{\da\dbe}B(\7T)=\5l_{\da\dbe}g(\5\xi,\7T)=0.
\label{b0cohom}
\eea
This follows from eq.\ (6.8) of \cite{glusy} because the 
$\5D_\da$-representation on the $\7T^\Tindex$ has QDS structure 
in the terminology
used there: indeed, the $\5D_\da$-representation
decomposes into the singlets $\varphi^\indec$ and $\7\lambda^i$
and infinitely many (D)-doublets given by
$((\7\nabla^+_+)^q\hatlambda{}^i,-(\7\nabla^+_+)^q\7F^{i(-)})$,
$((\7\nabla^+_+)^q\7F^{i(+)},2\Ii(\7\nabla^+_+)^{q+1}\7\lambda^i)$,
$((\7\nabla^+_+)^q\phi^\Index,(\7\nabla^+_+)^q\hatpsi{}^\Index$),
$((\7\nabla^+_+)^q\7\psi^\Index,(\7\nabla^+_+)^q\7H^\Index
-\Ii(\7\nabla^+_+)^{q+1}\phi^\Index)$,
$((\7\nabla^+_+)^q\5\varphi^\indec,(\7\nabla^+_+)^q\hatchi{}^\indec)$,
$((\7\nabla^+_+)^q\7\chi^\indec,-2\Ii(\7\nabla^+_+)^{q+1}\varphi^\indec)$
(where $q=0,1,\dots$).
(\ref{bcohom}) follows straightforwardly from (\ref{b0cohom})
by a standard argument: let $f$ be a $\5b$-cocycle; one can decompose
$f$ according to $f=\sum_{k\geq k_0} f_k$
into parts $f_k$ with definite degree $k$ in the $\7T$'s where
$k_0$ denotes the lowest nonvanishing degree of the decomposition; 
$\5b f=0$ implies
$\5b_0 f_{k_0}=0$; (\ref{b0cohom})
implies $f_{k_0}=A_{k_0}(\varphi,\7\lambda)
+\5D^2B_{k_0}(\7T)+\5b_0 g_{k_0}(\5\xi,\7T)$ for some
$A_{k_0}(\varphi,\7\lambda)$, $B_{k_0}(\7T)$ and 
$g_{k_0}(\5\xi,\7T)$ that are $\5l_{\da\dbe}$-invariant;
one now considers $f'=f-A_{k_0}(\varphi,\7\lambda)
-\5\cD^2B_{k_0}(\7T)-\5b g_{k_0}(\5\xi,\7T)$; by construction $f'$ is
$\5b$-invariant and its decomposition starts at some degree $k'_0>k_0$;
one repeats the arguments for $f'$ and goes on
until one has proved (\ref{bcohom}) with
$A=A_{k_0}+A_{k'_0}+\dots$, $B=B_{k_0}+B_{k'_0}+\dots$ and
$g=g_{k_0}+g_{k'_0}+\dots$

Armed with (\ref{delta-}) and
(\ref{bcohom}), one can prove the results (\ref{cocsusy}) 
for $g\geq 4$ as in section 6 of \cite{glusy}
(the arguments used in \cite{glusy} go through
in $\inv$ because all
relevant operations used there commute with the $\delta_i$) 
and for $g=3$ as the corresponding
result in appendix
E of \cite{sugra}.

(ii) The nontriviality and inequivalence of
the cocycles 1, $\4H^\Index$, $\4F^{\ifree}$ and
$\4H^{\Index} \4H^{\Jndex}$, $\Index< \Jndex$, in
$H(\susy,\inv)$ follows already from
(\ref{s<3,6}), (\ref{s<3,7}) and (\ref{s<3,8}).
The nontriviality and inequivalence 
of the $\cF^{i_A}$ in $H(\susy,\inv)$ can be
shown as the corresponding result in appendix E of \cite{sugra}.
That $\cF^{i_A}$ is not equivalent to
$\4F^{\ifree}$ or
$\4H^{\Index} \4H^{\Jndex}$ can be deduced from its
$\ext$-exactness, $\cF^{i_A}=\ext\7C^{i_A}$:
$k_{\ifree}\4F^{\ifree}+
\sfrac 12 k_{\Index\Jndex}\4H^\Index \4H^\Jndex+
k_{i_A}\cF^{i_A}=\susy\omega$ for some $\omega\in\inv$
implies $k_{\ifree}\4F^{\ifree}+\sfrac 12 k_{\Index\Jndex}\4H^\Index \4H^\Jndex
=\ext(\omega-k_{i_A}\7C^{i_A})$ and thus $k_{\ifree}=k_{[\Index\Jndex]}=0$
by (\ref{s<3,8}).
This proves the assertion on the cocycles
1, $\4H^\Index$, $\4F^{\ifree}$,
$\4H^{\Index} \4H^{\Jndex}$ and $\cF^{i_A}$ in part (ii).

Next we shall prove the assertion on
$\cO R(\7T)$.
Assume that $\cO R(\7T)=\susy\omega$
for some  $\omega\in\inv$. Notice
that $\omega$ is defined only up to the addition of 
$\susy$-cocycles in $\inv$. In particular we are free
to add terms $\susy f$ with $f\in\inv$
to $\omega$ 
(owing to $\susy^2f=0$ for $f\in\inv$).
Since $\omega$ has ghost number 2 in this case,
it decomposes into $\omega_0+\omega_1+\omega_2$ where subscripts
indicate the $\7c$-degree. One has
$\omega_0=\xi^\alpha\xi^\beta\omega_{\alpha\beta}(\7T)+
\xi^\alpha\5\xi^\da\omega_{\alpha\da}(\7T)
+\5\xi^\da\5\xi^\dbe\omega_{\da\dbe}(\7T)$.
The part of $\cO R(\7T)=\susy\omega$ with $\7c$-degree 0 reads
$0=\delta_0\omega_0+\delta_-\omega_1$. This gives
in particular
$\5b[\5\xi^\da\5\xi^\dbe\omega_{\da\dbe}(\7T)]=0$.
(\ref{bcohom}) implies thus
$\5\xi^\da\5\xi^\dbe\omega_{\da\dbe}(\7T)=\5b\eta$
for some $\eta=\5\xi^\da\eta_\da(\7T)\in\inv$. Without
loss of generality one can thus set $\omega_{\da\dbe}(\7T)$
to zero because the latter can be removed by subtracting
$\susy\eta$ from $\omega$.
Analogously one concludes that one can also set 
$\omega_{\alpha\beta}(\7T)$ to zero.
One is then left with $\omega_0=
\xi^\alpha\5\xi^\da\omega_{\alpha\da}(\7T)$ which
can also be set to zero because it can be removed from
$\omega$ by adding $\susy a$ with $a=
(\Ii/4)\7c^{\alpha\da}\omega_{\alpha\da}(\7T)$.
Hence, one can assume $\omega=\omega_1+\omega_2$
with $\delta_-\omega_1=0$.
Using (\ref{delta-}) and that $\omega$ has ghost number 2,
one concludes
$\omega_1=8\vartheta\Omega_1+8\5\vartheta\5\Omega_2$ for
some functions $\Omega_1^\alpha$ and $\5\Omega_2^\da$ of the $\7T$'s
that are gauge invariant and transform under Lorentz transformations
as indicated by their indices
[$\omega_1$ cannot contain a $\delta_-$-exact piece because it
has ghost number 2 and $\7c$-degree 1 and thus depends linearly
on the supersymmetry ghosts whereas $\delta_-$-exact terms
depend at least quadratically on them].
The part of $\cO R(\7T)=\susy\omega$ with $\7c$-degree 1
reads $4\Theta R(\7T)=\delta_0\omega_1+\delta_-\omega_2$.
This gives in particular $\5b\Omega_1^\alpha=0$.
Using (\ref{bcohom}) one
concludes $\Omega_1^\alpha=
A^\alpha_1(\varphi,\7\lambda)+\5\cD^2B^\alpha_1(\7T)$
for some functions $A^\alpha_1(\varphi,\7\lambda)$ and $B^\alpha_1(\7T)$.
Analogously one concludes 
$\5\Omega_2^\da(\cT)=
\5A^\da_2(\5\varphi,\hatlambda)+\cD^2\5B^\da_2(\7T)$.
Using this in
$4\Theta R(\7T)=\delta_0\omega_1+\delta_-\omega_2$
one obtains
$4\Theta R(\7T)+8\vartheta b\Omega_1+8\5\vartheta\5b\5\Omega_2
=\delta_-\omega_2$. By means of the $\delta_-$-cohomology
one concludes
$R(\7T)=\cD \Omega_1+\5\cD\5\Omega_2$
and $\omega_2=-\Ii \7c_{\alpha\da}\7c^\da{}_\beta\cD^\alpha\Omega^\beta_1
-\Ii\7c^\alpha{}_\da\7c_{\alpha\dbe}\5\cD^\da\5\Omega^\dbe_2$
[$\omega_2$ cannot contain a $\delta_-$-exact piece because it
has ghost number 2 and $\7c$-degree 2 and thus does not depend
on the supersymmetry ghosts].
Conversely, $R(\7T)=\cD \Omega_1+\5\cD\5\Omega_2$ with
$\Omega_1^\alpha$ and $\5\Omega_2^\da$ as above
implies $\cO R(\7T)=\susy (\omega_1+\omega_2)$
with $\omega_1$ and $\omega_2$ as above.

The assertion on $\cP \Omega_1+\5\cP \5\Omega_2$
is proved in an analogous manner by
decomposing the equation $\cP \Omega_1+\5\cP \5\Omega_2=\susy\omega$
into parts with different $\7c$-degrees. 
The parts with $\7c$-degree 0 and 1 read
$0=\delta_0\omega_0+\delta_-\omega_1$ and
$0=\delta_+\omega_0+\delta_0\omega_1+\delta_-\omega_2$, respectively.
As in the investigation of the case $G=3$ in \cite{sugra} one
concludes from these equations that, up to trivial terms,
one has $\omega_0=0$,
$\omega_1=32\Theta X(\cT)$, 
$\omega_2=8\Ii \7c^{\da\alpha}(\5\vartheta_\da \cD_\alpha
+\vartheta_\alpha\5\cD_\da)X$ for some $X=X(\7T)\in\inv$.
Using this in the
equation with $\7c$-degree 2 contained
in $\cP \Omega_1+\5\cP \5\Omega_2=\susy\omega$, one
obtains $4\Ii\vartheta\vartheta(\Omega_1-\5\cD^2X)
-4\Ii\5\vartheta\5\vartheta(\5\Omega_2+\cD^2X)=
-8\Ii\vartheta_\alpha\5\vartheta_\da [\cD^\alpha,\5\cD^\da]X
+\delta_-\omega_3$. By means of
the $\delta_-$-cohomology one concludes
$\Omega_1=\5\cD^2X$, $\5\Omega_2=-\cD^2X$ and
$\omega_3=-4\Ii\,\Xi_{\alpha\da} [\cD^\alpha,\5\cD^\da]X$.
Furthermore, using the algebra (\ref{alg}), one verifies that
$\ext(\omega_1+\omega_2+\omega_3)=
(\cP\5\cD^2-\5\cP\cD^2)X$ and concludes that
$\cP \Omega_1+\5\cP \5\Omega_2=\susy\omega$ is indeed
equivalent to $\Omega_1=\5\cD^2X$, $\5\Omega_2=-\cD^2X$
(I note that one has $\omega_1+\omega_2+\omega_3=8\cO X$).
\QED

\proof{prouseful}
(i) $\ext f_4=0$ implies $\delta_-f_4=0$ because $\delta_-$ is the
only part of $\ext$ that lowers the $\7c$-degree.
(\ref{delta-}) implies that the cohomology of $\delta_-$
is trivial for $\7c$-degrees larger than 2 because
$\vartheta^\alpha$ and $\5\vartheta^\da$ are anticommuting quantities.
We conclude that $f_4$ is $\delta_-$-exact, i.e.,
$f_4=\delta_-\eta_5$. This implies $f_4=0$ because
$\eta_5$ has $\7c$-degree 5 and thus vanishes.

(ii) Existence of $\eta_4$: 
$\ext (f_3+f_4)=0$ implies
$\delta_-f_3=0$, from which we conclude
$f_3=\delta_-\eta_4$ by means of
(\ref{delta-}). Consider now $f':=f_3+f_4-\ext\eta_4$: 
it is an $\ext$-closed function with $\7c$-degree 4.
By means of part (i) we conclude that $f'$ vanishes
and thus that $f_3+f_4=\ext\eta_4$.
Uniqueness of $\eta_4$: $f_3+f_4=\ext\eta_4$
and $f_3+f_4=\ext\eta'_4$ imply $\ext(\eta_4-\eta'_4)=0$
and thus $\eta_4-\eta'_4=0$ according to part (i).
\QED

\proof{proq}
We first recall the well-known 
Chern-Simons-polynomials 
$q_\Lindex(A,F)=m(\Lindex)\int_0^1 dt\ \TR(AF_t^{m(\Lindex)-1})$
with $F_t=tF+(t^2-t)A^2$ for
Yang-Mills connection forms $A$ and the curvature forms $F=dA+A^2$.
They satisfy
$dq_\Lindex(A,F)=\TR F^{m(\Lindex)}$, 
see, e.g., \cite{zumino}.
Now, written in terms of matrices,
eq.\ (\ref{s7C}) reads $\ext\7C=-\7C^2+\cF$
which arises from $F=dA+A^2$ by
substituting $\ext$, $\7C$ and $\cF$ for $d$,
$A$ and $F$, respectively. The action
of $\ext$ on polynomials
in the $\7C$ and $\cF$ is thus isomorphic to the
action of $d$ on polynomials
in the $A$ and $F$ (this statement refers to
the free differential algebras of
$\7C$'s and $\cF$'s, and $A$'s and $F$'s, respectively). We conclude
\bea
\ext q_\Lindex(\7C,\cF)=\TR\cF^{m(\Lindex)},\quad
q_\Lindex(\7C,\cF)=m(\Lindex)\int_0^1 dt\ \TR(\7C\cF_t^{m(\Lindex)-1}),
\label{CS}
\eea
for all $\Lindex=1,\dots,\rankg$.
We shall now show that $\TR\cF^{m(\Lindex)}$ vanishes for
$m(\Lindex)>3$. This follows from the explicit form
of $\cF$: eq.\ (\ref{cF}) gives
\bea
\cF=\7F-\Ii\vartheta\7\lambda-\Ii\5\vartheta\hatlambda,\quad
\7F=\sfrac 12\7c^\mu\7c^\nu\7F_{\mu\nu}^iT_i^{(\Lindex)},
\label{cF2}
\eea
with $\vartheta^\alpha$ and $\5\vartheta^\da$ as in
(\ref{delta-}). Since $\vartheta^\alpha$ and $\5\vartheta^\da$
have $\7c$-degree 1,
$\TR\cF^{m(\Lindex)}$ contains only terms with
$\7c$-degree $\geq m(\Lindex)$. In particular it thus vanishes
for $m(\Lindex)>4$ (since the spacetime dimension is 4 and the
translation ghosts anticommute).
For $m(\Lindex)=4$ one obtains:
\bea
\TR\cF^4=\TR(\vartheta\7\lambda+\5\vartheta\hatlambda)^4.
\label{m=4}
\eea
Since the $\vartheta^\alpha$ and $\5\vartheta^\da$ are four anticommuting
quantities, (\ref{m=4}) is proportional
to the product of all these four quantities and thus to
$\vartheta\vartheta\5\vartheta\5\vartheta$.
The latter object has $\7c$-degree 4, is bilinear both in the
$\xi$'s and the $\5\xi$'s, and Lorentz-invariant;
hence it is proportional to $\Xi\xi\xi\5\xi\5\xi$
which vanishes owing to $\xi\xi=\5\xi\5\xi=0$ (as the
supersymmetry ghosts commute).
We conclude that (\ref{m=4}) vanishes and thus
\bea
\TR\cF^{m(\Lindex)}=0\quad{{\mathrm for}}\quad m(\Lindex)\geq 4.
\label{m>=4}
\eea
Eqs.\ (\ref{CS}) and (\ref{m>=4}) yield $\ext\7q_\Lindex=0$
for $m(\Lindex)\geq 4$.

For $m(\Lindex)=3$, (\ref{CS}) gives explicitly:
\bea
\ext \,[\TR(\7C\cF^2-\sfrac 12\7C^3\cF+\sfrac 1{10}\7C^5)]=\TR\cF^3.
\label{m=31}
\eea
Evaluation of the right hand side gives, using
(\ref{cF2}) and the facts that
$\vartheta^\alpha$ and $\5\vartheta^\da$ anticommute and have $\7c$-degree 1:
\bea
\TR\cF^3&=&f_3+f_4,
\nonumber\\
f_3&=&\Ii\,\TR(\vartheta\7\lambda+\5\vartheta\hatlambda)^3=
-\sfrac{3\Ii}{2}
\TR(\5\vartheta\5\vartheta\,\vartheta\7\lambda\,\hatlambda\hatlambda
+\vartheta\vartheta\,\5\vartheta\hatlambda\,\7\lambda\7\lambda),
\label{m=32}
\eea
where $f_4$ has $\7c$-degree 4.
Owing to $\ext(\TR\cF^3)=0$ 
(which follows from (\ref{m=31}) because of $\ext^2=0$),
we can apply part (ii) of lemma \ref{prouseful} to (\ref{m=32})
and conclude:
\bea
\TR\cF^3=
\ext\,[-3\Ii\,\Xi\,\TR(\xi\7\lambda\hatlambda\hatlambda
+\5\xi\hatlambda\7\lambda\7\lambda)],
\label{m=33}
\eea
where we used that 
\bea
\delta_-(2\,\Xi\,\xi^\alpha)=\5\vartheta\5\vartheta\vartheta^\alpha,
\quad
\delta_-(2\,\Xi\,\5\xi_\da)=\vartheta\vartheta\5\vartheta_\da
\label{id1}
\eea
which can be directly verified.
Eqs.\ (\ref{m=31}) and (\ref{m=33}) yield $\ext\7q_\Lindex=0$
for $m(\Lindex)=3$ with $\7q_\Lindex$ as in (\ref{m=3}).
\QED

\proof{pros} 
Let us first describe the general strategy of the proof.
The derivation of the lemma starts off from
our result that the part $f_\om$ of
an $\ext$-cocycle takes the form
$f^\Pindex P_\Pindex(\theta,\7R)$ where
$f^\Pindex$ are representatives of $H(\susy,\inv)$
given by lemma \ref{prosusy}. 
The seven types of representatives given in the lemma
correspond to different representatives $f^\Pindex$
of $H(\susy,\inv)$.
The proof of the lemma comprises two aspects:
(i) determination of those functions $f^\Pindex P_\Pindex(\theta,\7R)$
which can be completed to inequivalent $\ext$-cocycles;
(ii) explicit computation of these $\ext$-cocycles.
These two aspects
can be treated largely independently, i.e., basically one can
carry out (i) without sophisticated computation.
The computations (ii) concern above all the explicit 
determination of the $X$-functions given in the lemma and
are partly quite involved. We shall only sketch the computation of the
functions $X^\Index_\Lindex$. The other  $X$'s can be
analogously derived.

(i) is carried out using results derived above,
in particular lemma \ref{prosusy} 
(which implies already the existence of $X$-functions
with the desired properties
as we shall show in the course of the proof), 
and results on the standard (non-extended)
BRST cohomology.
The latter can be employed here because $\ext$ includes the
standard (non-extended) BRST differential $\gauge$ 
(the $\gauge$-transformations arise from the $\ext$-transformations
by setting all constant ghosts 
$c^\mu$, $c^{\mu\nu}$, $\xi^\alpha$, $\5\xi^\da$ to zero).
A necessary condition for a function $f\in\Space$ to be an $\ext$-cocycle
is thus that it gives a solution to ``complete'' 
descent equations for $\gauge$ and $d$ (where ``complete'' means 
descent equations involving a volume form)
after 
setting $c^{\mu\nu}$, $\xi^\alpha$ and $\5\xi^\da$ to zero
and substituting $dx^\mu$ for $c^\mu$ (the reason is that
the substitution $c^\mu\rightarrow c^\mu+dx^\mu$ promotes
an $\ext$-cocycle to an $(\ext+d)$-cocycle, i.e., to a solution
of complete descent equations for $\ext$ and $d$, see section
\ref{DESCENT}).
In particular this implies that a function
$f^\Pindex P_\Pindex(\theta,\7R)$ cannot be completed
to an $\ext$-cocycle if there is no corresponding
solution of complete descent equtions for $\gauge$ and $d$.
Actually this statement can be refined because
the relevant descent equations for $\gauge$ and $d$
are those in the subspace of Poincar\'e invariant forms%
\footnote{Poincar\'e invariance of a $p$-form
$dx^{\mu_1}\dots dx^{\mu_p}\omega_{[\mu_1\dots\mu_p]}$ means here that
the coefficient functions $\omega_{[\mu_1\dots\mu_p]}$ do not
depend explicitly on the spacetime coordinates and transform
covariantly
under Lorentz transformations according to their
indices $[\mu_1\dots\mu_p]$.
}
with an arbitrary dependence on the Lorentz-$\theta$'s%
\footnote{Since the Lorentz-$\theta$'s are $\ext$-closed and $d$-closed,
they can appear arbitrarily in the solutions of the descent equations.}
(for the argument
applies equally to an extended BRST-differential
which involves the Poincar\'e transformations in addition
to $\gauge$ and arises from $\ext$ by setting only the constant supersymmetry
ghosts to zero).  

We shall now spell out the arguments
more specifically, and separately for the various
types of representatives:
\ben

\item[(1)]
The representatives $f^{(1)}$ arise from functions
$k^\Pindex P_\Pindex(\theta,\7R)$ with
complex numbers $k^{\Pindex}$,
i.e., they involve the constant
representatives of $H(\susy,\inv)$ ($g=0$ in Eq.\ (\ref{cocsusy})).
Functions $P(\theta,\7R)$ give rise to solutions 
of complete 
non-supersymmetric descent equations in four dimensions only if
they do not depend on the Abelian $\7C$'s, nor on
the Yang-Mills-$\theta$'s with
$m(\Lindex)=2$, nor on the $\7R$'s.
The reason is that Abelian ghosts, $R$'s
or Yang-Mills-$\theta$'s with
$m(\Lindex)=2$ lead to
obstructions to the ``lift''\footnote{See section 9.3 of \cite{report}
for the terminology and a general discussion of
``lifts'' in the context
of descent equations.} of 
0-forms $P(\theta_C,\Lth,R)$ to solutions of complete
descent equations where
the $\theta_C$'s are the Yang-Mills-$\theta$'s
with $C$'s in place of $\7C$'s
(i.e., $(\theta_C)_\Lindex\propto \TR C^{2m(\Lindex)-1}$):
if $P$ depends on Abelian ghosts,
the obstruction is encountered
at form-degree 2 and given by
$F^{i_A}\6P(\theta_C,\Lth,R)/\6C^{i_A}$ where
$F^{i_A}=dA^{i_A}$ with $A^{i_A}=dx^\mu A_\mu^{i_A}$;
if $P$ does not depend on Abelian ghosts but on
$R$'s, the obstruction is encountered
at form-degree 3 and given by $-\Ii H^\Index\6P(\theta_C,\Lth,R)/\6R^\Index$
where $H^\Index=dB^\Index$ with 
$B^\Index=(1/2)dx^\mu dx^\nu B_{\mu\nu}^\Index$;
if $P$ neither depends on Abelian ghosts nor on $R$'s
but on Yang-Mills-$\theta$'s with
$m(\Lindex)=2$, the obstruction is encountered
at form-degree 4 and given by
$\sum_{\Lindex:m(\Lindex)=2}\TR F^2\,\6P(\theta_C,\Lth)/
\6(\theta_C)_\Lindex$
where $F=dA+A^2$, $A=dx^\mu A_\mu^i T_i^{(\Lindex)}$.
Hence, in order that
$P$ can be lifted to a complete solution of the
non-supersymmetric descent equations, it
must only depend on the $\theta$'s with
$m(\Lindex)>2$ or the Lorentz-$\theta$'s.
Since the former can be completed to $\ext$-invariant $\7q$'s,
see lemma \ref{proq}, and the latter are already $\ext$-invariant,
one arrives at the representatives $f^{(1)}$.

\item[(2,3)]
The representatives $f^{(2)}$ and $f^{(3)}$ arise from functions
$k^\Pindex_\Index\4H^\Index P_\Pindex(\theta,\7R)$ with
complex numbers $k^\Pindex_{\Index}$,
i.e., they involve the
representatives $\4H^\Index$
of $H(\susy,\inv)$.
These functions give rise to Poincar\'e invariant solutions 
of the complete non-supersymmetric 
descent equations in four dimensions 
only if they
do not depend on the Abelian $\7C$'s
and at most linearly on the $\7R$'s through
terms $\4H^\Index P^{(2)}_\Index(\theta)
+\4H^{[\Index}\7R^{\Jndex]}P^{(3)}_{[\Index\Jndex]}(\theta)$.
This can be shown by arguments analogous to those
used in section 13 of \cite{report} where
the standard BRST cohomology
for free Abelian gauge fields was investigated.
The latter leads, among other things, to
the BRST-invariant forms $\star F^IP_I(C)$
where $F^I=dA^I$ are Abelian field strength 2-forms,
$\star F^I$ their Hodge duals and $P_I(C)$ polynomials
in the Abelian ghosts. The requirement that such
forms can be lifted to Poincar\'e invariant solutions of complete
descent
equations leads to the equation 
$\6_{(I}P_{J)}(C)=0$ 
whose general solution is
$P_I(C)=\6_IP(C)$ where $\6_I=\6/\6C^I$, see Eqs.\ (13.18) 
and (13.19) of \cite{report}.
The standard BRST cohomology
for the theories under study contains, among other things, the
BRST invariant forms $\star H^\Index P_\Index(R,\theta_C,\Lth)$.
The requirement that such
forms can be lifted to Poincar\'e invariant solutions of complete  descent
equations in four dimensions
leads analogously to $\6_{(\Index} P_{\Jndex)}(R,\theta_C,\Lth)=0$ 
and $\6_{i_A}P_\Index(R,\theta_C,\Lth)=0$ 
where $\6_\Index=\6/\6R^\Index$ and $\6_{i_A}=\6/\6C^{i_A}$ 
(otherwise the lift would be obstructed at
form-degree 3 by
$F^{i_A}\star H^\Index \6_{i_A}P_\Index(R,\theta_C,\Lth)$
or at form-degree 4 by
$H^\Index\star H^\Jndex\6_{\Index} P_{\Jndex}(R,\theta_C,\Lth)$).
Now, in contrast to the $C^I$, the $R^\Index$ are
commuting variables and thus the 
general solution to the first condition, 
which has the structure of Killing vector equations in
a flat space with coordinates $R^\Index$, is
$P_\Index(R,\theta_C,\Lth)=P^{(2)}_\Index(\theta_C,\Lth)
+\7R^{\Jndex}P^{(3)}_{[\Index\Jndex]}(\theta_C,\Lth)$
while the second condition imposes in addition
that $P^{(2)}_\Index$ and $P^{(3)}_{[\Index\Jndex]}$
do not depend on the Abelian ghosts.
Let us now separately show how this leads to the
representatives $f^{(2)}$ and $f^{(3)}$.

$f^{(2)}$ derives from $\4H^\Index P^{(2)}_\Index(\theta)$.
Completing the Yang-Mills-$\theta$'s
to the $\7q$'s yields
$\4H^\Index P^{(2)}_\Index(\7q,\Lth)$
(as $P^{(2)}_\Index(\theta)$ does not depend on the Abelian $\7C$'s).
Since the $\4H$'s, $\7q$'s with $m(\Lindex)>2$ 
and Lorentz-$\theta$'s are $\ext$-invariant, 
one has
\bea
\ext\Big[\4H^\Index P^{(2)}_\Index(\7q,\Lth)\Big]=
-\4H^\Index\!\!\!\sum_{\Lindex:m(\Lindex)=2}\!\!\!
\TR(\cF^2)\,
\frac{\6P^{(2)}_\Index(\7q,\Lth)}{\6\7q_\Lindex}\,
\label{2.1}
\eea
where we used 
that $\ext\7q_\Lindex=\TR\cF^2$ for $m(\Lindex)=2$.
The functions $\4H^\Index\,\TR(\cF^2)$ which
occur on the right hand side of (\ref{2.1})
are $\ext$-closed elements of $\inv$ with ghost number 5.
According to lemma \ref{prosusy} they are thus $\ext$-exact
in $\inv$, i.e., there are functions $X^\Index_\Lindex$ such that
\bea
\4H^\Index\,\TR(\cF^2)=\ext X^\Index_\Lindex\, ,\quad
X^\Index_\Lindex\in\inv.
\label{2.0}
\eea 
Such functions are given explicitly in the lemma. Of course, they are
determined only up to $\ext$-cocycles in $\inv$. 
However, this arbitrariness is irrelevant because adding
such $\ext$-cocycles to $X^\Index_\Lindex$ results at most
in adding a solution $f^{(7)}$ and an $\ext$-coboundary
to $f^{(2)}$ [since $X^\Index_\Lindex$ has ghost number
4, lemma \ref{prosusy} implies that it is determined up to
an $\ext$-cocycle of the form 
$\cP\Omega^\Index_{\Lindex,1}+\5\cP\5\Omega^\Index_{\Lindex,2}
+\ext h^\Index_\Lindex$ (with $h^\Index_\Lindex\in\inv$)
which gives rise to a representative $f^{(7)}$ up to 
an $\ext$-coboundary, see item (7) below]%
\footnote{For analogous reasons
the arbitrariness in the $X$'s occurring in
other representatives does not matter.}.
The explicit computation of functions $X^\Index_\Lindex$ 
is sketched at the end of
the proof.
Using (\ref{2.0}), the
right hand side of (\ref{2.1}) gives
\bea
\lefteqn{
-\4H^\Index\!\!\!\sum_{\Lindex:m(\Lindex)=2}\!\!\!
\TR(\cF^2)\,
\frac{\6P^{(2)}_\Index(\7q,\Lth)}{\6\7q_\Lindex}
=
-\!\!\!\sum_{\Lindex:m(\Lindex)=2}\!\!\!
(\ext X^\Index_\Lindex)\,
\frac{\6P^{(2)}_\Index(\7q,\Lth)}{\6\7q_\Lindex}
}
\nonumber\\
&&=-\ext\!\!\!\sum_{\Lindex:m(\Lindex)=2}\!\!\!
X^\Index_\Lindex\,
\frac{\6P^{(2)}_\Index(\7q,\Lth)}{\6\7q_\Lindex}
+\!\!\!\sum_{\Lindex:m(\Lindex)=2\atop \Lindex':m(\Lindex')=2}\!\!\!
X^\Index_\Lindex\,\TRprime(\cF^2)\,
\frac{\6^2P^{(2)}_\Index(\7q,\Lth)}{\6\7q_{\Lindex'}\6\7q_\Lindex}\, .
\label{2.2}
\eea
Since both $X^\Index_\Lindex$
and $\TRprime(\cF^2)$ contain only terms with $\7c$-degrees $\geq 2$,
the last term in (\ref{2.2}) has $\7c$-degree 4.
Combining
(\ref{2.1}) and (\ref{2.2}) and using part (i) of lemma \ref{prouseful},
one concludes
that this term vanishes and that $f^{(2)}$
is $\ext$-closed:
\bea
\ext f^{(2)}=
\!\!\!\sum_{\Lindex:m(\Lindex)=2\atop \Lindex':m(\Lindex')=2}\!\!\!
X^\Index_\Lindex\,\TRprime(\cF^2)\,
\frac{\6^2P^{(2)}_\Index(\7q,\Lth)}{\6\7q_{\Lindex'}\6\7q_\Lindex}
=0.
\label{2.3}
\eea

$f^{(3)}$ derives similarly from
$\4H^{\Index}\7R^{\Jndex}P^{(3)}_{[\Index\Jndex]}(\theta)$.
Again, we complete the Yang-Mills-$\theta$'s 
in $P^{(3)}_{[\Index\Jndex]}$ 
to the $\7q$'s and compute the $\ext$-variation
of the resultant function:
\bea
\ext\Big[
\4H^{\Index}\7R^{\Jndex}P^{(3)}_{[\Index\Jndex]}(\7q,\Lth)
\Big]=
-\Big[
\4H^{[\Index}\cH^{\Jndex]}+
\4H^{[\Index}\7R^{\Jndex]}\!\!\!\sum_{\Lindex:m(\Lindex)=2}\!\!\!
\TR(\cF^2)\,
\frac{\6}{\6\7q_\Lindex}
\Big]
P^{(3)}_{[\Index\Jndex]}(\7q,\Lth).
\label{3.1}
\eea
The function $\4H^{[\Index}\cH^{\Jndex]}$ is an $\ext$-closed
element of $\inv$ with ghost number 4. According to lemma
\ref{prosusy} it could only be nontrivial
in $\inv$ if it were
equivalent to a function
of the form $\cP\Omega_1+\5\cP\5\Omega_2$. However, this
is not the case:
$\4H^{[\Index}\cH^{\Jndex]}$ is quadratic in the fields of
the linear multiplets and
has dimension 0 (using dimension
assignments as in the proof of lemma \ref{proW});
hence, in order to be equivalent to a function
$\cP\Omega_1+\5\cP\5\Omega_2$,
$\Omega_1$ or $\Omega_2$ would have to be given by
$\5\cD^2B(\7T)$ where $B(\7T)$ would be quadratic
in the fields of
the linear multiplets and have dimension 2;
hence $B(\7T)$ would be a bilinear function in the
$\phi^\Index$; however, owing to the antisymmetry
of $\4H^{[\Index}\cH^{\Jndex]}$ 
in the indices $\Index$ and $\Jndex$, this
bilinear function would have to be proportional 
to $\phi^{[\Index}\phi^{\Jndex]}$
which vanishes since the $\phi$'s commute.
Hence, the functions $\4H^{[\Index}\cH^{\Jndex]}$
are $\ext$-exact in $\inv$, i.e., there are
functions $X^{[\Index\Jndex]}\in\inv$ such that
\bea
\4H^{[\Index}\cH^{\Jndex]}=\ext X^{[\Index\Jndex]},\quad
X^{[\Index\Jndex]}\in\inv.
\label{3.2}
\eea
Such functions are given explicitly in the lemma.
Using (\ref{3.2}) and (\ref{2.0}), the right hand side
of (\ref{3.1}) gives
\bea
\lefteqn{
-\Big[
\4H^{[\Index}\cH^{\Jndex]}+
\4H^{[\Index}\7R^{\Jndex]}\!\!\!\sum_{\Lindex:m(\Lindex)=2}\!\!\!
\TR(\cF^2)\,
\frac{\6}{\6\7q_\Lindex}
\Big]
P^{(3)}_{[\Index\Jndex]}(\7q,\Lth)
}
\nonumber\\
&&=
-\Big[(\ext X^{[\Index\Jndex]})+
\!\!\!\sum_{\Lindex:m(\Lindex)=2}\!\!\!
(\ext X^{[\Index}_\Lindex)\7R^{\Jndex]}\,
\frac{\6}{\6\7q_\Lindex}
\Big]
P^{(3)}_{[\Index\Jndex]}(\7q,\Lth)
\nonumber\\
&&=
-\ext\Big\{\Big[
X^{[\Index\Jndex]}+
\!\!\!\sum_{\Lindex:m(\Lindex)=2}\!\!\!
X^{[\Index}_\Lindex\7R^{\Jndex]}\,
\frac{\6}{\6\7q_\Lindex}
\Big]P^{(3)}_{[\Index\Jndex]}(\7q,\Lth)\Big\}
+Z^{(3)},
\label{3.3}
\eea
where
\bea
Z^{(3)}
&=&
\!\!\!\sum_{\Lindex:m(\Lindex)=2}\!\!\!
[
-X^{[\Index\Jndex]}\,\TR(\cF^2)+X^{[\Index}_\Lindex\cH^{\Jndex]}
]\,
\frac{\6P^{(3)}_{[\Index\Jndex]}(\7q,\Lth)}{\6\7q_\Lindex}
\nonumber\\
&&
+\!\!\!\sum_{\Lindex:m(\Lindex)=2\atop\Lindex':m(\Lindex')=2}\!\!\!
X^{[\Index}_\Lindex\7R^{\Jndex]}\,\TRprime(\cF^2)\,
\frac{\6^2P^{(3)}_{[\Index\Jndex]}(\7q,\Lth)}{\6\7q_{\Lindex'}\6\7q_\Lindex}
\, .
\label{z3}
\eea
It is easy to verify that
$Z^{(3)}$ has $\7c$-degree 4 (in particular
the term with $\7c$-degree 3 in $X^{[\Index}_\Lindex\cH^{\Jndex]}$
vanishes as it is proportional to $\phi^{[\Index}\phi^{\Jndex]}$).
Analogously to (\ref{2.3}), we thus conclude
from (\ref{3.1}) and (\ref{3.3}), using again
part (i) of lemma \ref{prouseful}:
\bea
\ext f^{(3)}=Z^{(3)}=0.
\label{3.4}
\eea

\item[(4)]
The representatives $f^{(4)}$ arise from functions
$k^\Pindex_{\ifree}\4F^{\ifree} P_\Pindex(\theta,\7R)$ with
complex numbers $k^\Pindex_{\ifree}$,
i.e., they involve the
representatives $\4F^\ifree$
of $H(\susy,\inv)$.
These functions give rise to solutions 
of complete non-supersymmetric 
Poincar\'e invariant descent equations in four dimensions 
only if they are of the form
$\4F^{\ifree}\6P^{(4)}(\theta,\7R)/\6\7C^{\ifree}$
where $P^{(4)}(\theta,\7R)$ can depend on all
$\theta$'s except for those Abelian $\7C$'s which are not
contained in $\{\7C^{\ifree}\}$, see section 13 of \cite{report} and
section 8 of \cite{BBH2} (and also the brief discussion in
item (2,3) above).
Proceeding as in item (2,3), one obtains
\bea
\lefteqn{
\ext\Big[
\4F^\ifree\,
\frac{\6P^{(4)}(\7q,\7C_\free,\Lth,\7R)}{\6\7C^{\ifree}}
\Big]
=
}
\nonumber\\
&\displaystyle{
\4F^\ifree\Big[
\cF^\jfree\,\frac{\6}{\6\7C^{\jfree}}
+\cH^\Index\,\frac{\6}{\6\7R^\Index}
+\!\!\!\sum_{\Lindex:m(\Lindex)=2}\!\!\! 
\TR(\cF^2)\,\frac{\6}{\6\7q_\Lindex}
\Big]
\frac{\6P^{(4)}(\7q,\7C_\free,\Lth,\7R)}{\6\7C^{\ifree}}\, .
}&
\quad
\label{4.1}
\eea
Notice that the first term on the right hand side actually
contains only the
antisymmetrized products $\4F^{[\ifree}\cF^{\jfree]}$
because $\6^2P^{(4)}/\6\7C^{\jfree}\6\7C^{\ifree}$ is
antisymmetric in $\ifree$ and $\jfree$
owing to the odd Grassmann parity of the $\7C$'s.
Using lemma \ref{prosusy} one concludes that the
$\ext$-closed
functions $\4F^{[\ifree}\cF^{\jfree]}$,
$\4F^\ifree\cH^\Index$ and $\4F^\ifree\TR(\cF^2)$ 
which occur in (\ref{4.1}) are
$\ext$-exact in $\inv$:
$\4F^\ifree\cH^\Index$ and $\4F^\ifree\TR(\cF^2)$ have
ghost  number 5 and 6, respectively, and are therefore
$\ext$-exact in $\inv$ by the result of lemma \ref{prosusy}
for $g\geq 5$; the $\ext$-exactness of
$\4F^{[\ifree}\cF^{\jfree]}$ in $\inv$ is seen using
arguments as in the text after Eq.\ (\ref{3.1}):
$\4F^{[\ifree}\cF^{\jfree]}$
has ghost number 4 and dimension 0 and is quadratic in
the $\7T$'s of the super-Yang-Mills multiplets which
implies that it cannot be equivalent to a function
$\cP\Omega_1+\5\cP\5\Omega_2$ because $\Omega_1$ or
$\Omega_2$ would have to be proportional to
$\7\lambda^{[\ifree}\7\lambda^{\jfree]}$ which
vanishes.
Hence lemma \ref{prosusy} implies
that there are functions $X^{[\ifree\jfree]}$,
$X^{\ifree\Index}$ and $X^\ifree_\Lindex$ such that
\bea
\ba{ll}
\4F^{[\ifree}\cF^{\jfree]}=-\ext X^{[\ifree\jfree]},&
X^{[\ifree\jfree]}\in\inv\, ,
\\
\4F^\ifree\cH^\Index=-\ext X^{\ifree\Index},&
X^{\ifree\Index}\in\inv\, ,
\\
\4F^\ifree\TR(\cF^2)=-\ext X^\ifree_\Lindex\, ,&
X^\ifree_\Lindex\in\inv\, .
\ea
\label{4.2}
\eea
Such functions are explicitly 
given in the lemma. 
Using the same reasoning that led us to Eqs.\ (\ref{2.3}) and
(\ref{3.4}) one concludes from
(\ref{4.1}) and (\ref{4.2}) by means of part (i) of lemma \ref{prouseful}
that $f^{(4)}$ is $\ext$-closed:
\bea
\ext f^{(4)}=0.
\label{4.3}
\eea

\item[(5)] 
The representatives $f^{(5)}$ arise from functions
$k^\Pindex_{\Index\Jndex}\4H^\Index\4H^\Jndex 
P_{\Pindex}(\theta,\7R)$ with
complex numbers $k^\Pindex_{\Index\Jndex}$,
i.e., they involve the
representatives $\4H^\Index\4H^\Jndex$
of $H(\susy,\inv)$.
In order to give rise to Poincar\'e invariant solutions of the complete
non-supersymmetric descent equations, these
functions must not involve Abelian $\7C$'s because
otherwise the lifts of the BRST invariant 2-forms
$k^\Pindex_{\Index\Jndex}(\star H^\Index)(\star H^\Jndex) 
P_{\Pindex}(\theta_C,C_\Abel,\Lth,\7R)$ to 
Poincar\'e invariant solutions of the descent equations
were obstructed by the 4-forms
$k^{\Pindex}_{\Index\Jndex}(\star H^\Index)(\star H^\Jndex) F^{i_A}
\6P_{\Pindex}(\theta_C,C_\Abel,\Lth,\7R)/\6C^{i_A}$.
This leaves us in this case with functions
$\4H^\Index\4H^\Jndex P^{(5)}_{[\Index\Jndex]}(\theta,\7R)$
where $P^{(5)}_{[\Index\Jndex]}(\theta,\7R)$ does not depend on
Abelian $\7C$'s. The antisymmetry of 
$P^{(5)}_{[\Index\Jndex]}$ in $\Index$ and $\Jndex$
is due to the fact that the $\4H$'s anticommute since they are
Grassmann odd. Substituting the $\7q$'s for the corresponding
$\theta$'s and computing the $\ext$-variation of the resultant
function gives
\bea
\lefteqn{
\ext\Big[
\4H^\Index\4H^\Jndex P^{(5)}_{[\Index\Jndex]}(\7q,\Lth,\7R)
\Big]
}
\nonumber\\
&&=\4H^\Index\4H^\Jndex\Big[
\cH^\Kdex\,\frac{\6}{\6\7R^\Kdex}
+\!\!\!\sum_{\Lindex:m(\Lindex)=2}\!\!\! 
\TR(\cF^2)\,
\frac{\6}{\6\7q_\Lindex}
\Big]
P^{(5)}_{[\Index\Jndex]}(\7q,\Lth,\7R).
\label{5.1}
\eea
The functions $\4H^\Index\4H^\Jndex\cH^\Kdex$ and
$\4H^\Index\4H^\Jndex\,\TR(\cF^2)$ are $\ext$-closed
elements of $\inv$ with ghost numbers 5 and 6, respectively.
Using
lemma \ref{prosusy} we conclude that they are $\ext$-exact in $\inv$.
Hence there are functions $X^{[\Index\Jndex]\Kdex}$
and $X^{[\Index\Jndex]}_\Lindex$ such that
\bea
\4H^\Index\4H^\Jndex\cH^\Kdex=-\ext X^{[\Index\Jndex]\Kdex},\quad
\4H^\Index\4H^\Jndex\,\TR(\cF^2)=-\ext X^{[\Index\Jndex]}_\Lindex,\quad
X^{[\Index\Jndex]\Kdex},X^{[\Index\Jndex]}_\Lindex\in\inv\, .
\label{5.2}
\eea
Such functions are explicitly given in the lemma.
Using (\ref{5.2}), we obtain:
\bea
\lefteqn{
\4H^\Index\4H^\Jndex\Big[
\cH^\Kdex\,\frac{\6}{\6\7R^\Kdex}
+\!\!\!\sum_{\Lindex:m(\Lindex)=2}\!\!\! 
\TR(\cF^2)\,
\frac{\6}{\6\7q_\Lindex}
\Big]
P^{(5)}_{[\Index\Jndex]}(\7q,\Lth,\7R)
}
\nonumber\\
&&=
-\Big[
(\ext X^{[\Index\Jndex]\Kdex})\,\frac{\6}{\6\7R^\Kdex}
+\!\!\!\sum_{\Lindex:m(\Lindex)=2}\!\!\! 
(\ext X^{[\Index\Jndex]}_\Lindex)\,\frac{\6}{\6\7q_\Lindex}
\Big]P^{(5)}_{[\Index\Jndex]}(\7q,\Lth,\7R)
\nonumber\\
&&=
-\ext\Big[X^{[\Index\Jndex]\Kdex}\,
\frac{\6P^{(5)}_{[\Index\Jndex]}(\7q,\Lth,\7R)}{\6\7R^\Kdex}
+\!\!\!\sum_{\Lindex:m(\Lindex)=2}\!\!\! 
X^{[\Index\Jndex]}_\Lindex\,
\frac{\6P^{(5)}_{[\Index\Jndex]}(\7q,\Lth,\7R)}{\6\7q_\Lindex}
\Big]+Z^{(5)},\quad
\label{5.3}
\eea
where
\bea
Z^{(5)}=
X^{[\Index\Jndex]\Kdex}\,\ext\,
\frac{\6P^{(5)}_{[\Index\Jndex]}(\7q,\Lth,\7R)}{\6\7R^\Kdex}
-\!\!\!\sum_{\Lindex:m(\Lindex)=2}\!\!\! 
X^{[\Index\Jndex]}_\Lindex\,\ext\,
\frac{\6P^{(5)}_{[\Index\Jndex]}(\7q,\Lth,\7R)}{\6\7q_\Lindex}\ .
\label{5.4}
\eea
$Z^{(5)}$ contains only terms with $\7c$-degrees 3 and 4
because $X^{[\Index\Jndex]\Kdex}$ and $X^{[\Index\Jndex]}_\Lindex$
contain only terms with $\7c$-degrees $\geq 2$ and the $\ext$-variations
in (\ref{5.4}) contain only terms with $\7c$-degrees $\geq 1$.
Furthermore $Z^{(5)}$ is $\ext$-closed because it is $\ext$-exact
as one sees from (\ref{5.3}) and
(\ref{5.1}). According to
part (ii) of lemma \ref{prouseful}, $Z^{(5)}$ is thus the
$\ext$-variation of a function with $\7c$-degree 4 
whose $\delta_-$-variation
is equal to the part of $Z^{(5)}$ with $\7c$-degree 3.
Explicitly one obtains
\bea
Z^{(5)}=-\ext\Big[
X^{[\Index\Jndex](\Kdex\Ldex)}\,
\frac{\6^2P^{(5)}_{[\Index\Jndex]}(\7q,\Lth,\7R)}{\6\7R^\Ldex\6\7R^\Kdex}
\Big],
\label{5.5}
\eea
with $X^{[\Index\Jndex](\Kdex\Ldex)}$ as given in the lemma.
(\ref{5.1}), (\ref{5.3}) and (\ref{5.5}) show that
$f^{(5)}$ is $\ext$-closed:
\bea
\ext f^{(5)}=0.
\label{5.6}
\eea

\item[(6)]
The representatives $f^{(6)}$ arise from functions
$(\cO R^\Pindex(\7T))P_\Pindex(\theta,\7R)$, i.e., they involve
representatives $\cO R(\7T)$
of $H(\susy,\inv)$.
Substituting the $\7q$'s for the corresponding
$\theta$'s and computing the $\ext$-variation of the resultant
function gives
\bea
\lefteqn{
\ext\Big[ (\cO R^\Pindex(\7T))\,P_\Pindex(\7q,\7C_\Abel,\Lth,\7R)\Big]
=
}
\nonumber\\
&\displaystyle{
-(\cO R^\Pindex(\7T))
\Big[
\cF^{i_A}\,\frac{\6}{\6\7C^{i_A}}
+\cH^\Index\,\frac{\6}{\6\7R^\Index}
+\!\!\!\sum_{\Lindex:m(\Lindex)=2}\!\!\! 
\TR(\cF^2)\,\frac{\6}{\6\7q_\Lindex}
\Big]P_\Pindex(\7q,\7C_\Abel,\Lth,\7R).
}&\quad\quad
\label{6.1}
\eea
$(\cO R^\Pindex(\7T))\cF^{i_A}$, $(\cO R^\Pindex(\7T))\cH^\Index$ and
$(\cO R^\Pindex(\7T))\,\TR(\cF^2)$ are $\ext$-closed elements
of $\inv$ with ghost numbers 5, 6 and 7, respectively.
According to lemma \ref{prosusy} they are thus $\ext$-exact
in $\inv$, i.e. there are functions 
$X^{\Pindex i_A}$, $X^{\Pindex\Index}$ and $X^\Pindex_\Lindex$ such that
\bea
\ba{ll}
(\cO R^\Pindex(\7T))\cF^{i_A}=\ext X^{\Pindex i_A},&
X^{\Pindex i_A}\in\inv\, ,
\\
(\cO R^\Pindex(\7T))\cH^\Index=\ext X^{\Pindex\Index},&
X^{\Pindex\Index}\in\inv\, ,
\\
(\cO R^\Pindex(\7T))\,\TR(\cF^2)=\ext X^\Pindex_\Lindex,&
X^\Pindex_\Lindex\in\inv\, .
\ea
\label{6.2}
\eea
Such functions are explicitly given in the lemma.
Using the same reasoning that led us to Eqs.\ (\ref{2.3}) and
(\ref{3.4}) one concludes from
(\ref{6.1}) and (\ref{6.2}) by means of part (i) of lemma \ref{prouseful}
that $f^{(6)}$ is $\ext$-closed:
\bea
\ext f^{(6)}=0.
\label{6.3}
\eea

\item[(7)]
The representatives $f^{(7)}$ arise from functions
$(\cP\Omega^\Pindex_1+\5\cP\5\Omega^\Pindex_2)
P_\Pindex(\theta,\7R)$, i.e., they involve
representatives $(\cP\Omega_1+\5\cP\5\Omega_2)$
of $H(\susy,\inv)$.
Proceeding as in item (6) we obtain
\bea
\lefteqn{
\ext\Big[ (\cP\Omega^\Pindex_1+\5\cP\5\Omega^\Pindex_2)\,
P_\Pindex(\7q,\7C_\Abel,\Lth,\7R)\Big]
=
}
\nonumber\\
&\displaystyle{
(\cP\Omega^\Pindex_1+\5\cP\5\Omega^\Pindex_2)
\Big[
\cF^{i_A}\,\frac{\6}{\6\7C^{i_A}}
+\cH^\Index\,\frac{\6}{\6\7R^\Index}
+\!\!\!\sum_{\Lindex:m(\Lindex)=2}\!\!\! 
\TR(\cF^2)\,\frac{\6}{\6\7q_\Lindex}
\Big]P_\Pindex(\7q,\7C_\Abel,\Lth,\7R).
}&\quad\quad
\label{7.1}
\eea
(\ref{7.1}) contains
only terms with $\7c$-degrees 3 and 4 because
$\cP\Omega^\Pindex_1+\5\cP\5\Omega^\Pindex_2$, $\cF^{i_A}$, $\cH^\Index$
and $\TR(\cF^2)$ contain only terms with
$\7c$-degrees $\geq 2$, 1, 1 and 2, respectively. The
terms with $\7c$-degree 2 in $\cP\Omega_1+\5\cP\5\Omega_2$
are $4\Ii(\vartheta\vartheta\Omega_1-\5\vartheta\5\vartheta\5\Omega_2)$,
the terms with $\7c$-degree 1 in
$\cF^{i_A}$ and $\cH^\Index$ are 
$-\Ii(\vartheta\7\lambda+\5\vartheta\hatlambda)^{i_A}$ and
$-2\Theta\phi^\Index$, respectively.
Owing to $\Theta=\vartheta\xi=\5\xi\5\vartheta$ and
$\vartheta^\alpha\vartheta^\beta\vartheta^\gamma=
\5\vartheta^\da\5\vartheta^\dbe\5\vartheta^\dg=0$ (the latter holds
since the $\vartheta^\alpha$ and $\5\vartheta^\da$ anticommute)
the terms with $\7c$-degree 3 in (\ref{7.1}) are
\[
f_3=4(\vartheta\vartheta\Omega^\Pindex_1\5\vartheta\hatlambda{}^{i_A}
-\5\vartheta\5\vartheta\5\Omega^\Pindex_2\vartheta\7\lambda^{i_A})\,
\frac{\6P_\Pindex(\theta_C,R)}{\6C^{i_A}}\ .
\]
Part (ii) of lemma \ref{prouseful} implies thus that
(\ref{7.1}) is the $\ext$-variation of the function with
$\7c$-degree 4 whose $\delta_-$-variation equals
$f_3$. Using (\ref{id1}) it is easy to identify this function:
\begin{align}
f_3&=-\delta_-\Big[ 8\,\Xi\,(\hatlambda{}^{i_A}\5\xi\,\Omega^\Pindex_1
    +\xi\7\lambda^{i_A}\5\Omega^\Pindex_2)\,
    \frac{\6P_\Pindex(\theta_C,R)}{\6C^{i_A}}\Big]
\nonumber\\
  &=-\delta_-\Big[ 8\,\Xi\,(\hatlambda{}^{i_A}\5\xi\,\Omega^\Pindex_1
    +\xi\7\lambda^{i_A}\5\Omega^\Pindex_2)\,
    \frac{\6P_\Pindex(\7q,\7C_\Abel,\Lth,\7R)}{\6\7C^{i_A}}\Big] .
\label{7.2}
\end{align}
Hence we conclude from (\ref{7.1}), (\ref{7.2}) and part (ii)
of lemma \ref{prouseful} that $f^{(7)}$ is $\ext$-closed:
\bea
\ext f^{(7)}=0.
\label{7.3}
\eea

\item[(8)]
The cocycles which arise from functions
$k^\Pindex_{i_A}\cF^{i_A} P_\Pindex(\theta,\7R)$
containing the representatives $\cF^{i_A}$
of $H(\susy,\inv)$
are equivalent to linear combinations of
representatives $f^{(6)}$ and $f^{(7)}$, as will be now shown.
The lift of the BRST-invariant 2-form
$F^{i_A} P_{i_A}(\theta_C,\Lth,R)$
to a Poincar\'e invariant solution 
of complete non-supersymmetric 
descent equations in four dimensions
is obstructed by the 4-form $F^{i_A}F^{j_A} \6_{j_A}
P_{i_A}(\theta_C,\Lth,R)$ where $\6_{i_A}=\6/\6C^{i_A}$.
This obstruction is absent only if $\6_{(j_A}
P_{i_A)}(\theta_C,\Lth,R)=0$. The general solution
to this condition is $P_{i_A}(\theta_C,\Lth,R)=
\6_{i_A}P(\theta_C,\Lth,R)$, see section 13.2.2 of \cite{report}.
Hence functions
$k^\Pindex_{i_A}\cF^{i_A} P_\Pindex(\theta,\7R)$
can give rise to $\ext$-cocycles only if
they are of the form
$\cF^{i_A}\6P(\theta,\7R)/\6\7C^{i_A}$.
Proceeding as in the other cases, one obtains
\bea
\lefteqn{
\ext\Big[
\cF^{i_A}\,
\frac{\6P(\7q,\7C_\Abel,\Lth,\7R)}{\6\7C^{i_A}}
\Big]
=
}
\nonumber\\
&\displaystyle{
\cF^{i_A}\Big[
\cH^\Index\,\frac{\6}{\6\7R^\Index}
+\!\!\!\sum_{\Lindex:m(\Lindex)=2}\!\!\! 
\TR(\cF^2)\,\frac{\6}{\6\7q_\Lindex}
\Big]
\frac{\6P(\7q,\7C_\Abel,\Lth,\7R)}{\6\7C^{i_A}}\, .
}&
\quad
\label{8.1a}
\eea
The functions $\cF^{i_A}\cH^\Index$ and $\cF^{i_A}\,\TR(\cF^2)$
are $\ext$-coycles in $\inv$ with ghost numbers 5 and 6, respectively,
and thus $\ext$-exact in $\inv$ according to lemma \ref{prosusy}.
Hence there are functions $X^{i_A\Index}$ and $X^{i_A}_\Lindex$
such that
\bea
\cF^{i_A}\cH^\Index=-\ext X^{i_A\Index},\quad
\cF^{i_A}\,\TR(\cF^2)=-\ext X^{i_A}_\Lindex,\quad
X^{i_A\Index},X^{i_A}_\Lindex\in\inv.
\label{8.1}
\eea
Owing to $\cH^\Index=-(1/2)\cO\phi^\Index$,
the first equation in (\ref{8.1}) is just a special case
of the first equation in (\ref{6.2}). $X^{i_A\Index}$
can thus be obtained from $X^{\Pindex i_A}$  
by choosing $R^\Pindex(\cT)\equiv \phi^\Index/2$.
The computation of $X^{i_A}_\Lindex$ is very similar to
the computation of the functions $X^{\ifree}_\Lindex$
which satisfy the last equation in (\ref{4.2}).
One obtains
\begin{align}
X^{i_A\Index}&=-\Ii\,\Xi_\mu\,\7\lambda^{i_A}\sigma^\mu\5\xi\,\phi^\Index
+\Xi\,(\sfrac 14 \phi^\Index\cD\7\lambda^{i_A}
+\7\lambda^{i_A}\7\psi^\Index)-\CC,
\nonumber\\
X^{i_A}_\Lindex&=
-\Ii\,\Xi\,[\hatlambda{}^{i_A}\5\xi\,\TR(\7\lambda\7\lambda)
+2\7\lambda^{i_A\alpha}\,\TR(\7\lambda_\alpha\hatlambda\5\xi)]+\CC
\label{8.2}
\end{align}
Proceeding as
in items (2,3) and (4), one obtains $\ext$-cocycles $f^{(8)}$:
\bea
\ext f^{(8)}=0,\quad
f^{(8)}=\Big[
\cF^{i_A}+X^{i_A\Index}\,\frac{\6}{\6\7R^\Index}
+\!\!\!\sum_{\Lindex:m(\Lindex)=2}\!\!\! 
X^{i_A}_\Lindex\,\frac{\6}{\6\7q_\Lindex}
\Big]
\frac{\6P(\7q,\7C_\Abel,\Lth,\7R)}{\6\7C^{i_A}}\ .
\label{8.3}
\eea
To show that $f^{(8)}$ is equivalent to a linear combination
of representatives $f^{(6)}$ and $f^{(7)}$, as we have asserted above,
we shall use the following relation:
\bea
\TR(\cF^2)=-\sfrac{\Ii}8\cP\,\TR(\7\lambda\7\lambda)
+\sfrac{\Ii}8\5\cP\,\TR(\hatlambda\hatlambda)
+\ext \left[\Xi_\mu\,\TR(\7\lambda\sigma^\mu\hatlambda)\right].
\label{8.4}
\eea
Furthermore
we recall that one has
\bea
\lefteqn{
\ext P(\7q,\7C_\Abel,\Lth,\7R)=
}
\nonumber\\
&&
\Big[
\cF^{i_A}\frac{\6}{\6\7C^{i_A}}
+\cH^\Index\,\frac{\6}{\6\7R^\Index}
+\!\!\!\sum_{\Lindex:m(\Lindex)=2}\!\!\! 
\TR(\cF^2)\,\frac{\6}{\6\7q_\Lindex}
\Big]
P(\7q,\7C_\Abel,\Lth,\7R).
\label{8.5}
\eea
(\ref{8.3}), (\ref{8.4}) and (\ref{8.5}) give [one
may use 
part (i) of lemma \ref{prouseful} to verify this]:
\bea
\lefteqn{
f^{(8)}-
\ext\Big[ P(\7q,\7C_\Abel,\Lth,\7R)-
\!\!\!\sum_{\Lindex:m(\Lindex)=2}\!\!\!
\Xi_\mu\,\TR(\7\lambda\sigma^\mu\hatlambda)\,
\frac{\6P(\7q,\7C_\Abel,\Lth,\7R)}{\6\7q_\Lindex}
\Big]
}
\nonumber\\
&=
\displaystyle{
\Big[
-\cH^\Index
+X^{i_A\Index}\,\frac{\6}{\6\7C^{i_A}}
+\!\!\!\sum_{\Lindex:m(\Lindex)=2}\!\!\!
Y^\Index_\Lindex\,\frac{\6}{\6\7q_\Lindex}
+Y^{\Index\Jndex}\,\frac{\6}{\6\7R^\Jndex}
\Big]
\frac{\6P(\7q,\7C_\Abel,\Lth,\7R)}{\6\7R^\Index}
}
&
\nonumber\\
&
\displaystyle{
+\!\!\!\sum_{\Lindex:m(\Lindex)=2}\!\!\!
\Big[\{\sfrac{\Ii}8\cP\,\TR(\7\lambda\7\lambda)
-\sfrac{\Ii}8\5\cP\,\TR(\hatlambda\hatlambda)\}
+Y^{i_A}_\Lindex\,\frac{\6}{\6\7C^{i_A}}
\Big]
\frac{\6P(\7q,\7C_\Abel,\Lth,\7R)}{\6\7q_\Lindex} ,
}
&\quad
\label{8.6}
\eea
where
\beann
&
Y^\Index_\Lindex=4\,\Xi\,\phi^\Index\,\TR(\xi\lambda\,\hatlambda\5\xi),
\quad
Y^{\Index\Jndex}=\Ii\,\Xi\,\phi^\Jndex
(\xi\7\psi^\Index-\hatpsi{}^\Index\5\xi),
&
\\
&
Y^{i_A}_\Lindex=\Ii\,\Xi\,\{
\hatlambda{}^{i_A}\5\xi\,\TR(\7\lambda\7\lambda)
-\xi\7\lambda^{i_A}\,\TR(\hatlambda\hatlambda)\}.
&
\eeann
Recall that $\{P_\Pindex(\theta,\7R)\}$ denotes a basis
for the monomials in the $\theta$'s and $\7R$'s.
Hence one has 
\[
\frac{\6P(\theta,\7R)}{\6\7R^\Index}=
k^\Pindex_\Index P_\Pindex(\theta,\7R),\quad
\frac{\6P(\theta,\7R)}{\6\theta_\Lindex}
=k^{\Lindex\Pindex}P_\Pindex(\theta,\7R)
\]
for some complex numbers $k^\Pindex_\Index$ and $k^{\Lindex\Pindex}$.
(\ref{8.6}) shows explicitly that
$f^{(8)}$ is indeed equivalent to a linear combination of representatives 
$f^{(6)}$ and $f^{(7)}$ arising from the particular choices
\[
R^\Pindex(\7T)=\sfrac 12 k^\Pindex_\Index\phi^\Index,\quad
\Omega_1^\Pindex=\!\!\!\sum_{\Lindex:m(\Lindex)=2}\!\!\!
\sfrac{\Ii}8 k^{\Lindex\Pindex}\,\TR(\7\lambda\7\lambda),\quad
\5\Omega_2^\Pindex
=-\!\!\!\sum_{\Lindex:m(\Lindex)=2}\!\!\!
\sfrac{\Ii}8 k^{\Lindex\Pindex}\,\TR(\hatlambda\hatlambda).
\]

\item[(9)]
To illustrate the derivation of the $X$-functions,
let me finally describe in some detail 
how one derives $X^\Index_\Lindex$
satisfying (\ref{2.0}). 
(\ref{s<3}) and (\ref{cF2}) yield
\bea
\4H^\Index\,\TR(\cF^2)=
(\Ii\hatpsi\5\xi-\Ii\xi\7\psi+\7c^\mu \7H_\mu)^\Index\,
\TR(\7F-\Ii\vartheta\7\lambda-\Ii\5\vartheta\hatlambda)^2.
\label{2.4}
\eea
We start from the terms with lowest $\7c$-degree in this expression.
They have $\7c$-degree 2 and are given by
\bea
&&
-\Ii(\hatpsi\5\xi-\xi\7\psi)^\Index\,
\TR(\vartheta\7\lambda+\5\vartheta\hatlambda)^2
\nonumber\\
&&=
-\Ii(\hatpsi\5\xi-\xi\7\psi)^\Index\,
[-\sfrac 12\vartheta\vartheta\,\TR(\7\lambda\7\lambda)
+2\vartheta^\alpha\5\vartheta^\da\,\TR(\7\lambda_\alpha\hatlambda_\da)
-\sfrac 12\5\vartheta\5\vartheta\,\TR(\hatlambda\hatlambda)].
\quad
\label{2.5}
\eea
(\ref{2.5}) is $\delta_-$-closed since it is the term of
lowest $\7c$-degree of an $\ext$-cocycle.
The $\delta_-$-cohomology (\ref{delta-})
shows that it contains precisely
two terms which are not $\delta_-$-exact.
These are $(\Ii/2)\hatpsi{}^\Index\5\xi
\vartheta\vartheta\,\TR(\7\lambda\7\lambda)$
and
$(-\Ii/2)\xi\7\psi^\Index\5\vartheta\5\vartheta\,\TR(\hatlambda\hatlambda)$
which are easily seen to be $\5b$-exact and $b$-exact, respectively
(owing to $b\phi=\xi\7\psi$,
$\5b\phi=\hatpsi\5\xi$, $b\hatlambda=\5b\7\lambda=0$).
Using that one has $\ext f=(\delta_-+b+\5b+\7c^\mu\7\nabla_\mu)f$
for $f\in\inv$, we can thus write these terms as
\bea
&\sfrac {\Ii}2\hatpsi{}^\Index\5\xi\vartheta\vartheta\,
\TR(\7\lambda\7\lambda)
=(\ext-b-\7c^\mu\7\nabla_\mu)\,
[\sfrac {\Ii}2\vartheta\vartheta\phi^\Index\,\TR(\7\lambda\7\lambda)],
&
\nonumber\\
&-\sfrac {\Ii}2\xi\7\psi^\Index
\5\vartheta\5\vartheta\,\TR(\hatlambda\hatlambda)=
(\ext-\5b-\7c^\mu\7\nabla_\mu)\,
[-\sfrac {\Ii}2\5\vartheta\5\vartheta\phi^\Index\,\TR(\hatlambda\hatlambda)].
&
\label{2.6}
\eea
Using (\ref{2.6}) in (\ref{2.4}), one obtains
\bea
&&\4H^\Index\,\TR(\cF^2)-\ext X^\Index_{\Lindex,2}=
f^\Index_{\Lindex,2}+f^\Index_{\Lindex,3}+f^\Index_{\Lindex,4}
\label{2.7a}\\
&&X^\Index_{\Lindex,2}=
\sfrac {\Ii}2\vartheta\vartheta\phi^\Index\,\TR(\7\lambda\7\lambda)
-\sfrac {\Ii}2\5\vartheta\5\vartheta\phi^\Index\,\TR(\hatlambda\hatlambda),
\label{2.7}
\eea
where $f^\Index_{\Lindex,2}$, $f^\Index_{\Lindex,3}$ and
$f^\Index_{\Lindex,4}$ have $\7c$-degree
2, 3 and 4, respectively, and $f^\Index_{\Lindex,2}$ is $\delta_-$-exact,
i.e., there is some $X^\Index_{\Lindex,3}$ such that
\bea
\delta_-X^\Index_{\Lindex,3}=f^\Index_{\Lindex,2}\, .
\label{2.8a}
\eea
[Of course,
$X^\Index_{\Lindex,3}$ is only defined up to
the $\delta_-$-variation of some function
with $\7c$-degree 4.]
Explicitly one obtains:
\begin{align}
f^\Index_{\Lindex,2}&=
\Ii\,\xi\7\psi^\Index\,
[-\sfrac 12\vartheta\vartheta\,\TR(\7\lambda\7\lambda)
+2\vartheta^\alpha\5\vartheta^\da\,\TR(\7\lambda_\alpha\hatlambda_\da)]
-\sfrac {\Ii}2\vartheta\vartheta\, b\,[\phi^\Index\,\TR(\7\lambda\7\lambda)]
+\CC
\nonumber\\
&=
\Ii\,\xi\7\psi^\Index\,
[-\vartheta\vartheta\,\TR(\7\lambda\7\lambda)
+2\vartheta^\alpha\5\vartheta^\da\,\TR(\7\lambda_\alpha\hatlambda_\da)]
-\sfrac {\Ii}2\vartheta\vartheta\, \phi^\Index\,\xi\cD\,
\TR(\7\lambda\7\lambda)
+\CC
\nonumber\\
X^\Index_{\Lindex,3}&=
-\Ii\,\Xi_\mu( \5\xi\5\sigma^\mu\7\psi^\Index
+\sfrac 12 \phi^\Index\,\5\xi\5\sigma^\mu\cD)\,
\TR(\7\lambda\7\lambda)
-\Ii\,\Xi_\mu\,\xi\7\psi^\Index\,\TR(\7\lambda\sigma^\mu\hatlambda)
+\CC
\quad
\label{2.8}
\end{align} 
Using (\ref{2.8a}) in (\ref{2.7a}), one obtains
\bea
\4H^\Index\,\TR(\cF^2)-\ext(X^\Index_{\Lindex,2}+X^\Index_{\Lindex,3})=
\4f^\Index_{\Lindex,3}+\4f^\Index_{\Lindex,4}\, ,
\label{2.9}
\eea
where $\4f^\Index_{\Lindex,3}=f^\Index_{\Lindex,3}
-(b+\5b)X^\Index_{\Lindex,3}$ and
$\4f^\Index_{\Lindex,4}=f^\Index_{\Lindex,4}
-\7c^\mu\7\nabla_\mu X^\Index_{\Lindex,3}$.
(\ref{2.9}) is an
$\ext$-closed sum of
terms with $\7c$-degrees 3 and 4. According to
part (ii) of lemma \ref{prouseful} it is thus the
$\ext$-variation of a function $X^\Index_{\Lindex,4}$ with
$\7c$-degree 4 that
fulfills $\delta_-X^\Index_{\Lindex,4}=\4f^\Index_{\Lindex,3}$.
We conclude that
\bea
\4H^\Index\,\TR(\cF^2)-\ext(X^\Index_{\Lindex,2}+X^\Index_{\Lindex,3})=
\ext X^\Index_{\Lindex,4}\, ,
\label{2.9a}
\eea
where $\delta_-X^\Index_{\Lindex,4}=\4f^\Index_{\Lindex,3}$.
The explicit computation gives
\bea
X^\Index_{\Lindex,4}=\Xi\,(
\sfrac 12\7\psi^\Index\cD
+\sfrac 18\phi^\Index\cD^2
)\,
\TR(\7\lambda\7\lambda)
-\sfrac 12\Xi\,\7H_\mu^\Index\,\TR(\7\lambda\sigma^\mu\hatlambda)+\CC
\label{2.10}
\eea
(\ref{2.9a}) yields $X^\Index_\Lindex=
X^\Index_{\Lindex,2}+X^\Index_{\Lindex,3}+X^\Index_{\Lindex,4}$.
\QED
\een


\providecommand{\href}[2]{#2}\begingroup\raggedright\endgroup

\end{document}